\documentclass[10pt,final,journal,twocolumn]{IEEEtran}

\usepackage{times}
\usepackage{epsfig}
\usepackage{graphicx}
\usepackage{amsmath}
\usepackage{amssymb}
\usepackage{booktabs}
\usepackage{algorithm}
\usepackage{algpseudocode}
\usepackage{multirow}
\usepackage{makecell}
\usepackage[dvipsnames]{xcolor}

\newtheorem{thm}{\textbf{Theorem}}[section]
\newtheorem{mydef}{\textbf{Definition}}[section]

\begin{document}

\title{Framelet Representation of Tensor Nuclear Norm for Third-Order Tensor Completion
\thanks{This work is supported by the National Natural Science Foundation of China (61772003, 61876203, and 61702083), HKRGC GRF (12306616, 12200317, 12300218, and 12300519), HKU Grant (104005583), and China Postdoctoral Science Foundation  (2017M610628 and 2018T111031).}
\thanks{T.-X. Jiang is with FinTech Innovation Center, Financial Intelligence and Financial Engineering Research Key Laboratory of Sichuan province, School of Economic Information Engineering, Southwestern University of Finance and Economics, Chengdu, Sichuan, China (e-mail: taixiangjiang@gmail.com). }
\thanks{Michael K. Ng is with Department of Mathematics, The University of Hong Kong, Pokfulam, Hong Kong (e-mail: mng@maths.hku.hk).}
\thanks{X.-L. Zhao, and Ting-Zhu Huang are with the Research Center for Image and Vision Computing, School of Mathematical Sciences, University of Electronic Science and Technology of China, Chengdu 611731, P.R.China (e-mail: xlzhao122003@163.com; tingzhuhuang@126.com).}
}

\author{Tai-Xiang Jiang,
Michael K. Ng,
Xi-Le Zhao,
and Ting-Zhu Huang}
\maketitle
\begin{abstract}
The main aim of this paper is to develop a framelet representation of the tensor nuclear norm for
third-order tensor recovery.
In the literature, the tensor nuclear norm can be computed by using tensor
singular value decomposition based on the discrete Fourier transform matrix, and tensor completion can be performed by the minimization of the tensor nuclear norm which is the relaxation of
the sum of matrix ranks from all Fourier transformed matrix frontal slices.
These Fourier transformed matrix frontal slices are obtained by applying the discrete Fourier transform
on the tubes of the original tensor.
In this paper, we propose to employ the framelet representation of each tube
so that a framelet transformed tensor can be constructed. Because of framelet basis redundancy,
the representation of each tube is sparsely represented. When the matrix slices of
the original tensor are highly correlated, we expect the corresponding
sum of matrix ranks from all framelet transformed matrix frontal slices would be small,
and the resulting tensor completion can be performed much better.
The proposed minimization model is convex and global minimizers can be obtained.
Numerical results on several types of multi-dimensional data (videos, multispectral images, and magnetic resonance imaging data) have tested and
shown that the proposed method outperformed the other testing methods.
\end{abstract}

\section{Introduction}
As a high order extension of matrix, the tensor is an important data format for multi-dimensional data applications, such as color image and video processing \cite{sobral2017matrix,Korah2007TIP,Liu2013PAMItensor}, hyperspectral data recovery and fusion \cite{yang2020remote,dian2019hyperspectral,deng2019fusion}, personalized web search \cite{Sun2005web,lima2017cellular}, high-order web link analysis \cite{Kolda2005Datamining}, magnetic resonance imaging (MRI) data recovery \cite{MRITV}, and seismic data reconstruction \cite{Kreimer2012HSVDtensor}.
Owing to the objective restrictions, for example, the imaging condition for the visual data acquiring and the limitation of the transmission bandwidth, the multi-dimensional data in many applications are incomplete or grossly corrupted.
This motivates us to perform tensor completion \cite{Liu2013PAMItensor} or tensor robust principal component analysis (RPCA) \cite{yang2020low}, in which how to characterize and utilize the internal structural information of these multidimensional data is of crucial importance.

For the matrix processing, low-rank models can effectively and efficiently handle two-dimensional data of various sources \cite{candes2012exact,candes2011robust}.
Generalized from matrix format, a tensor is able to contain more essentially structural information, being a powerful tool for dealing with multi-modal and multi-relational data \cite{song2016sublinear}.
Unfortunately, it is not easy to directly extend the low-rankness from the matrix to tensors.
More precisely, there is not an exact (or unique) definition for the tensor's rank.
In the past decades, the most popular rank definitions are the CANDECOMP/PARAFAC (CP)-rank \cite{acar2011scalable,tichavsky2017numerical} and the Tucker-rank \cite{li2018low,li2017mr} (or denoted as ``$n$-rank'' in \cite{gandy2011tensor}).
The CP-rank is based on the CP decomposition, however, computing the CP-rank of a given tensor is NP-hard \cite{hillar2013most}. The Tucker-rank is based on the Tucker decomposition, in which the tensor is unfolded along each mode unavoidably destroying the intrinsic structures of the tensor.

In this paper, we investigate the newly emerged tensor rank definitions, i.e., the tensor multi-rank and the tensor tubal-rank, which are computable and induced from the tensor singular value decomposition (t-SVD). The t-SVD is initially proposed by Braman {\em et al.} \cite{braman2010third} and Kilmer {\em et al.} \cite{kilmer2011factorization}, based on the tensor-tensor product (denoted as t-prod), in which the third-order tensors are operated integrally avoiding the loss of information inherent in matricization or flattening of the tensor \cite{kilmer2013third}. Meanwhile, the t-SVD has shown its superior performance in capturing the spatial-shifting correlation that is ubiquitous in real-world data \cite{martin2013order,braman2010third,kilmer2011factorization}.
Although the t-SVD is initially designed for third-order tensors, it has been extended to high order tensors with arbitrary dimensions \cite{martin2013order,zheng2019mixed}.

In \cite{kernfeld2015tensor}, Kernfeld {\em et al.} note
that the t-prod is based on a convolution-like operation, which can be implemented using the discrete Fourier transform (DFT).
Then, given a third-order tensor $\mathcal{X}\in\mathbb{R}^{n_1\times n_2\times n_3}$, its Fourier transformed (along the third mode) tensor is denoted as $\widehat{\mathcal{X}}\in\mathbb{R}^{n_1\times n_2\times n_3}$ and its tensor multi-rank is a vector with the $i$-th element equal to the rank of $i$-th frontal slice of $\widehat{\mathbf{\mathcal{X}}}$ \cite{zhang2014novel}. The tensor nuclear norm (TNN) of $\mathcal{X}$ is subsequently defined and it equals to the sum of the nuclear norm of $\widehat{\mathbf{\mathcal{X}}}$'s frontal slices and is the relaxation of the sum of matrix ranks from all $\widehat{\mathbf{\mathcal{X}}}$'s slices.
By minimizing the TNN, Zhang {\em et al.} \cite{zhang2014novel} build the low-rank tensor completion model and provided theoretical performance bounds for third-order tensor recovery from limited sampling.
Lu {\em et al.} \cite{lu2016tensor} utilize the TNN\footnote{In \cite{lu2016tensor}, the TNN is defined with a factor $1/n_3$.} for the tensor RPCA. Similar researches, which adopt the TNN for multi-dimensional data recovery, can be found in \cite{jiang2017exact,lu2019tensor,hu2017twist}.

Other than the Fourier transform, Kernfeld {\em et al.} find that the t-prod, together with the tensor decomposition scheme, can be defined via any invertible transform, for instance, the discrete cosine transform (DCT).
Namely, the t-prod can be implemented by the matrices' product after the invertible transformation along the third mode.
Xu {\em et al.} \cite{xu2018cosine} validate that, when minimizing the DCT based TNN for the tensor completion problem, the DCT is superior to the DFT in terms of the preservation of the head and the tail frontal slices, because of its mirror boundary condition. Corroborative results can be found in \cite{lu2019low,lu2019exact2}, which demonstrates that any invertible linear transform can be applied to induce the TNN for the tensor completion task.
Coincidentally, Song {\em et al.} \cite{song2019robust} find that the corresponding transformed tubal-rank could be approximately smaller with an appropriate unitary transform, for instance, the Haar wavelet transform, and they prove that one can recover a low transformed tubal-rank tensor exactly with overwhelming probability provided that its transformed tubal rank is sufficiently small and its corrupted entries are reasonably sparse.

The tensor data recovery within the t-SVD framework can be viewed as finding a low-rank approximation in the transformed domain. Therefore, if the transformed tensor could be approximately lower-rank, minimizing the corresponding TNN, namely the TNN defined based on the transformation, would be more effective for the recovery \cite{song2019robust}. In \cite{song2019robust,lu2019low,lu2019exact2}, the authors establish elegant theoretical results based on the unitary transform or the invertible linear transform. However, the requirement of the invertibility prevents their results from other non-invertible (or semi-invertible) transformations, which could bring in redundancy. We note that redundancy in the transformation is important
as such transformed coefficients can contain information of missing data in the original domain, see
for example the work by Cai {\em et al.} \cite{cai10}.

In this paper, we suggest to use the tight wavelet frame (framelet) as the transformation within the t-SVD framework. Because of framelet basis redundancy, the representation of each tube is sparsely represented.
We expect when each matrix slices of the original tensor, the corresponding sum of matrix ranks from all framelet transformed matrix slices would be small.
As an example, we illustrate this motivation by using magnetic resonance image (MRI) of size $142 \times 178 \times 121$, multispectral image (MSI) of size $512 \times 512 \times 31$ and video data of size $144 \times 176 \times 100$ to demonstrate their rank reduction via framelet transformation\footnote{The piece-wise cubic B-spline is used to generate framelet system.} to the Fourier transformation.
Note that for real imaging data, each transformed matrix frontal slice is not an exact low-rank matrix, but it is close to a low-rank matrix. There are many small singular values of each transformed matrix frontal slice.
We show in Table \ref{lowerrank} that the mean value of the matrix ranks of $\mathcal{X}(:,:,i)$ (the $i$-th transformed matrix frontal slice).
Here we discard the singular values of transformed matrix frontal slice when they are smaller than the truncation parameter, and the truncated rank of transformed matrix slice is obtained.
It is clear that the mean value of such truncated matrix ranks by using framelet transformation is lower than that by using the Fourier transformation.
When a framelet transformed tensor is close to a low-rank tensor compared with the use of the Fourier transform, it is expected that the resulting tensor completion can be performed much better in practice.
The framelet based TNN (F-TNN) minimization models are subsequently formulated for the low-rank tensor completion (LRTC) and tensor RPCA. The proposed minimization models are convex and global minimizers can be obtained via the alternating direction multipliers method (ADMM) \cite{boyd2011distributed} with a theoretical convergence guarantee.
We conduct numerical experiments on various types of multi-dimensional imaging data and the results verify that our framelet based method outperforms the compared methods.

\begin{table}[!t]\label{lowerrank}
\renewcommand\arraystretch{0.9}\setlength{\tabcolsep}{4pt}\scriptsize\centering
\caption{The mean value of all the truncated transformed matrix slices ranks by using
the FFT and the framelet transform for MRI, MSI and video data sets.}
\begin{tabular}{ccccccc}
\toprule

\multirow{3}{*}{Data} &\multirow{3}{*}{Parameter $\epsilon$} &FFT & Framelet & \multirow{3}{*}{Reduction}\\
                                        &          &Multi-rank  & Multi-rank & \\
                                        &          &(mean value) & (mean value) & \\\midrule
\multirow{3}{*}{MRI}    & 0.02  & 101.0  & 77.8  & 23.3  \\
                        & 0.01  & 120.1  & 94.1  & 25.9  \\
                        & 0.005 & 131.9  & 108.9  & 23.0  \\\midrule
\multirow{3}{*}{Video}  & 0.02  & 106.7  & 74.5  & 32.2  \\
                        & 0.01  & 122.7  & 92.2  & 30.5  \\
                        & 0.005 & 132.6  & 108.5  & 24.1  \\\midrule
\multirow{3}{*}{MSI}    & 0.02  & 83.8  & 46.1  & 37.7  \\
                        & 0.01  & 132.8  & 77.8  & 55.0  \\
                        & 0.005 & 218.0  & 136.0  & 82.0  \\
 \bottomrule
\end{tabular}
\label{MRIpara}
\end{table}
%
%

\subsection{Contributions}
The main contributions can be summarised as follows.
\textbf{(i)} We suggest the framelet transform within the t-SVD framework and proposed a tensor completion model, which minimizes the framelet representation of the tensor nuclear norm.
\textbf{(ii)} To tackle the non-invertible framelet transform based models, we develop alternating direction multipliers method (ADMM) based algorithms with guaranteed convergence, and we test our method on various types of multi-dimensional data. The outperformance of our method further corroborates the usage of framelet.

The outline of this paper is given as follows.
In Section \ref{Sec:Pre}, some preliminary background on tensors and the framelet is given.
The main results, including the  proposed model and algorithm, are presented in Section \ref{Sec:Model}.
Experimental results are reported in Section \ref{Sec:Exp}.
Finally, Section \ref{Sec:Con} draws some conclusions.

\section{Preliminaries}\label{Sec:Pre}
This section provides the basic ingredients to induce the proposed method. We firstly give the basic tensor notations and then introduce the t-SVD framework, which has been proposed in \cite{kilmer2013third,kilmer2011factorization,zhang2014novel,lu2016tensor}. We restate them here at the readers' convenience.
Next, the basics of framelet are briefly presented.

\subsection{Tensor Notations And Definitions}

Generally, a third-order tensor is denoted as $\mathbf{\mathcal{X}}\in \mathbb{R}^{n_{1}\times n_2\times n_{3}}$, and $x_{i,j,k}$ is its $(i,j,k)$-th component.
We use $\mathcal{X}^{(k)}$ or $\mathcal{X}(:,:,k)$ to denote the $k$-th frontal slice of a third-order tensor $\mathbf{\mathcal{X}}\in\mathbb{R}^{n_1\times n_2\times n_3}$.

{\begin{mydef}[tensor mode-3 unfolding and folding \cite{kolda2009tensor}]
The mode-$3$ unfolding of a tensor $\mathcal{X}\in\mathbb{R}^{n_1\times n_2\times n_3}$ is denoted as a matrix $\mathbf X_{(3)}\in \mathbb{R}^{n_3\times n_1n_2}$,
where the tensor's $(i,j,k)$-th element maps to the matrix's $(k,l)$-th element satisfying {$l=(j-1)n_1+i$}.
The mode-3 unfolding operator and its inverse are respectively denoted as ${\tt{unfold}}_3$ and ${\tt{fold}}_3$, and they satisfy $\mathcal{X}={\tt fold}_{3}({\tt unfold}_{3}(\mathcal{X})) = {\tt fold}_{3}(\mathbf X_{(3)})$.
\end{mydef}}

{
\begin{mydef}[mode-3 tensor-matrix product \cite{kolda2009tensor}] The mode-3 tensor-matrix product of a tensor $\mathcal{X} \in \mathbb{R}^{n_1\times n_2\times n_3}$ with a matrix $\mathbf A\in\mathbb{R}^{m\times n_3}$ is denoted by ${\mathcal{X}}\times_3\mathbf A$ and is of size $n_1\times n_2\times m$.
Elementwise, we have
\begin{equation}
(\mathcal{X}\times_3 \mathbf A)_{i,j,k}=\sum_{n=1}^{n_3}x_{i,j,n}\cdot a_{k,n}.
\end{equation}
The mode-3 tensor-matrix product can also be expressed in terms of the mode-3 unfolding
\begin{equation}\nonumber
\mathcal{Y}=(\mathcal{X}\times_3 \mathbf A)\quad \Leftrightarrow \quad \mathbf Y_{(3)}=\mathbf A\cdot\text{unfold}_3(\mathcal{X}).
\end{equation}
\end{mydef}}

The one-dimensional DFT on a vector ${\mathbf{x}}\in\mathbb{R}^n$, denoted as $\mathbf{\bar{x}}$, is given by $\mathbf{\bar{x}} = \mathbf F_n\mathbf x \in \mathbb{C}^n$,
where $\mathbf F_n\in\mathbb{C}^{n\times n}$ is the DFT matrix.
In this paper, we use $\widehat{\mathcal{X}}$ to denote the transformed tensor by performing one-dimensional DFT along the mode-3 fibers (tubes) of $\mathcal{X}$. By using the DFT matrix $\mathbf{F}_{n_3}\in\mathbb{C}^{n_3\times n_3}$, we have $$
\widehat{\mathbf{\mathcal{X}}} =\mathcal{X}\times_3\mathbf F_{n_3} = {\tt fold}_3\left( \mathbf{F}_{n_3}{\tt unfold}_3(\mathcal{X})\right)\in \mathbb{C}^{n_{1}\times n_2\times n_{3}}.$$

\begin{mydef}[tensor conjugate transpose \cite{kilmer2013third}]
The conjugate transpose of a tensor $\mathbf{\mathcal{A}}\in \mathbb{C}^{n_{2}\times n_1\times n_{3}}$ is tensor $\mathbf{\mathcal{A}}^\text{\rm H}\in \mathbb{C}^{n_{1}\times n_2\times n_{3}}$ obtained by conjugate transposing each of the frontal slice and then reversing the order of transposed frontal slices 2 through $n_3$, i.e., $
\left(\mathbf{\mathcal{A}}^\text{\rm H}\right)^{(1)}=\left(\mathbf{\mathcal{A}}^{(1)}\right)^\text{\rm H}$ and $\left(\mathbf{\mathcal{A}}^\text{\rm H}\right)^{(i)}=\left(\mathbf{\mathcal{A}}^{(n_3+2-i)}\right)^\text{\rm H}$ ($i=2,\cdots,n_3$).
\end{mydef}


\begin{mydef}[t-prod \cite{kilmer2013third}]\label{Def:1}
The tensor-tensor-product (t-prod) $\mathbf{\mathcal{C}}=\mathbf{\mathcal{A}}*\mathbf{\mathcal{B}}$
of $\mathbf{\mathcal{A}}\in \mathbb{R}^{n_{1}\times n_2\times n_{3}}$ and $\mathbf{\mathcal{B}}\in \mathbb{R}^{n_{2}\times n_4\times n_{3}}$ is a tensor of size
$n_1\times n_4 \times n_3$, where the $(i,j)$-th tube $\mathbf{c}_{ij:}$ is given by
\begin{equation}
\mathbf{c}_{ij:} = \mathbf{\mathcal{C}}(i,j,:) = \sum_{k=1}^{n_2}\mathbf{\mathcal{A}}(i,k,:)*\mathbf{\mathcal{B}}(k,j,:)
\end{equation}
where $*$ denotes the circular convolution between two tubes of same size.
\label{def:tprod}
\end{mydef}

\begin{mydef}[identity tensor \cite{kilmer2013third}]\label{Def:2}
The identity tensor $\mathbf{\mathcal{I}}\in \mathbb{R}^{n_{1}\times n_1\times n_{3}}$ is the tensor whose first frontal slice is the $n_1\times n_1$ identity matrix, and whose other frontal slices are all
zeros.
\end{mydef}


\begin{mydef}[orthogonal tensor \cite{kilmer2013third}]\label{Def:3}
A tensor $\mathbf{\mathcal{Q}} \in \mathbb{C}^ {n_{1} \times n_1\times n_{3}}$  is orthogonal if it satisfies
\begin{equation}
\mathbf{\mathcal{Q}}^\text{\rm H}*\mathbf{\mathcal{Q}}=\mathbf{\mathcal{Q}}*\mathbf{\mathcal{Q}}^\text{\rm H}=\mathbf{\mathcal{I}}.
\end{equation}
\end{mydef}


\begin{mydef}[f-diagonal tensor \cite{kilmer2013third}]\label{Def:1}
A tensor $\mathbf{\mathcal{A}}$ is called f-diagonal if each frontal slice $\mathbf{\mathcal{A}}^{(i)}$ is a diagonal matrix.
\end{mydef}

\begin{figure}[hbtp]
  \centering
  \includegraphics[width=0.85\linewidth]{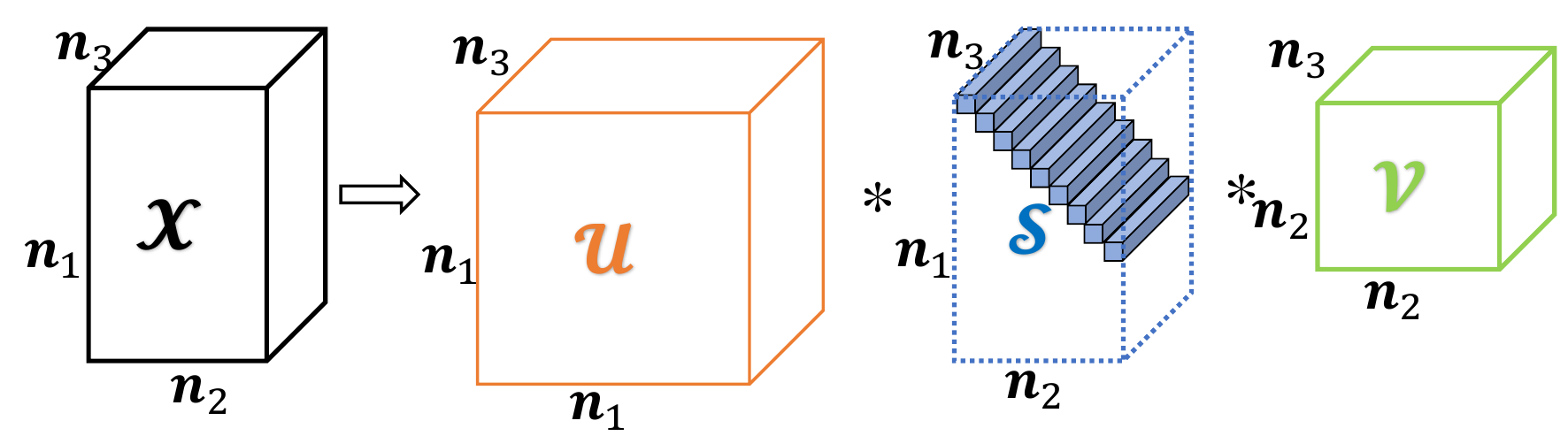}
  \caption{The t-SVD of an $n_1 \times n_2 \times n_3$ tensor.}
  \label{tsvd}
\end{figure}

\begin{thm}[t-SVD \cite{kilmer2013third,kilmer2011factorization}]
For $\mathbf{\mathcal{A}}\in \mathbb{R}^{n_{1}\times n_2\times n_{3}}$, the t-SVD of $\mathbf{\mathcal{A}}$ is given by
\begin{equation}
\mathbf{\mathcal{A}}=\mathbf{\mathcal{U}}*\mathbf{\mathcal{S}}*\mathbf{\mathcal{V}}^\text{\rm H}
\end{equation}
where $\mathbf{\mathcal{U}}\in \mathbb{R}^{n_{1}\times n_1\times n_{3}}$ and $\mathbf{\mathcal{V}}\in \mathbb{R}^{n_{2}\times n_2\times n_{3}}$ are orthogonal tensors, and $\mathbf{\mathcal{S}}\in \mathbb{R}^{n_{1}\times n_2\times n_{3}}$ is an f-diagonal tensor.
\end{thm}
The t-SVD is illustrated in Figure \ref{tsvd}.

\begin{mydef}[tensor tubal-rank and multi-rank \cite{zhang2014novel}]\label{Def:tubal}
The tubal-rank of a tensor $\mathbf{\mathcal{A}}\in\mathbb{R}^{n_1\times n_2\times n_3}$, denoted as $\text{rank}_t(\mathbf{\mathcal{A}})$, is defined to be the number of non-zero singular tubes of $\mathbf{\mathcal{S}}$, where $\mathbf{\mathcal{S}}$ comes from the t-SVD of $\mathbf{\mathcal{A}}$: $\mathbf{\mathcal{A}}=\mathbf{\mathcal{U}}*\mathbf{\mathcal{S}}*\mathbf{\mathcal{V}}^\top$.
That is
\begin{equation}
\text{rank}_t(\mathbf{\mathcal{A}})=\#\{i:\mathbf{\mathcal{S}}(i,:,:)\neq0\}.
\end{equation}

The tensor multi-rank of $\mathbf{\mathcal{A}}\in\mathbb{R}^{n_1\times n_2\times n_3}$ is a vector, denoted as $\text{rank}_r (\mathcal{A})\in\mathbb{R}^{n_3}$, with the $i$-th element equals to the rank of $i$-th frontal slice of $\widehat{\mathbf{\mathcal{A}}}$.
\end{mydef}

\begin{mydef}[block diagonal form \cite{zhang2014novel}]\label{Def:bldg}
Let $\overline{\mathbf{\mathcal{A}}}$ denote the block-diagonal matrix of the tensor $\widehat{\mathbf{\mathcal{A}}}$
in the Fourier domain, i.e.,
\begin{equation}
\begin{aligned}
\overline{\mathcal{A}}&\triangleq {\tt blockdiag}(\widehat{\mathbf{\mathcal{A}}})\\
&\triangleq
\left [
\begin{tabular}{cccc}
$\widehat{\mathbf{\mathcal{A}}}^{(1)}$&&&\\
&$\widehat{\mathbf{\mathcal{A}}}^{(2)}$ &&\\
&&$\ddots$ &\\
&&&$\widehat{\mathbf{\mathcal{A}}}^{(n_3)}$
\end{tabular}\right]
\in\mathbb{C}^{n_1n_3\times n_2n_3},
\end{aligned}
\end{equation}
\end{mydef}
where $\widehat{\mathcal{A}}^{(k)}=\widehat{\mathcal{A}}(:,:,k)$ is the $k$-th slice of $\widehat{\mathcal{A}}$ for $k = 1,2,\cdots,n_3$.

It is not difficult to find that $\overline{\mathcal{{A}}^\text{\rm H}}=\overline{\mathcal{{A}}}^\text{\rm H}$, i.e.,  the block diagonal form of a tensor's conjugate transpose equals to the matrix conjugate transpose of the tensor's block diagonal form. Further more, for any tensor $\mathbf{\mathcal{A}}\in \mathbb{R}^{n_{1}\times n_2\times n_{3}}$ and $\mathbf{\mathcal{B}}\in \mathbb{R}^{n_{2}\times n_4\times n_{3}}$, we have
\[
\mathbf{\mathcal{A}}*\mathbf{\mathcal{B}}=\mathbf{\mathcal{C}} \Leftrightarrow \overline{\mathcal{A}}\cdot\overline{\mathcal{B}}=\overline{\mathcal{{C}}},
\]
where $\cdot$ is the matrix product.

\begin{mydef}[tensor-nuclear-norm (TNN) \cite{zhang2014novel}]
The tensor nuclear norm of a tensor $\mathbf{\mathcal{A}}\in \mathbb{R}^{n_{1}\times n_2\times n_{3}}$, denoted as $\|\mathbf{\mathcal{A}}\|_{\text{\rm TNN}}$, is defined as
\begin{equation}
\begin{aligned}
\|\mathbf{\mathcal{A}}\|_{\text{TNN}}\triangleq\|\overline{\mathbf{\mathcal{A}}}\|_{*},
\end{aligned}
\label{tnn}
\end{equation}
\end{mydef}
where $\|\cdot\|_*$ refers to the matrix nuclear norm. For a matrix $\mathbf X\in\mathbb{C}^{m\times n}$, $\|\mathbf X\|_* = \sum_i^{\min\{m,n\}}\sigma_i$, where $\sigma_i$ is the $i$-th singular value of $\mathbf X$.
The TNN can be computed via the summation of the matrix nuclear norm of Fourier transformed tensor's slices, which are also the blocks of $\overline{\mathbf{\mathcal{A}}}$. That is
$\|\mathbf{\mathcal{A}}\|_{\text{TNN}}=\sum\limits_{i=1}^{n_3}\|\widehat{\mathbf{\mathcal{A}}}^{(i)}\|_*$.

We summary the frequent used notations in Table \ref{notations}.
\begin{table}[htbp]\label{notations}
\renewcommand\arraystretch{1.3}\setlength{\tabcolsep}{2pt}
\caption{Tensor notations}
 \begin{tabular}{p{0.27\columnwidth} p{0.70\columnwidth}}
  \toprule
Notation &  Explanation \\
  \midrule
$\mathbf{\mathcal{X}},\mathbf{X},\mathbf{x},x$
                                & Tensor, matrix, vector, scalar.\\
\multirow{2}{*}{$*$}
                                & The tensor-tensor product or the circular convolution between vectors.\\
\multirow{2}{*}{$\mathcal{X}(:,:,k)$ (or $\mathcal{X}^{(k)}$)}         & The $k$-th frontal slice of a third-order tensor $\mathbf{\mathcal{X}}\in\mathbb{R}^{n_1\times n_2\times n_3}$.\\
${\tt fold}_3$ (${\tt unfold}_3$)
                                & The fold (or unfold) operation along the third mode.\\
$\mathbf{{X}}_{(3)}$
                                & The mode-3 unfolding of a tensor $\mathbf{\mathcal{X}}$.\\
$\widehat{\mathcal{X}}$
                                & The Fourier transformed (along the third mode) tensor.\\
\multirow{2}{*}{$\text{rank}_r (\mathcal{A})$}
                                & The multi-rank of a tensor $\mathbf{\mathcal{X}}$ and its $i$-th element equals to $\text{rank}(\widehat{\mathcal{X}}^{(k)})$.\\
\multirow{2}{*}{$\left\|\mathbf{\mathcal{X}}\right\|_\text{TNN}$}
                                & The tensor nuclear norm of a tensor $\mathbf{\mathcal{X}}$ and it equals to the sum of the nuclear norms of $\widehat{\mathcal{X}}$'s slices. \\
\bottomrule
 \end{tabular}
 \end{table}
\subsection{Framelet}\label{Framelet}
A tight frame is defined as a countable set $X\subset L_2(\mathbb{R}$) with the property that $\forall f\in L_2(\mathbb{R})$,
$f=\sum\limits_{g\in X}\langle f,g\rangle.$
This is equivalent to that $\forall f\in L_2(\mathbb{R})$, we have
\begin{equation*}
\Vert{f}\Vert_{L_2(\mathbb{R})}^2=\sum\limits_{g\in X}\vert\langle f,g\rangle\vert^2,
\end{equation*}
where $\langle \cdot ,\cdot\rangle$ is the inner product in $L^2(\mathbb{R})$, and $\Vert \cdot \Vert_{L^2(\mathbb{R})}=\langle \cdot ,\cdot\rangle^ \frac {1}{2}$.

For given $\Psi :=\{\psi_1,\psi_2,\cdots,\psi_r\}\subset L^2(\mathbb{R})$, the affine (or wavelet) system is defined by the collection of the dilations and the shifts of $\Psi$ as
$X(\Psi):=\{\psi_{l,j,k} : 1\le l\le r; j,k\in\mathbb{Z}\}$,
where $\psi_{l,j,k}:=2^{j/2}\psi_l(2^j\cdot \textnormal{-}k)$.
When $X(\Psi)$ forms a tight frame of $L^2(\mathbb{R})$, it is called a tight wavelet frame, and $\psi_l,l=1,2,\cdots,r$ are called the (tight) framelets.
In the numerical scheme of image processing, the framelet transform (decomposition operator) of a vector $\mathbf v\in \mathbb{R}^{n}$ can be represented by a matrix $\mathbf W \in\mathbb{R}^{wn\times n}$ is the framelet transform matrix constructed with $n$ filters and $l$ levels and $w=(n-1)l+1$. The processes of generating such matrices have been detailed in many
literatures such as \cite{cai10,JiangFramelet}.
We omit them here for readability.
Then the framelet transform of a discrete signal $\mathbf v\in \mathbb{R}^{n}$, can be written as $\mathbf u=\mathbf W\mathbf v \in \mathbb{R}^{wn}$.
Besides, the unitary extension principle (UEP) \cite{ron1997affine} asserts that $\mathbf W^\top\mathbf W\mathbf v=\mathbf v$, where $\mathbf W^\top $ indicates the inverse framelet transform. However, $\mathbf W\mathbf W^\top\mathbf u\neq\mathbf u$.
\vspace{-2mm}

\section{Main results}\label{Sec:Model}

In this section, we replace the Fourier transform by the framelet transform. The starting point of our idea is that the framelet transform would bring in redundancy and the transformed data is of lower multi-rank.
Then, we build the LRTC model and tensor RPCA model based on the framelet representation of the tensor nuclear norm and propose the ADMM based algorithms to optimize these models.
\vspace{-2mm}

\subsection{From DFT to The Framelet Transform}
For a three way tensor $\mathcal{X}\in\mathbb{R}^{n_1\times n_2 \times n_3}$, owing to the circular convolution in Def. \ref{def:tprod}, its t-SVD can be efficiently computed via the DFT.
Computing the one-dimensional DFT of a vector of length $n$ by using the DFT matrix costs $O(n^2)$, and the computational cost can be reduced to O($n\log n$) by employing the fast Fourier transform (FFT) technique \cite{bergland1969guided}.
Using the DFT matrix, for a tensor $\mathbf{\mathcal{X}}\in \mathbb{R}^{n_{1}\times n_2\times n_{3}}$, we can obtain its Fourier transformed tensor as
$$
\widehat{\mathbf{\mathcal{X}}} ={\tt fold}_3\left( \mathbf{F}_{n_3}\mathbf{X}_{(3)}\right)\in \mathbb{C}^{n_{1}\times n_2\times n_{3}},
$$
where $\mathbf{X}_{(3)}$ is the mode-3 unfolding of $\mathcal{X}$

Next, we will adopt the framelet transform as a substitute for the Fourier transform, and give the definition of the framelet representation of the tensor nuclear norm.
For simplicity, we denote the tensor after framelet transform along the third mode as
$$\mathcal{X}_{\mathbf W}={\tt fold}_3\left(\mathbf{W}\mathbf{X}_{(3)}\right)\in\mathbb{R}^{n_1\times n_2\times wn_3},$$
where $\mathbf W \in\mathbb{R}^{wn_3\times n_3}$ is the framelet transform matrix constructed with $n$ filters and $l$ levels and $w=(n-1)l+1$.
Considering the UEP property of the framelet transform, we have $\mathcal{X} ={\tt fold}_3(\mathbf{W}^\top{[\mathbf{X}_{\mathbf W}]}_{(3)})$, where ${[\mathbf{X}_{\mathbf W}]}_{(3)}= {\tt unfold}_3\left(\mathcal{X}_\mathbf{W}\right).$

Recalling Def. \ref{Def:tubal}, the tensor multi-rank is defined as a vector of the ranks of the frontal slices in the Fourier transform domain.
Therefore, the framelet based multi-rank is defined in the same manner as follows.

\begin{figure}[!t]
  \centering
  \includegraphics[width=8.1cm,height=5.85cm]{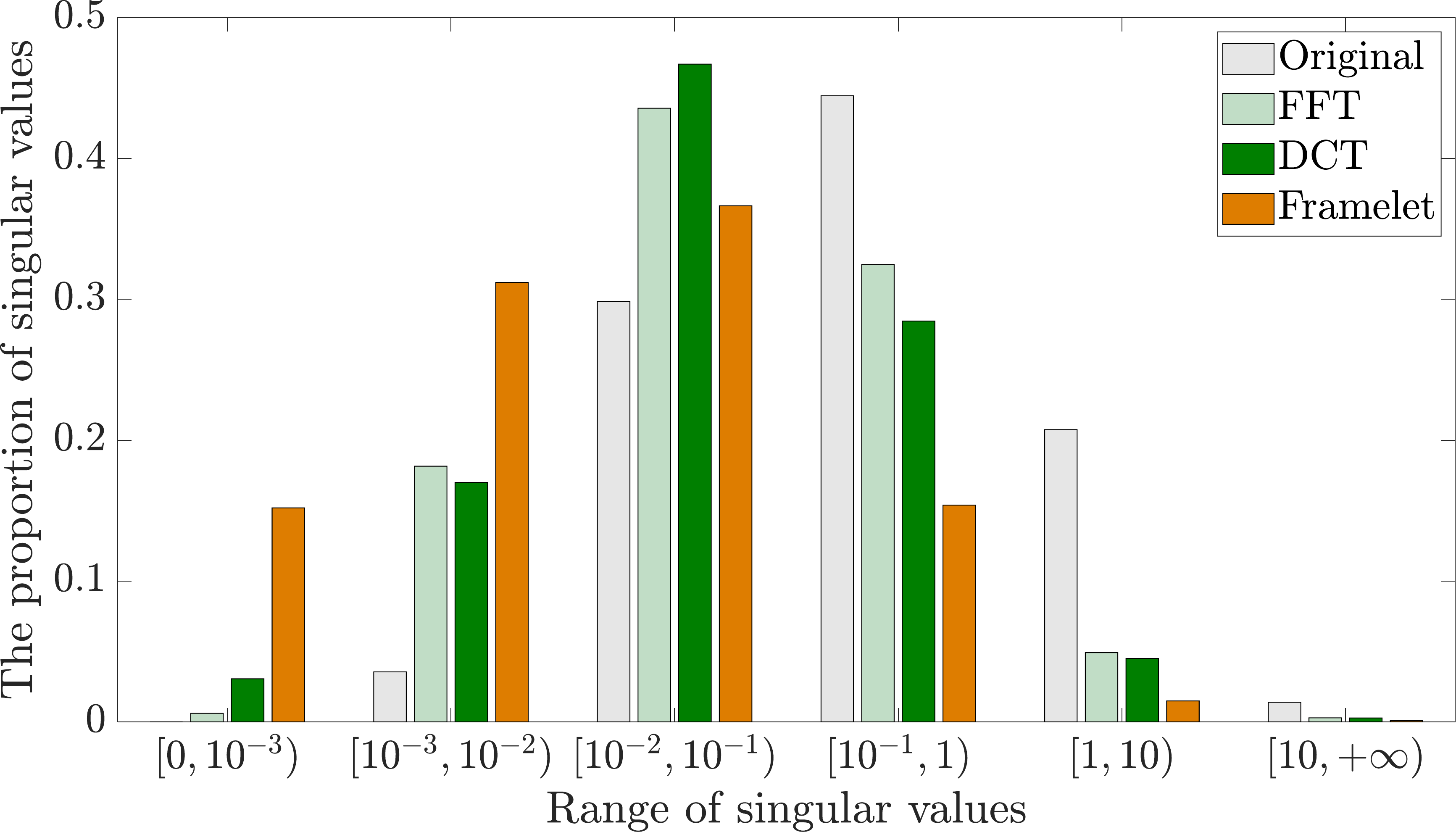}
  \caption{The distribution of singular values. Here, the singular values are obtained by conducting SVD on each frontal slice of the original tensor data or the transformed tensors.}
  \label{sva1}
\end{figure}

\begin{mydef}[Framelet based multi-rank]
The framelet based multi-rank  of a tensor $\mathbf{\mathcal{X}}\in \mathbb{R}^{n_{1}\times n_2\times n_{3}}$ is defined as a vector $\mathbf r_w\in\mathbb{R}^{wn_3}$ with the $i$-th elements $\mathbf r_w(i) = \text{rank}(\mathcal{X}_{\mathbf W}(:,:,i))$ for $i = 1,2,\cdots,wn_3$.
\end{mydef}

Here we have replaced the Fourier transform by the framelet and defined the framelet based multi-rank.
As mentioned before, the framelet transformed tensor can be of lower (framelet based) multi-rank.
To understand this in-depth, we give some empirically numerical analyses on the singular values of the frontal slices of the transformed tensors. Here, taking the video data “news”\footnote{Data available at { http://trace.eas.asu.edu/yuv/}.} as an example, the original video data is denoted as $\mathcal{X}\in\mathbb{R}^{144\times176\times100}$ and its Fourier, DCT, and framelet transformed tensors are denoted as $\widehat{\mathcal{X}}$, $\mathcal{X}_\text{DCT}$\footnote{$\mathcal{X}_\text{DCT}$ is obtained by replacing the DFT with DCT, being similar to $\mathcal{X}_{\mathbf W}$.}, and $\mathcal{X}_{\mathbf W}$, respectively.
In Figure \ref{sva1}, we exhibit the distributions of the singular values of the frontal slices
of $\mathcal{X}$, the Fourier transformed tensors $\widehat{\mathcal{X}}$, the DCT transformed tensor $\mathcal{X}_\text{DCT}$, and
the framelet transformed tensors
$\mathcal{X}_{\mathbf W}$\footnote{The piece-wise cubic B-spline is used to generate framelet system.}.
In Figure \ref{sva1}, we show the proportion of the number of singular values of
transformed matrix frontal slices in each magnitude interval.
It can be found in the figure that a large proportion
of the singular values of the framelet transformed data appears
in the interval of $[0,10^{-2}]$ compared with the original video
data, the Fourier transformed tensor $\widehat{\mathcal{X}}$, and the DCT transformed tensor $\mathcal{X}_\text{DCT}$.
This phenomenon brings in an advantage that the data can be better approximated with lower rank via the framelet representation. In Section \ref{Sec:Exp}, we will illustrate tensor completion and tensor RPCA can be obtained
by using the framelet representation.

\subsection{Framelet Based TNN}

Using the DFT matrix $\mathbf F_{n_3}$, the tensor nuclear norm in \eqref{tnn} of a tensor $\mathcal{X}\in \mathbb{R}^{n_{1}\times n_2\times n_{3}}$ can be expressed as
\begin{equation}
\begin{aligned}
\|\mathbf{\mathcal{X}}\|_{\text{TNN}}&=\|\overline{\mathbf{\mathcal{X}}}\|_{*}=\sum\limits_{i=1}^{n_3}\|\widehat{\mathbf{\mathcal{X}}}^{(i)}\|_*\\
&= \sum\limits_{i=1}^{n_3}\|\left[{\tt fold}_3\left(\mathbf{F}_{n_3}\mathbf{X}_{(3)}\right)\right](:,:,k)\|_*,
\label{TNN}
\end{aligned}
\end{equation}
where $\mathbf{X}_{(3)}$ is the mode-3 unfolding of $\mathcal{X}$.
\begin{mydef}[Framelet based TNN (F-TNN)]
Similarly, the framelet representation of the tensor nuclear norm can be formulated as\end{mydef}
\begin{equation}
\begin{aligned}
\|\mathbf{\mathcal{X}}\|_{\text{F-TNN}}
&=\|{\tt{blockdiag}}(\mathcal{X}_{\mathbf W})\|_*= \sum\limits_{k=1}^{wn_3}\|\mathcal{X}_{\mathbf W}(:,:,k)\|_*\\
&=\sum\limits_{k=1}^{wn_3}\|\left[{\tt fold}_3\left( \mathbf W \mathbf X_{(3)}\right)\right](:,:,k)\|_*,
\end{aligned}\label{FTNN}
\end{equation}
{where $\mathbf W \in\mathbb{R}^{wn_3\times n_3}$ is the framelet transform matrix.}

It is not difficult to obtain that the F-TNN is a convex envelope of the $\ell_1$ norm of the framelet based multi-rank.

\subsection{Tensor Completion via Minimizing F-TNN}

Based on the proposed framelet based TNN, our tensor completion model, which is convex, is formulated as
\begin{equation}
\begin{aligned}
\min\limits_{\mathcal{X}} \quad& \|\mathbf{\mathcal{X}}\|_{\text{F-TNN}}\\
\text{s.t.}\quad& \mathcal{X}_{\Omega}=\mathcal{O}_{\Omega},
\end{aligned}
\label{Model_1}
\end{equation}
where $\mathcal{O}\in\mathbb{R}^{n_1\times n_2 \times n_3}$ is the incomplete observed data, and $\Omega$ is the set of indexes of the observed entries.
$\mathcal{X}_{\Omega}=\mathcal{O}_{\Omega}$ constrains that the entries of $\mathcal{X}$ should agree with $\mathcal{O}$ in $\Omega$.

The next part gives the solving algorithm for our tensor completion model \eqref{Model_1}.
Let
\begin{equation}\centering
\mathcal{I}_\Phi(\mathbf{\mathcal{X}})=\left\{
\begin{aligned}
&0,\quad       & \mathbf{\mathcal{X}}\in\Phi,\\
&\infty, &\text{otherwise},
\end{aligned}
\right.
\end{equation}
where $\Phi :=\{\mathbf{\mathcal{X}}\in\mathbb{R}^{n_1\times n_2 \times n_3}, \mathbf{\mathcal{X}}_{\Omega}=\mathbf{\mathcal{O}}_{\Omega}\}$.

Thus, the problem (\ref{Model_1}) can be rewritten as
\begin{equation}
\min\limits_{\mathbf{\mathcal{X}}} \quad\mathcal{I}_\Phi(\mathbf{\mathcal{X}})+\sum\limits_{k=1}^{wn_3}\|\mathcal{X}_{\mathbf W}(:,:,k)\|_{*}\\
\label{A_un}
\end{equation}
Then, the minimization problem (\ref{A_un}) can be efficiently solved via ADMM \cite{boyd2011distributed}.

After introducing the auxiliary variable $\mathcal{V}\in\mathbb{R}^{n_1\times n2\times wn_3}$, the problem (\ref{A_un}) can be rewritten as the following unconstraint problem
\begin{equation}
\begin{aligned}
\min\limits_{\mathbf{\mathcal{X}}} \quad&\mathcal{I}_\Phi(\mathcal{X}) +\sum\limits_{k=1}^{wn_3}\|\mathcal{V}(:,:,k)\|_{*}\\
\text{s.t.}\quad    & \mathcal{V} = \mathcal{X}_{\mathbf W}.
\end{aligned}
\label{A_AU}
\end{equation}


The augmented Lagrangian function of (\ref{A_AU}) is given by
\begin{equation}
\begin{aligned}
L_\beta(\mathcal{X}, \mathcal{V},\Lambda)=&\mathcal{I}_\Phi(\mathcal{X}) +\sum\limits_{k=1}^{wn_3}\|\mathcal{V}(:,:,k)\|_{*}\\
&+ \frac{\beta}{2}\|\mathcal{X}_{\mathbf{W}}-\mathcal{V}+\frac{\Lambda}{\beta}\|_F^2 \\
 \end{aligned}
 \label{A_AUG}
\end{equation}
where $\Lambda\in\mathbb{R}^{n_1\times n_2\times wn_3}$ is the Lagrangian multiplier, $\beta$ is the penalty parameter for the violation of the linear constraints.
In the scheme of the ADMM, we update each variable alternately.

\textbf{$\mathcal{V}$ sub-problem}:
The $\mathcal{V}$ at $t$-th iteration is
\begin{equation}
\begin{aligned}
\mathcal{V}^{t+1} = &\arg\min\limits_{\mathcal{V}}\  \sum\limits_{k=1}^{wn_3}\|\mathcal{V}(:,:,k)\|_{*}   + \frac{\beta}{2}\|\mathcal{X}_{\mathbf{W}}^t-\mathcal{V}+\frac{\Lambda^t}{\beta}\|_F^2\\
 \end{aligned}
 \label{V1_1}
\end{equation}
Then, \eqref{V1_1} can be decomposed into $wn_3$ subproblems and it is easy to obtain the closed form solution of these sub-problems with the singular value thresholding (SVT) operator \cite{cai2010singular}.
Hence, we update $\mathcal{V}$ as
\begin{equation}
\begin{aligned}
\mathcal{V}^{t+1}(:,:,k) = {\tt SVT}_{\frac{1}{\beta}}\left(\mathcal{X}^{t}_{\mathbf{W}}(:,:,k)+\frac{\Lambda^t(:,:,k)}{\beta}\right),
 \end{aligned}
 \label{V1_2}
\end{equation}
where $k = 1,2\cdots,wn_3$. The complexity of computing $\mathcal{V}$ at each iteration is $O(wn_1n_2n_3 \min(n_1n_2))$.

\textbf{$\mathcal{X}$ sub-problem}:
For convenience, the subproblem of optimizing $L_\beta$ with respect to $\mathcal{X}$ at $t$-th iteration is written in the matrix format as (recalling that $\mathcal{X}_{\mathbf W} = {\tt fold}_3\left(\mathbf{W}\mathbf{X}_{(3)}\right)$)
\begin{equation}
\begin{aligned}
\mathbf{X}^{t+1} = \arg\min\limits_{\mathbf{X}}\ \mathcal{I}_\Phi(\mathcal{X})
+ \frac{\beta}{2}\|\mathbf{WX}-\mathbf{V}^{t+1}_{(3)}+\frac{\Lambda^t_{(3)}}{\beta}\|_F^2,
 \end{aligned}
 \label{X_1}
\end{equation}
where $\mathbf{V}^{t+1}_{(3)} = {\tt unfold}_3(\mathcal{V}^{t+1})$ and $\Lambda^t_{(3)} = {\tt unfold}_3(\Lambda^t)$.
To optimize \eqref{X_1}, we first solve the following equation
\begin{equation}
\begin{aligned}
\mathbf{W^\top W}\mathbf{X}_{(3)} = & \mathbf{W^\top}\left(\mathbf{V}^{t+1}_{(3)}-\frac{\Lambda^t_{(3)}}{\beta}\right).\\
 \end{aligned}
 \label{X_2}
\end{equation}
Thus, considering that $\mathbf{W^\top W}\mathbf{X}_{(3)} = \mathbf{X}_{(3)}$ (the UEP property of the framelet transformation), we have

\begin{equation}
\begin{aligned}
\mathcal{X}^{t+1} = \mathcal{P}_{\Omega^C}\left({\tt fold}_3 (\mathbf{W^\top}(\mathbf{V}^{t+1}_{(3)}-\frac{\Lambda^t_{(3)}}{\beta}))\right)+\mathcal{P}_{\Omega}\left(\mathcal{O}\right),
 \end{aligned}
 \label{X_4}
\end{equation}
where $\mathcal{P}_{\Omega}(\cdot)$ is the projection function that keeps the entries of $\cdot$ in $\Omega$ while making others be zeros, and $\Omega^{c}$ denotes the complementary set of $\Omega$. Meanwhile, we have $\mathcal{X}^{t+1}_{\mathbf W} = {\tt fold}_3(\mathbf{W}\mathbf{X}^{t+1}_{(3)})$. The complexity of computing $\mathcal{X}$ is $O(wn_1n_2n_3^2)$ at each iteration.

\textbf{Updating the multiplier}:
The multiplier $\Lambda$ can be updated by
\begin{equation}
\begin{aligned}
\Lambda^{t+1}  & = \Lambda^{t} +\beta \left(\mathcal{X}^{t+1}_{\mathbf W}-\mathcal{V}^{t+1}\right).\\
\end{aligned}
\label{M_up}
\end{equation}
Updating $\Lambda$ costs $O(wn_1n_2n_3)$ at each iteration.

Finally, our algorithm is summarized in Algorithm \ref{alg}. The total complexity of Algorithm \ref{alg} at each iteration is $O(wn_1n_2n_3(n_3+\min(n_1,n_2)))$.  The objective function of the proposed model in \eqref{Model_1} is convex. Our algorithm fits the standard ADMM framework and its convergence is theoretically guaranteed \cite{boyd2011distributed}.

\begin{algorithm}[htp]
\renewcommand\arraystretch{1.3}
\caption[Caption for LOF]{Tensor completion via minimizing F-TNN}
\begin{algorithmic}[1]
\renewcommand{\algorithmicrequire}{\textbf{Input:}} 
\Require
The observed tensor $\mathcal{O}\in\mathbb{R}^{n_1\times n_2\times n_3}$; Lagrange parameter $\beta$; convergence criteria $\epsilon$; maximum iteration $t_\text{max}$.
\renewcommand{\algorithmicrequire}{\textbf{Initialization:}} 
\Require The framelet transform matrix $\mathbf W$; $\mathcal{V}^{(0)}={\tt fold}_3(\mathbf{W}\mathbf O_{(3)})$;
$\mathcal{X}^{(0)} = \mathcal{O}$; $t = 0$.
\While {not converged and $t<t_\text{max}$}
\State Update $\mathcal{V}^{t+1}$ via Eq. \eqref{V1_2};
\State Update $\mathcal{X}^{t+1}$ via Eq. \eqref{X_4};
\State Update $\Lambda^{t+1}$ via Eq. (\ref{M_up});
\State Check the convergence conditions $\|\mathbf{\mathcal{V}}^{k+1}-\mathbf{\mathcal{V}}^{k}\|_\infty\leq\epsilon$ and $\| \mathbf{\mathcal{X}}^{k+1} -\mathbf{\mathcal{X}}^{k} \|_\infty\leq\epsilon$;
\State $t = t+1$.
\EndWhile
\renewcommand{\algorithmicrequire}{\textbf{Output:}}
\Require The reconstructed tensor $\mathcal{X}$.
\end{algorithmic}
\label{alg}
\end{algorithm}

\subsection{Tensor Robust Principal Components Analysis}
As aforementioned, another typical tensor recovery problem is the tensor RPCA problem, which aims to recover the tensor from
grossly corrupted observations.
Adopting the F-TNN to characterize the low-rank part, our tensor RPCA model is formulated as
\begin{equation}
\begin{aligned}
\min\limits_{\mathcal{L},\mathcal{S}} \quad& \|\mathbf{\mathcal{L}}\|_{\text{F-TNN}}+\lambda\|\mathcal{E}\|_1\\
\text{s.t.}\quad& \mathcal{L} +\mathcal{E}=\mathcal{O},
\end{aligned}
\label{Model_2}
\end{equation}
where $\mathcal{O}\in\mathbb{R}^{n_1\times n_2 \times n_3}$ is the observed data, $\mathcal{E}$ indicates the sparse part, $\|\mathcal{E}\|_1=\sum_{ijk}|\mathcal{E}_{i,j,k}|$, and $\lambda$ is a non-negative parameter.

For convenience, we introduce an auxiliary variable $\mathcal{V}\in\mathbb{R}^{n_1\times n2\times wn_3}$, and reformulate \eqref{Model_2} as
\begin{equation}
\begin{aligned}
\min\limits_{\mathcal{L},\mathcal{S},\mathcal{V}} \quad& \sum\limits_{k=1}^{wn_3}\|\mathcal{V}(:,:,k)\|_{*}+\lambda\|\mathcal{E}\|_1\\
\text{s.t.}\quad& \mathcal{L} +\mathcal{E}=\mathcal{O},\quad \mathcal{V}=\mathcal{L}_\mathbf{W},
\end{aligned}
\label{Model_2}
\end{equation}
where $\mathcal{L}_\mathbf{W}={\tt fold}_3\left(\mathbf{W}\mathbf{L}_{(3)}\right)\in\mathbb{R}^{n_1\times n_2\times wn_3}$ and $\mathbf W \in\mathbb{R}^{wn_3\times n_3}$ is the framelet transform matrix constructed with $n$ filters and $l$ levels ($w=(n-1)l+1$).

Similarly, we adopt ADMM to solve \eqref{Model_2}.
The augmented Lagrangian function of  \eqref{Model_2} is given as
\begin{equation}
\begin{aligned}
L_\beta(\mathcal{L},\mathcal{V},\mathcal{E},\Lambda)\hspace{-.5mm}=\hspace{-.5mm}&\sum\limits_{k=1}^{wn_3}\hspace{-.5mm}\|\mathcal{V}(:,:,k)\|_{*} \hspace{-.5mm}+\hspace{-.5mm}\frac{\beta}{2}\|\mathcal{L}_\mathbf{W}\hspace{-.5mm}-\hspace{-.5mm}\mathcal{V}\hspace{-.5mm}+\hspace{-.5mm}\frac{\Lambda_1}{\beta}\|_F^2\\
&+\lambda\|\mathcal{E}\|_1+ \frac{\beta}{2}\|\mathcal{O}-\mathcal{L}-\mathcal{E}+\frac{\Lambda_2}{\beta}\|_F^2
 \end{aligned}
 \label{A_AUG}
\end{equation}
where $\Lambda_1\in\mathbb{R}^{n_1\times n_2\times wn_3}$ and $\Lambda_2\in\mathbb{R}^{n_1\times n_2\times n_3}$ are the Lagrangian multiplier, and $\beta$ is a nonnegative parameter.
In the scheme of the ADMM, we update each variable alternately as:
\begin{equation}
\left\{\hspace{-1mm}
\begin{aligned}
\mathcal{V}^{t+1} \hspace{-1mm}&=\hspace{-.75mm} \arg\min\limits_{\mathcal{V}} \sum\limits_{k=1}^{wn_3}\|\mathcal{V}(:,:,k)\|_{*}   + \frac{\beta}{2}\|\mathcal{L}_\mathbf{W}^t-\mathcal{V}+\frac{\Lambda_1^t}{\beta}\|_F^2,\\
\mathcal{L}^{t+1} \hspace{-1mm}&=\hspace{-.75mm} \arg\min\limits_{\mathcal{L}}\frac{\beta}{2}\|\mathcal{L}_\mathbf{W}\hspace{-1mm}-\hspace{-1mm}\mathcal{V}^{t\hspace{-.5mm}+\hspace{-.5mm}1}\hspace{-1mm}+\hspace{-1mm}\frac{\Lambda_1^t}{\beta}\|_F^2
\hspace{-1mm}+\hspace{-1mm}\frac{\beta}{2}\|\mathcal{O}\hspace{-1mm}-\hspace{-1mm}\mathcal{L}\hspace{-1mm}-\hspace{-1mm}\mathcal{E}^t\hspace{-1mm}+\hspace{-1mm}\frac{\Lambda^t_2}{\beta}\|_F^2,\\
\mathcal{E}^{t+1} \hspace{-1mm}&=\hspace{-.75mm} \arg\min\limits_{\mathcal{E}}\lambda\|\mathcal{E}\|_1+ \frac{\beta}{2}\|\mathcal{O}-\mathcal{L}^{t+1}-\mathcal{E}+\frac{\Lambda^t_2}{\beta}\|_F^2,\\
\Lambda_1^{t+1} \hspace{-1mm}&=\hspace{-.75mm} \Lambda_1^{t} +\beta \left(\mathcal{L}^{t+1}_{\mathbf W}-\mathcal{V}^{t+1}\right),\\
\Lambda_2^{t+1} \hspace{-1mm}&=\hspace{-.75mm} \Lambda_2^{t} +\beta \left(\mathcal{O}-\mathcal{L}^{t+1}-\mathcal{E}^{t+1}\right).\\
 \end{aligned}\right.
 \label{RPCA-update}
\end{equation}

Specifically, the $\mathcal{V}$ subproblem in \eqref{RPCA-update} can be solved by
\begin{equation}
\mathcal{V}^{t+1}(:,:,k)={\tt SVT}_{\frac{1}{\beta}}\left(\mathcal{L}^{t}_{\mathbf{W}}(:,:,k)+\frac{\Lambda_1^t(:,:,k)}{\beta}\right),
\label{RPCA-update-V}
\end{equation}
for $k=1,2,\cdots,wn_3$. The complexity of updating $\mathcal{V}$ is $O(wn_1n_2n_3 \min(n_1n_2))$ at each iteration. The $\mathcal{L}$ subproblem is a least square problem and its solution can be obtained as
\begin{equation}
\mathcal{L}^{t+1} \hspace{-1mm}=\hspace{-.75mm} \frac{1}{2}{\tt fold}_3 \left(\mathbf{W}^\top(\mathbf{V}^{t+1}_{(3)}\hspace{-.75mm}-\hspace{-.75mm}\frac{{\Lambda^t_1}_{(3)}}{\beta})\right)\hspace{-.75mm}+\hspace{-.75mm} \frac{1}{2}\left(\mathcal{O}\hspace{-.75mm}-\hspace{-.75mm}\mathcal{E}^t\hspace{-.75mm}+\hspace{-.75mm}\frac{\Lambda^t_2}{\beta}\right).
 \label{RPCA-update-L}
\end{equation}
At each iteration, computing $\mathcal{L}$ costs $O(wn_1n_2n_3^2)$.
The $\mathcal{E}$ subproblem can be solved by
\begin{equation}
\mathcal{E}^{t+1}= {\tt Soft}_{\frac{\lambda}{\beta}}\left(\mathcal{O}-\mathcal{L}^{t+1}+\frac{\Lambda^t_2}{\beta}\right),
 \label{RPCA-update-E}
\end{equation}
where ${\tt Soft}_\tau (\cdot)$ is the tensor soft-thresholding operator, and ${\tt Soft}_\tau(\cdot)={\tt{sign}}(\cdot)\max(|\cdot|-\tau,0)$. Computing $\mathcal{E}$ and updating the multipliers $\Lambda_1$ cost $O(wn_1n_2n_3)$ at each iteration. While the computation complexity of updating $\Lambda_2$ is $O(n_1n_2n_3)$.
%

\begin{algorithm}[htp]
\caption[Caption for LOF]{Tensor RPCA via minimizing F-TNN}
\begin{algorithmic}[1]
\renewcommand{\algorithmicrequire}{\textbf{Input:}} 
\Require
The observed tensor $\mathcal{O}\in\mathbb{R}^{n_1\times n_2\times n_3}$; the Lagrange parameter $\beta$; the parameter $\lambda$; convergence criteria $\epsilon$; maximum iteration $t_\text{max}$.
\renewcommand{\algorithmicrequire}{\textbf{Initialization:}} 
\Require the framelet transform matrix $\mathbf W$; $\mathcal{V}^{(0)}={\tt fold}_3(\mathbf{W}\mathbf O_{(3)})$ and $\mathcal{E}^{(0)} = {\tt zeros}(n_1\times n_2\times n_3)$; $t = 0$.
\While {not converged and $t<t_\text{max}$}
\State Update $\mathcal{V}^{t+1}$ via Eq. \eqref{RPCA-update-V};
\State Update $\mathcal{L}^{t+1}$ via Eq. \eqref{RPCA-update-L};
\State Update $\mathcal{E}^{t+1}$ via Eq. \eqref{RPCA-update-E};
\State Update $\Lambda_1$ and $\Lambda_2$ via Eq. (\ref{M_up});
\State Check the convergence conditions $\|\mathbf{\mathcal{V}}^{k+1}-\mathbf{\mathcal{V}}^{k}\|_\infty\leq\epsilon$, $\|\mathbf{\mathcal{L}}^{k+1} -\mathbf{\mathcal{L}}^{k} \|_\infty\leq\epsilon$, and $\|\mathbf{\mathcal{E}}^{k+1} -\mathbf{\mathcal{E}}^{k} \|_\infty\leq\epsilon$;
\State $t = t+1$.
\EndWhile
\renewcommand{\algorithmicrequire}{\textbf{Output:}}
\Require The low-rank component $\mathcal{L}$ and the sparse component $\mathcal{E}$.
\end{algorithmic}
\label{alg2}
\end{algorithm}

The pseudo-code of our algorithm for tensor RPCA is summarized in Algorithm \ref{alg2}. At each iteration of Algorithm \ref{alg2}, it costs $O(wn_1n_2n_3(n_3+\min(n_1,n_2)))$. Likewise, Algorithm \ref{alg2} fits the standard ADMM framework and its convergence is theoretically guaranteed \cite{boyd2011distributed}.

\section{Numerical experiments}\label{Sec:Exp}

In this section, to illustrate the performance of the proposed method, we will exhibit the tensor completion experimental results on three typical kinds of third-order data, i.e., the MRI data, the MSI data, and the video data. Meanwhile, we conduct Three numerical metrics, consisting of the peak signal-to-noise ratio (PSNR), the structural similarity index (SSIM) \cite{ssim}, and the feature similarity index (FSIM) \cite{zhang2011fsim}  are selected to quantitatively measure the reconstructed results. On account of that the data are third-order tensors, we report the mean values of PSNR, SSIM, and FISM of all the frontal slices.

\begin{figure*}[htbp]
\centering\scriptsize\setlength{\tabcolsep}{1pt}
\renewcommand\arraystretch{0.9}
\begin{tabular}{cccccccccc}
&Observed&LRMC \cite{candes2009exact}&HaLRTC \cite{Liu2013PAMItensor}&TMac \cite{Xu2013Tmac}&TNN \cite{zhang2017exact}&PSTNN \cite{jiang2017PSTNN}&DCTNN \cite{lu2019low}& F-TNN& Ground truth\\
\rotatebox[origin=l]{90}{\quad\textbf{SR = 0.1}}&
\includegraphics[width=0.105\linewidth]{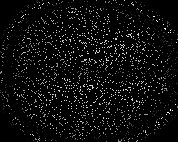} &
\includegraphics[width=0.105\linewidth]{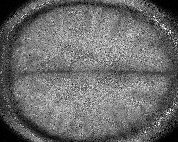} &
\includegraphics[width=0.105\linewidth]{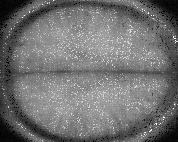} &
\includegraphics[width=0.105\linewidth]{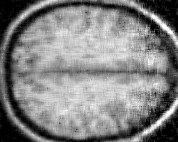} &
\includegraphics[width=0.105\linewidth]{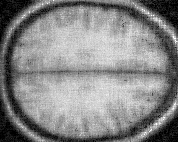} &
\includegraphics[width=0.105\linewidth]{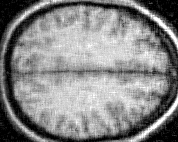} &
\includegraphics[width=0.105\linewidth]{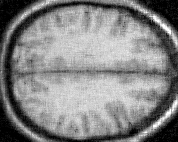} &
\includegraphics[width=0.105\linewidth]{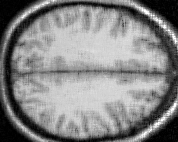} &
\includegraphics[width=0.105\linewidth]{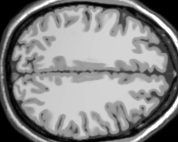} \\

\rotatebox[origin=l]{90}{\quad\textbf{SR = 0.2}}&
\includegraphics[width=0.105\linewidth]{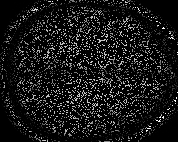} &
\includegraphics[width=0.105\linewidth]{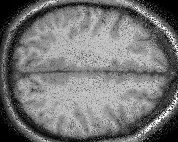} &
\includegraphics[width=0.105\linewidth]{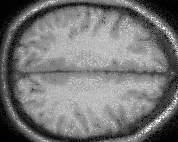} &
\includegraphics[width=0.105\linewidth]{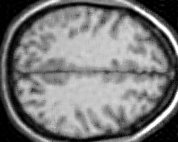} &
\includegraphics[width=0.105\linewidth]{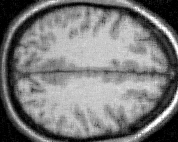} &
\includegraphics[width=0.105\linewidth]{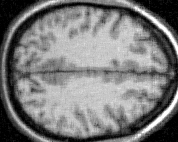} &
\includegraphics[width=0.105\linewidth]{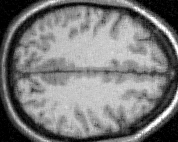} &
\includegraphics[width=0.105\linewidth]{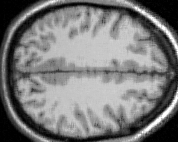} &
\includegraphics[width=0.105\linewidth]{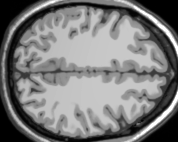} \\
\rotatebox[origin=l]{90}{\quad\textbf{SR = 0.3}}&
\includegraphics[width=0.105\linewidth]{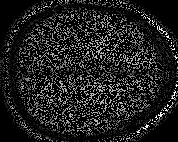} &
\includegraphics[width=0.105\linewidth]{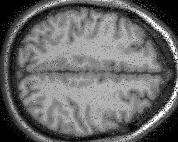} &
\includegraphics[width=0.105\linewidth]{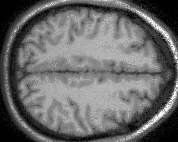} &
\includegraphics[width=0.105\linewidth]{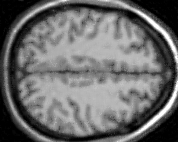} &
\includegraphics[width=0.105\linewidth]{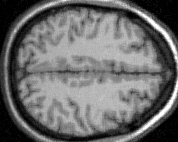} &
\includegraphics[width=0.105\linewidth]{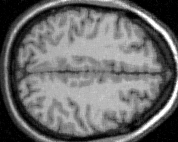} &
\includegraphics[width=0.105\linewidth]{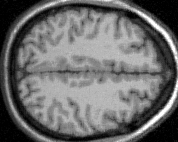} &
\includegraphics[width=0.105\linewidth]{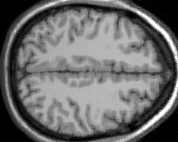} &
\includegraphics[width=0.105\linewidth]{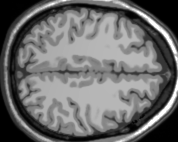} \\

\end{tabular}
\caption{The visual illustration of the results on the {\bf MRI} data by different methods with different sampling rates (SR). From left to right are the frontal slices of observed incomplete data, results by different methods and the ground truth, respectively. From top to bottom are respectively corresponding to the 106-th slice, the 110-th slice and the 115-th slice.}
\label{MRIframe}
\end{figure*}

\textbf{Experimental Settings}:
We generated the framelet system via the piece-wise cubic B-spline. 
If not specified, the framelet decomposition level $l$ is set as 4 ($l=2$ for the MSI data),
and the Lagrangian penalty parameter $\beta = 1$ for the tensor completion task and $\beta = 5$ when dealing with the tensor RPCA problems.
The maximum iteration $t_\text{max}$ and the convergence tolerance $\epsilon$ are chosen as $(t_\text{max},\epsilon) = (100,10^{-2})$ for the tensor completion and $(t_\text{max},\epsilon) = (200,10^{-3})$ for the tensor RPCA.
All the methods are implemented on the platform of Windows 10 and Matlab (R2017a) with an Intel(R) Core(TM) i5-4590 CPU at 3.30GHz and 16 GB RAM.


\begin{figure*}[!t]
\centering\scriptsize\setlength{\tabcolsep}{1pt}
\renewcommand\arraystretch{0.9}
\begin{tabular}{ccccccccccc}
&Observed&LRMC \cite{candes2009exact}&HaLRTC \cite{Liu2013PAMItensor}&TMac \cite{Xu2013Tmac}&TNN \cite{zhang2017exact}&PSTNN \cite{jiang2017PSTNN}&DCTNN \cite{lu2019low}& F-TNN& Ground truth\\
\rotatebox[origin=l]{90}{\quad ``beads''}&
\includegraphics[width=0.105\linewidth]{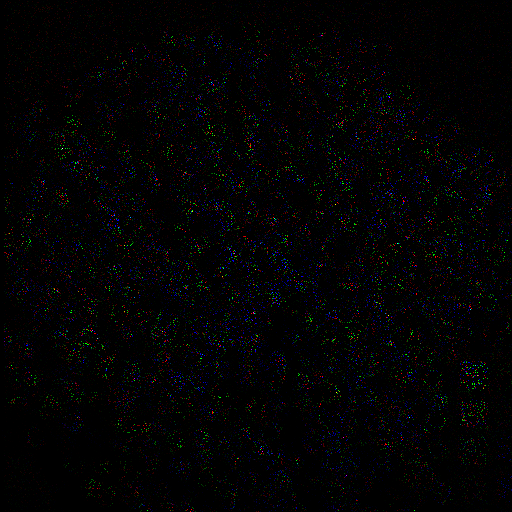} &
\includegraphics[width=0.105\linewidth]{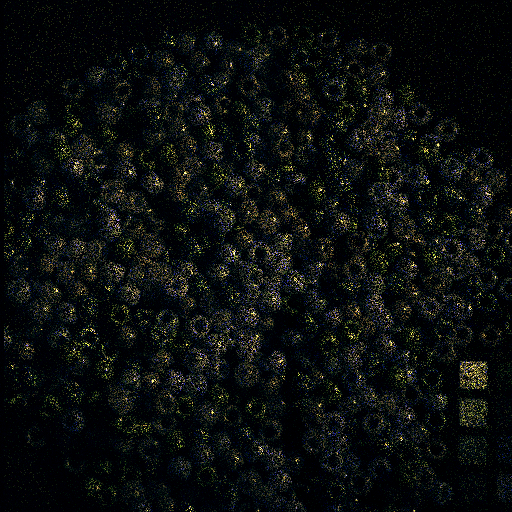} &
\includegraphics[width=0.105\linewidth]{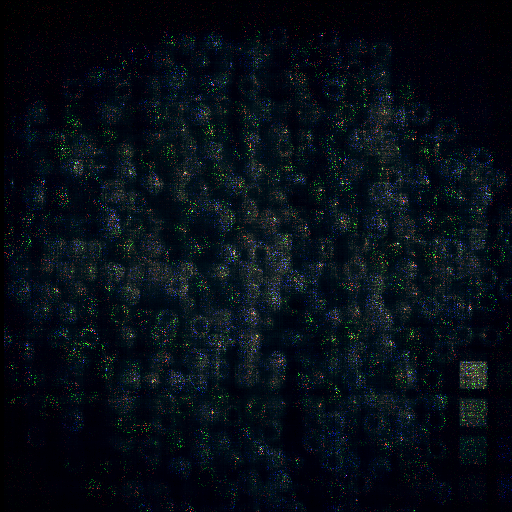} &
\includegraphics[width=0.105\linewidth]{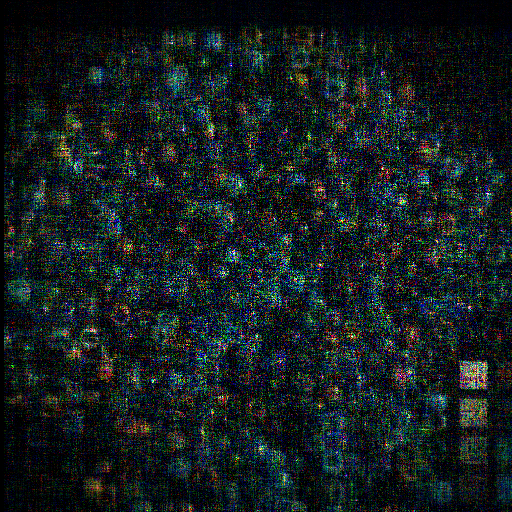} &
\includegraphics[width=0.105\linewidth]{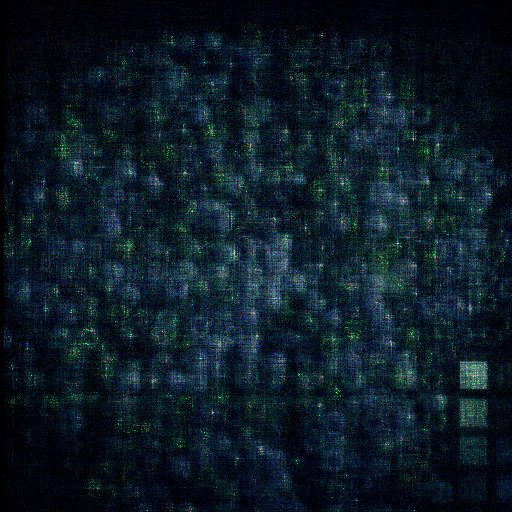} &
\includegraphics[width=0.105\linewidth]{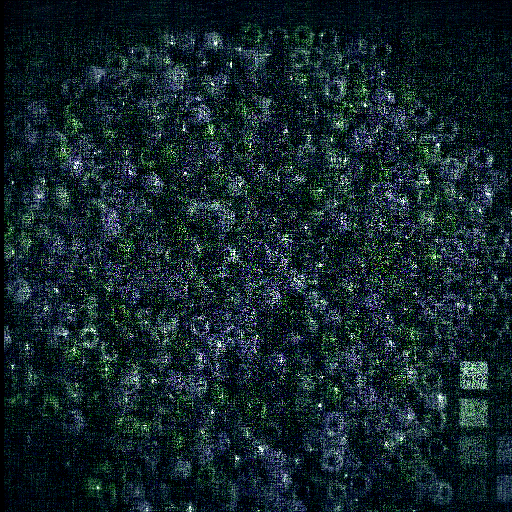} &
\includegraphics[width=0.105\linewidth]{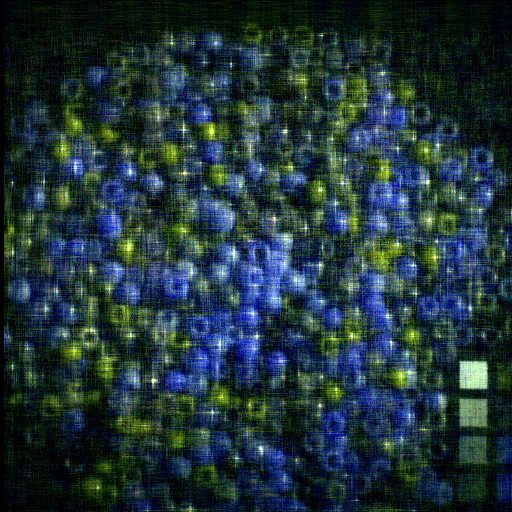} &
\includegraphics[width=0.105\linewidth]{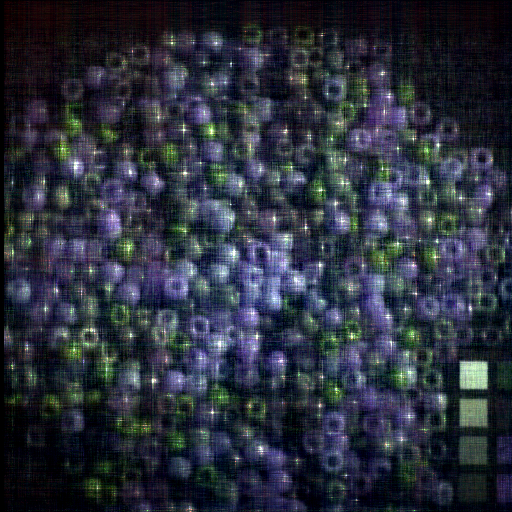} &
\includegraphics[width=0.105\linewidth]{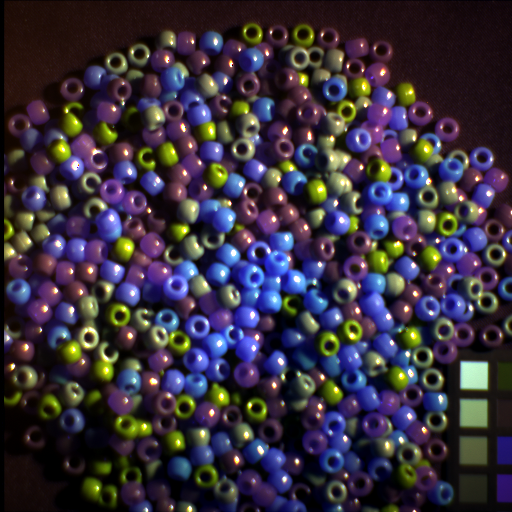} \\

\rotatebox[origin=l]{90}{\quad  ``cd''}&
\includegraphics[width=0.105\linewidth]{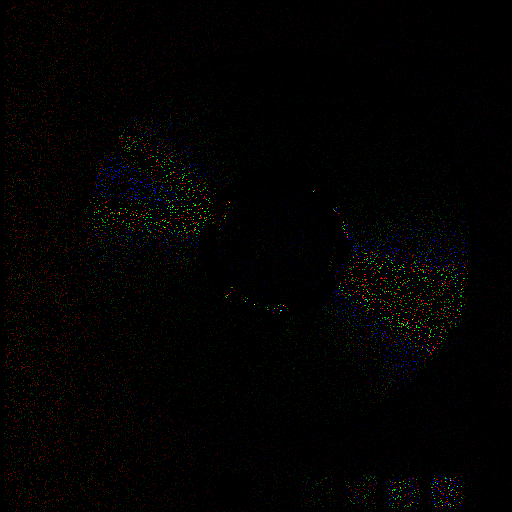} &
\includegraphics[width=0.105\linewidth]{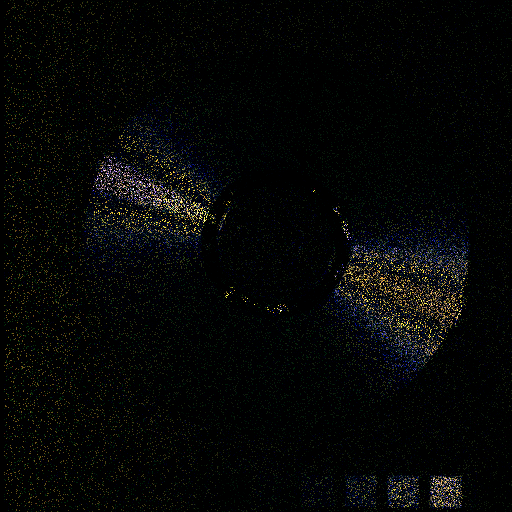} &
\includegraphics[width=0.105\linewidth]{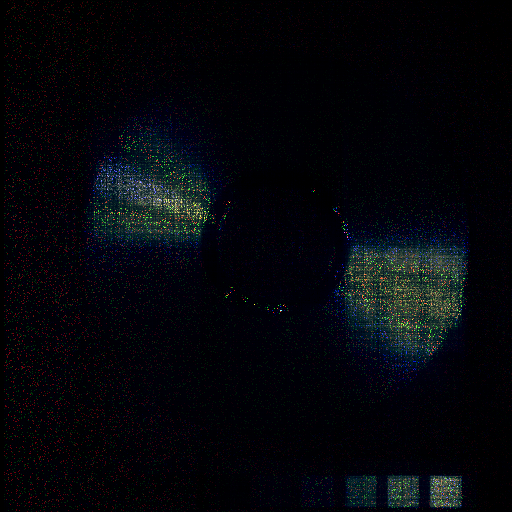} &
\includegraphics[width=0.105\linewidth]{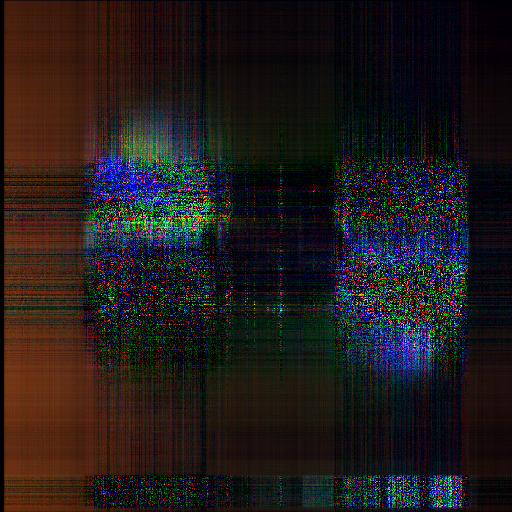} &
\includegraphics[width=0.105\linewidth]{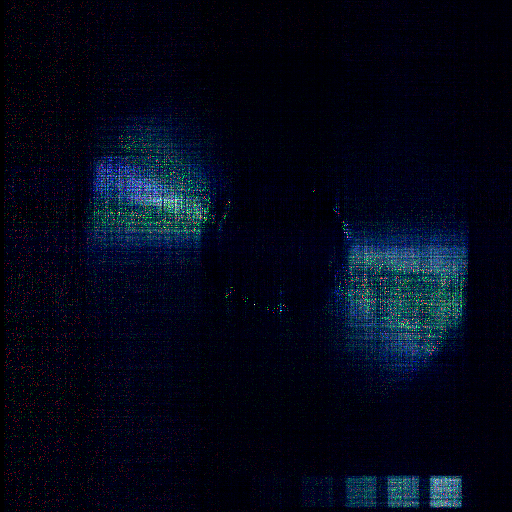} &
\includegraphics[width=0.105\linewidth]{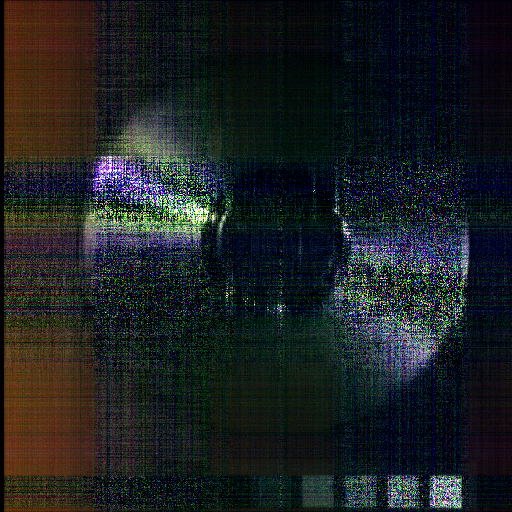} &
\includegraphics[width=0.105\linewidth]{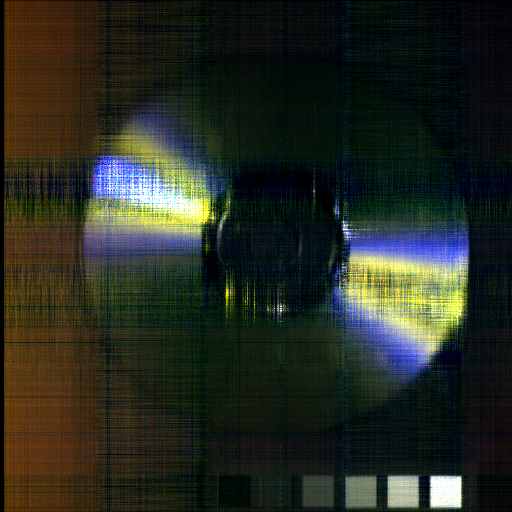} &
\includegraphics[width=0.105\linewidth]{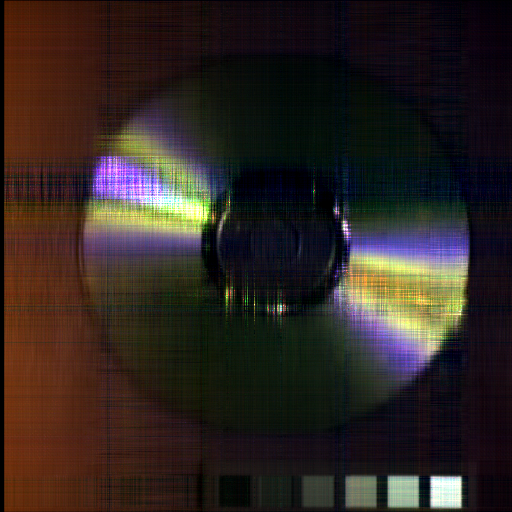} &
\includegraphics[width=0.105\linewidth]{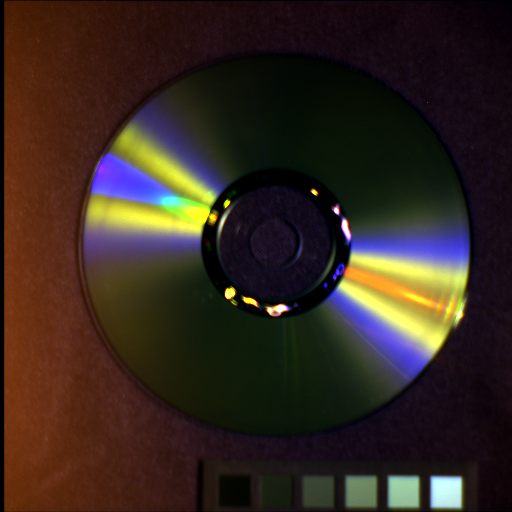} \\

\rotatebox[origin=l]{90}{  \quad``clay''}&
\includegraphics[width=0.105\linewidth]{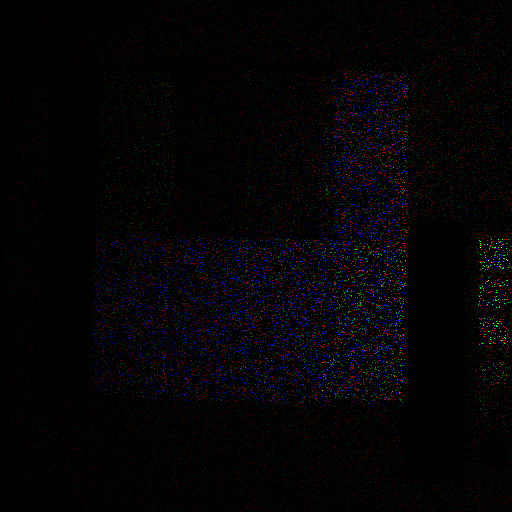} &
\includegraphics[width=0.105\linewidth]{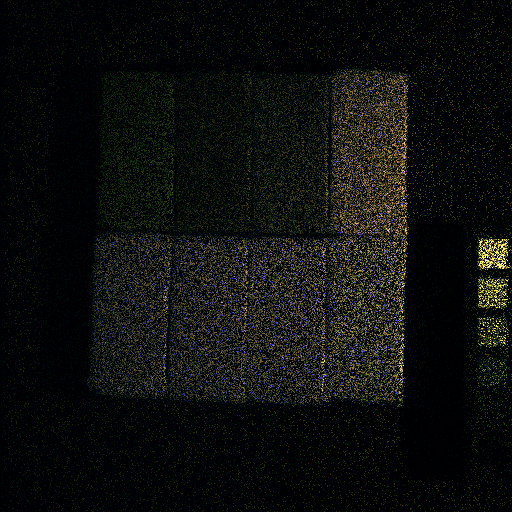} &
\includegraphics[width=0.105\linewidth]{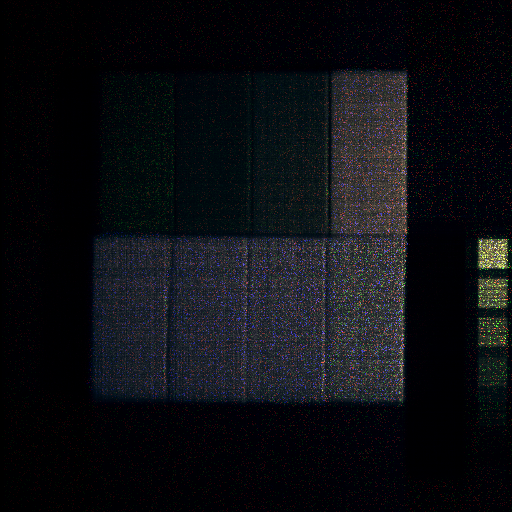} &
\includegraphics[width=0.105\linewidth]{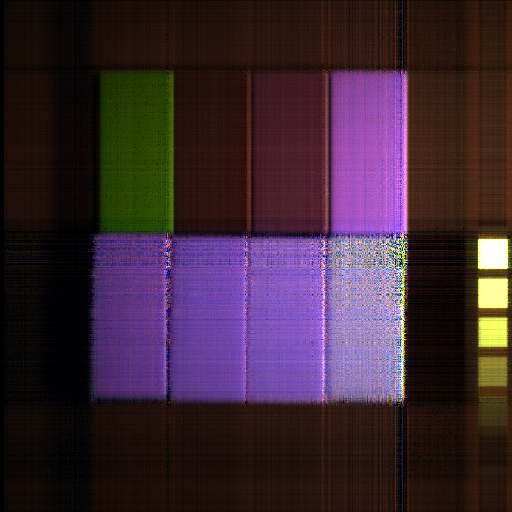} &
\includegraphics[width=0.105\linewidth]{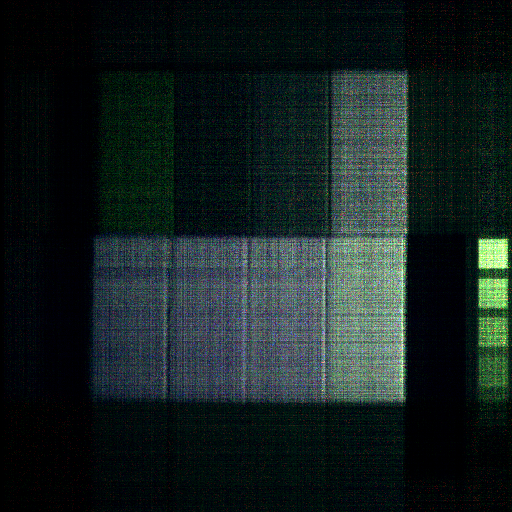} &
\includegraphics[width=0.105\linewidth]{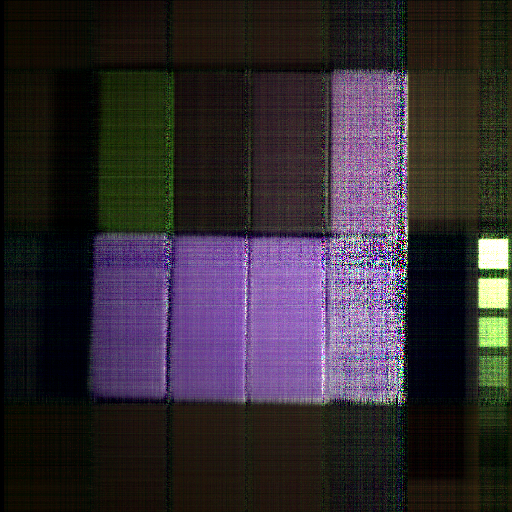} &
\includegraphics[width=0.105\linewidth]{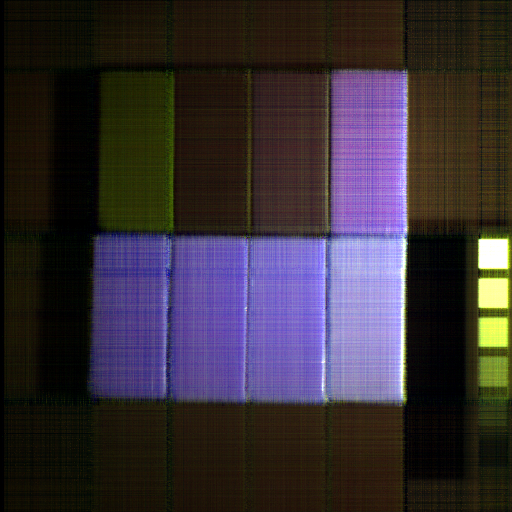} &
\includegraphics[width=0.105\linewidth]{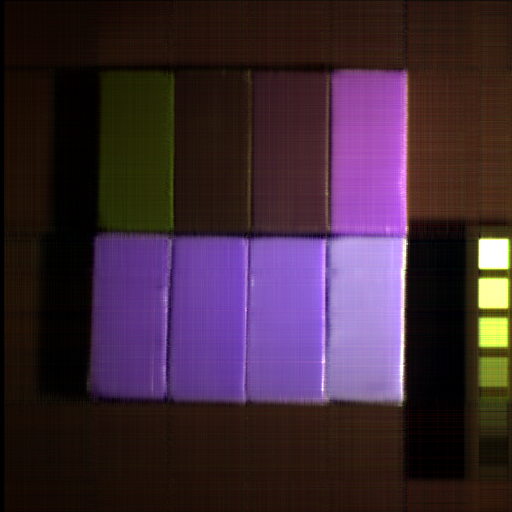} &
\includegraphics[width=0.105\linewidth]{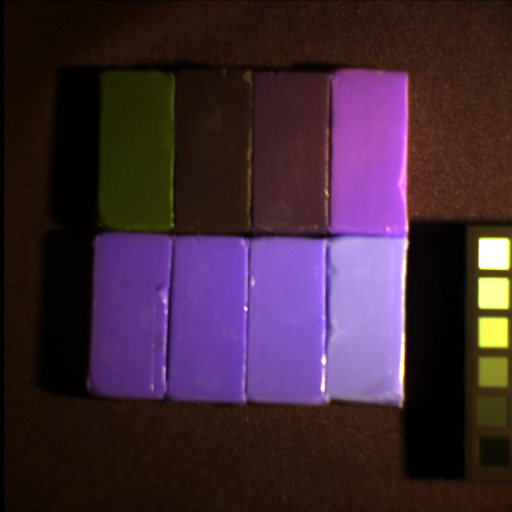} \\

\end{tabular}
\caption{The pseudo-color images (R-1 G-2 B-31) of the completion results on the {\bf MSI} data ``beads'' (top row), ``cd'' (mid row), and ``clay'' (bottom row) by different methods, respectively, with the sampling rate = 0.05. From left to right are the observed incomplete data, results by different methods and the ground truth, respectively. For better visualization, the intensity of the pixels are adjusted.}
\label{MSIframe}
\end{figure*}

\subsection{Tensor Completion}
We compare our F-TNN based tensor completion method with six methods, including a baseline low-rank matrix completion (LRMC) method \cite{candes2009exact},
two Tucker-rank based methods HaLRTC \cite{Liu2013PAMItensor} and TMac \cite{Xu2013Tmac}, a TNN based method \cite{zhang2017exact}, a non-convex method minimizing the partial sum of the TNN (PSTNN) \cite{jiang2017PSTNN}, the DCT based TNN method (denoted as DCTNN) \cite{lu2019low}. When employing LRMC, the input third-order tensor data is unfolded to a matrix along the third dimension.

\subsubsection{\bf MRI Data}
We evaluate the performance of the proposed method and the compared methods on the MRI data\footnote{http://brainweb.bic.mni.mcgill.ca/brainweb/selection\_normal.html.}, which is of size $142\times178\times121$.
As shown in Fig. \ref{MRIframe}, this is an MRI of the brain, which consists of abundant textures of the gray matter and the white matter.
The sampling rates (SR) are set as 10\%, 20\%, and 30\%.

Table \ref{MRI} shows the quantitative assessments of the results recovered by different methods.
Form Table \ref{MRI}, it can be found that the proposed method reaches the highest indices for different sampling rates.
The results by TMac and DCTNN alternatively rank the second-best place.
The margins between the results by our method and the second-best results are more than 1.3dB considering the PSNR, and 0.03 for the SSIM and FSIM.

\begin{table}[!t]\renewcommand\arraystretch{1}\setlength{\tabcolsep}{2pt}\scriptsize
\renewcommand\arraystretch{1}\centering
\caption{Quantitative comparisons of the {\bf MRI} data completion results by LRMC \cite{candes2009exact}, HaLRTC \cite{Liu2013PAMItensor}, TMac \cite{Xu2013Tmac}, TNN \cite{zhang2017exact}, PSTNN \cite{jiang2017PSTNN}, DCTNN \cite{lu2019low} and the proposed method. The \textbf{best} values and the \underline{second best} values are respectively highlighted by bolder fonts and underlines.}
\begin{tabular}{cccccccccccc}
\toprule
SR &Index  &Observed& LRMC & HaLRTC&TMac& TNN&    PSTNN&    DCTNN&    F-TNN\\           \midrule
\multirow{3}{*}{10\% }
& PSNR &  9.048  &  17.541  &  18.012  & \underline{24.866} &  21.855  &  24.578  &  24.716  & \bf 26.104 \\
& SSIM &  0.047  &  0.317  &  0.388  &  0.658  &  0.524  &  0.628  & \underline{0.659} & \bf 0.759 \\
& FSIM &   0.474  &   0.694  &   0.686  &   0.809  &   0.760  &   0.802  & \underline{0.817} & \bf 0.862 \\       \midrule
\multirow{3}{*}{20\% }
& PSNR &  9.561  &  22.781  &  23.404  & {28.523} &  27.301  &  28.566  &  \underline{28.595}  & \bf 30.207 \\
& SSIM &  0.073  &  0.590  &  0.657  & \underline{0.835} &  0.776  &  0.806  &  0.820  & \bf 0.886 \\
& FSIM &   0.523  &   0.813  &   0.823  & \underline{0.896} &   0.871  &   0.885  &   0.892  & \bf 0.925 \\       \midrule
\multirow{3}{*}{30\% }
& PSNR &  10.141  &  25.730  &  26.896  &  30.771  &  30.897  &  31.382  & \underline{31.547} & \bf 33.142 \\
& SSIM &  0.103  &  0.730  &  0.794  &  0.889  &  0.880  &  0.885  & \underline{0.896} & \bf 0.936 \\
& FSIM &   0.550  &   0.875  &   0.892  &   0.919  &   0.925  &   0.928  & \underline{0.935} & \bf 0.956 \\
\bottomrule
\end{tabular}
\label{MRI}
\end{table}

We illustrate one frontal slice of the results by different methods with different random sampling rates in Fig. \ref{MRIframe}. As shown in the top row of Fig. \ref{MRIframe}, when the sampling rate is 10\%, the proposed method accurately reconstructs the MRI data, with a clear margin of the gray matter and the white matter. When the sampling rate is 30\%, all the methods get good performances, and the white matter regions recovered by the proposed method and TMac are the visually best.

\subsubsection{MSI Data}
In this subsection, we evaluate the performance of our method and the compared methods on 32 MSIs \footnote{{http://www.cs.columbia.edu/CAVE/databases/multispectral/}.} from the CAVE databases \cite{yasuma2010generalized}.
The size of the MSIs is $512\times512\times31$, where the spatial resolution is $512\times512$ and the spectral resolution is 31. The sampling rates (SR) are set as 5\%, 10\%, and 20\%\footnote{For the MSI data, when the sampling rate is higher than 20\%, all the methods achieve very high performances and the results are very close to the ground truths. Therefore, we select the lower sampling rates to exhibit.}.

The average quantitative assessments of all the results by different methods are listed in Table \ref{MSI}.
We can find that the proposed method achieves the best performance while DCTNN obtains the second best-metrics.
When the sampling rate is 20\%, TMac, TNN, PSTNN, DCTNN, and the proposed method all have good performances.

\begin{table}[!t]
\renewcommand\arraystretch{1}\setlength{\tabcolsep}{2pt}\scriptsize
\centering
\caption{The average PSNR, SSIM and FSIM of the completion results on 32 {\bf MSIs} by LRMC \cite{candes2009exact}, HaLRTC \cite{Liu2013PAMItensor}, TMac \cite{Xu2013Tmac}, TNN \cite{zhang2017exact}, PSTNN \cite{jiang2017PSTNN}, DCTNN \cite{lu2019low} and the proposed method with different sampling rates. The \textbf{best} values and the \underline{second best} values are respectively highlighted by bolder fonts and underlines.}
\begin{tabular}{cccccccccc}                                                                              \toprule
SR &Index  &Observed& LRMC & HaLRTC&TMac&      TNN&    PSTNN&    DCTNN&    F-TNN\\                      \midrule
\multirow{3}{*}{5\% }
& PSNR &  14.718  &  16.687  &  17.831  &  25.633  &  21.863  &  23.073  & \underline{32.068} & \bf 33.536 \\
& SSIM &  0.231  &  0.588  &  0.661  &  0.794  &  0.729  &  0.771  & \underline{0.909} & \bf 0.930 \\
& FSIM &   0.697  &   0.773  &   0.799  &   0.871  &   0.836  &   0.856  & \underline{0.940} & \bf 0.955 \\       \midrule
\multirow{3}{*}{10\% }
& PSNR &  14.954  &  19.369  &  22.369  &  32.306  &  31.165  &  33.945  & \underline{37.870} & \bf 38.415 \\
& SSIM &  0.277  &  0.679  &  0.789  &  0.917  &  0.906  &  0.945  & \underline{0.974} & \bf 0.977 \\
& FSIM &   0.718  &   0.828  &   0.876  &   0.942  &   0.939  &   0.961  & \underline{0.981} & \bf 0.984 \\       \midrule
\multirow{3}{*}{20\% }
& PSNR &  15.464  &  24.581  &  33.004  &  38.258  &  40.077  &  41.944  & \underline{42.675} & \bf 43.557 \\
& SSIM &  0.368  &  0.783  &  0.940  &  0.973  &  0.983  &  0.988  &  \underline{0.992}  & \bf 0.993 \\
& FSIM &   0.740  &   0.892  &   0.963  &   0.979  &   0.987  &   0.991  & \underline{0.994} & \bf 0.995 \\
\bottomrule
\end{tabular}
\label{MSI}
\end{table}

The third dimension of the MSI represents the spectral information and facilitates a fine delivery of more faithful knowledge under real scenes \cite{xie2016multispectral}.
Therefore, in Fig. \ref{MSIframe}, we illustrate the pseudo-color images (Red-1 Green-2 Blue-31) of the results on the MSI data ``beads'', ``cd'', and ``clay'', with the sampling rate = 0.05.
From the similarity of the color between the results and the ground truth, we can recognize the spectral distortion.
From the first row of Fig. \ref{MSIframe}, we can see that, although DCTNN also obtains clear results on ``beads'' as our F-TNN, the result by DCTNN is spectrally distorted.
TMac performs well on ``clay'', however, undesirable artifacts can be found.
The superior of the proposed F-TNN is visually obvious, considering the reconstruction of the image and preservation of spectral information.

\subsubsection{Video Data}
\begin{figure*}[!t]
\centering\scriptsize\setlength{\tabcolsep}{1pt}
\renewcommand\arraystretch{0.9}
\begin{tabular}{cccccccccc}

&Observed&LRMC \cite{candes2009exact}&HaLRTC \cite{Liu2013PAMItensor}&TMac \cite{Xu2013Tmac}&TNN \cite{zhang2017exact}&PSTNN \cite{jiang2017PSTNN}&DCTNN \cite{lu2019low}& F-TNN& Ground truth\\

\rotatebox[origin=l]{90}{      \textbf{SR = 0.1}}&
\includegraphics[width=0.105\linewidth]{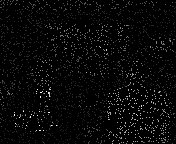} &
\includegraphics[width=0.105\linewidth]{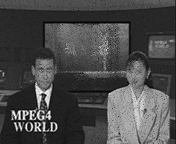} &
\includegraphics[width=0.105\linewidth]{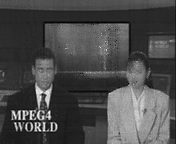} &
\includegraphics[width=0.105\linewidth]{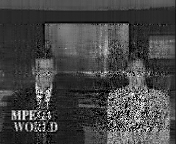} &
\includegraphics[width=0.105\linewidth]{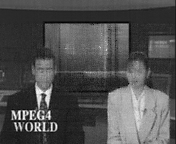} &
\includegraphics[width=0.105\linewidth]{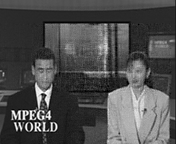} &
\includegraphics[width=0.105\linewidth]{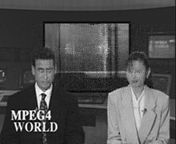} &
\includegraphics[width=0.105\linewidth]{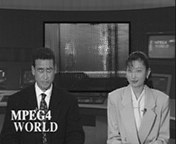} &
\includegraphics[width=0.105\linewidth]{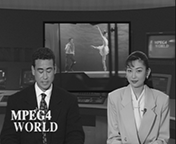} \\
\rotatebox[origin=l]{90}{      \textbf{SR = 0.2}}&
\includegraphics[width=0.105\linewidth]{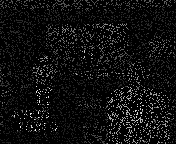} &
\includegraphics[width=0.105\linewidth]{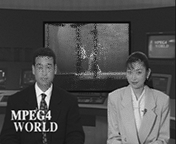} &
\includegraphics[width=0.105\linewidth]{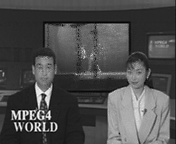} &
\includegraphics[width=0.105\linewidth]{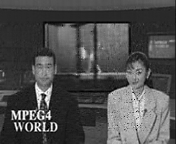} &
\includegraphics[width=0.105\linewidth]{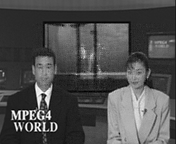} &
\includegraphics[width=0.105\linewidth]{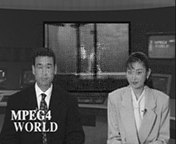} &
\includegraphics[width=0.105\linewidth]{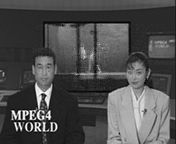} &
\includegraphics[width=0.105\linewidth]{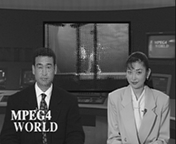} &
\includegraphics[width=0.105\linewidth]{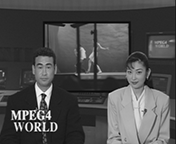} \\

\end{tabular}
\caption{The completion results on the {\bf video} data ``news'' different methods with different sampling rates. From left to right are the observed incomplete data, results by different methods and the ground truth, respectively. From top to bottom are respectively the 15-th frame and the 67-th frame.}
\label{videoframe}
\end{figure*}

In this subsection, 9 videos\footnote{http://trace.eas.asu.edu/yuv/.} (respectively named ``foreman'', ``hall'', ``carphone'', ``highway'', ``container'', ``claire'', ``news'', ``coastguard'' and ``suzie'')
with the size $144\times176\times100$ are selected as the ground truth third-order data. The contents of these videos are different, consisting of humans, roads, rivers, cars, boats, bridges, walls and so on. The scenarios in some videos (such as ``foreman'', ``coastguard'', ``suzie'', and ``highway'') are more dynamic while in others are more static.

\begin{table}[htbp]
\renewcommand\arraystretch{0.9}\setlength{\tabcolsep}{2pt}\scriptsize
\caption{The average PSNR, SSIM and FSIM of the completion results on 9 {\bf videos} by LRMC \cite{candes2009exact}, HaLRTC \cite{Liu2013PAMItensor}, TMac \cite{Xu2013Tmac}, TNN \cite{zhang2017exact}, PSTNN \cite{jiang2017PSTNN}, DCTNN \cite{lu2019low} and the proposed method with different sampling rates. The \textbf{best} values and the \underline{second best} values are respectively highlighted by bolder fonts and underlines.}
\centering
\begin{tabular}{cccccccccc}                                                                             \toprule
SR &Index  &Observed& LRMC & HaLRTC&TMac&      TNN&    PSTNN&    DCTNN&    F-TNN\\                     \midrule
\multirow{3}{*}{10\% }
& PSNR &  6.176  &  18.190  &  19.936  &  24.317  &  26.411  &  29.118  & \underline{29.246} & \bf 30.654 \\
& SSIM &  0.018  &  0.417  &  0.567  &  0.688  &  0.758  &  0.809  & \underline{0.819} & \bf 0.880 \\
& FSIM &   0.423  &   0.719  &   0.773  &   0.829  &   0.875  &   0.904  & \underline{0.909} & \bf 0.931 \\       \midrule
\multirow{3}{*}{20\% }
& PSNR &  6.687  &  29.315  &  30.150  &  30.250  &  31.329  &  32.012  & \underline{32.259} & \bf 33.568 \\
& SSIM &  0.031  &  0.851  &  0.871  &  0.868  &  0.871  &  0.876  & \underline{0.881} & \bf 0.927 \\
& FSIM &   0.413  &   0.928  &   0.927  &   0.921  &   0.934  &   0.937  & \underline{0.940} & \bf 0.957 \\       \midrule
\multirow{3}{*}{30\% }
& PSNR &  7.266  &  32.080  &  32.977  &  32.189  &  34.050  &  34.056  & \underline{34.434} & \bf 35.820 \\
& SSIM &  0.046  &  0.907  & \underline{0.917} &  0.910  &  0.915  &  0.912  &  0.915  & \bf 0.951 \\
& FSIM &   0.408  &   0.952  &   0.952  &   0.944  &   0.956  &   0.956  & \underline{0.958} & \bf 0.971 \\
\bottomrule

\end{tabular}
\label{videoa}
\end{table}

Table \ref{videoa} lists the average MPSNR, MSSIM, and MFSIM on these 9 videos with different sampling rates. For different sampling rates, our F-TNN obtains the results with the best quantitative metrics. When the sampling rates are 10\% and 20,  the performances of PSTNN and DCTNN are comparable. The DCTNN ranks second with the sampling rate of 30\%. Fig. \ref{videoframe} exhibits the frames of the results on the videos, ``news'' with sampling rates 10\% and 20\%.
The video ``news'' is captured by a static camera in a stationary scenery, and there are two dynamic parts, which are the two newscasters in the front position and a playing screen in the back, in this video. Thus, the scenario in this video contains both dynamic and static components. Most compared methods can reconstruct the static parts well while the proposed method obtains the best recovering performances on both the two newscasters (see their faces) and the dynamic screen.

\begin{figure}[!t]
\centering
\includegraphics[width=0.95\linewidth]{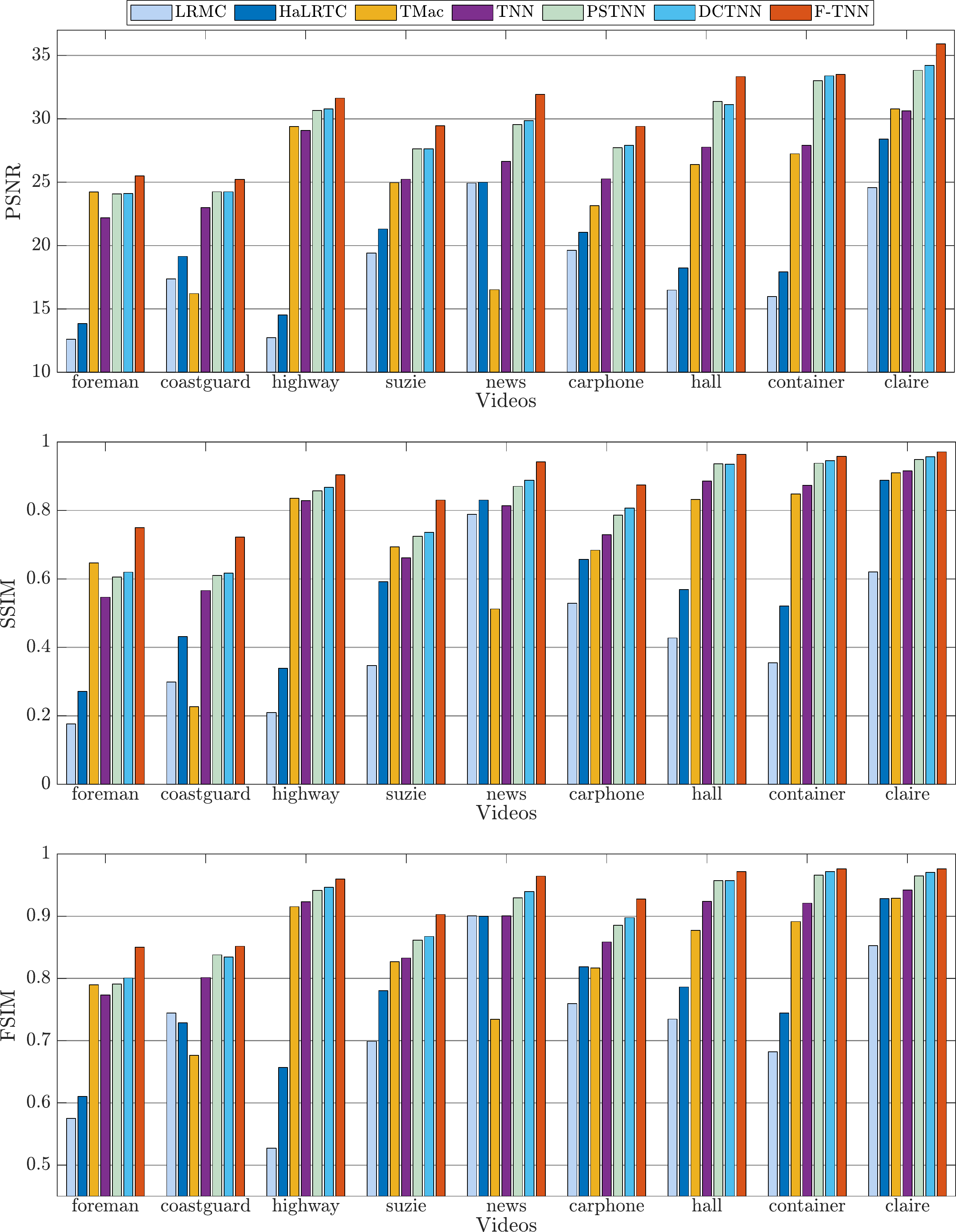}
\caption{The PNSR, SSIM, and FSIM of the results by different methods on all the video data with the sampling rate 10\%.}
\label{video_01}
\end{figure}

To further illustrate the performance of all the methods on different videos, in Fig. \ref{video_01} we exhibit the PSNR, SSIM, and FSIM on all the videos by all the methods when the sampling rate is 10\%.
From Fig. \ref{video_01}, it can be found that TMac is unstable with respect to different videos while other methods maintain better metrics when the video is more static. Although the scenario in ``highway'' is dynamic along the temporal direction, the contents in this video are not complicated. Therefore, many methods achieve good performances.
It can be observed that the proposed method obtains the highest PSNR, SSIM, and FISM on all the videos. This validates the robustness of our F-TNN.

\subsection{Tensor Robust Principal Component Analysis}
In this section, we test our F-TNN based TRPCA methods on two problems, i.e., color images recovery from observations corrupted by the salt-and-pepper noise, and the background subtraction for surveillance videos.
The compared methods consist of one matrix nuclear norm minimization based RPCA method (denoted as MRPCA) \cite{candes2011robust}, a sum of the nuclear norm minimization based tensor RPCA method (denoted as ``SNN'')\cite{goldfarb2014robust}, a TNN based tensor RPCA method \cite{lu2019tensor}, and a DCT transformed TNN based tensor RPCA method \cite{lu2019exact2}.
The $\ell_1$ norm is used to characterize the sparse component by all the compared methods. The balance parameter $\lambda$, which is added to the $\ell_1$ term, is manually selected for the best performances for all the methods. We list the settings of $\lambda$ in Table \ref{table-lambda} When implementing MRPCA, we unfold the observed data $\mathcal{O}$ along the third mode and input $\mathbf O_{(3)}$. For the image recovery, since that the framelet transformation matrix $\mathbf W$ requires the third dimension of the input data no less than 40, we shift the dimension of the observed image as $\hat{\mathcal{O}}\in\mathbb{R}^{n_2\times n_3\times n_1}$ via the Matlab command ``shiftdim($\cdot$,1)''.

\begin{table}[!t]
\renewcommand\arraystretch{1}\setlength{\tabcolsep}{8pt}\scriptsize
\renewcommand\arraystretch{1}\centering
\caption{The settings of the parameter $\lambda$ for all the methods, given the observation $\mathcal{O}\in\mathbb{R}^{n_1\times n_2\times n_3}$.}
%
\begin{tabular}{ccc}                                                          \toprule
Method  &  Image recovery  &  Background substraction\\           \midrule
MRPCA   & $ 1.5/\sqrt{n_1n_2}$           & $ 1/\sqrt{n_1n_2}$             \\
SNN     & $ 3/\sqrt{n_1n_2}$             & $ 0.5/\sqrt{\max(n_1,n_2)n_3}$ \\
TNN     & $ 3/\sqrt{\max(n_1,n_2)n_3}$   & $ 1/\sqrt{2\max(n_1,n_2)n_3}$  \\
DCTNN   & $ 2/\sqrt{\max(n_1,n_2)n_3}$   & $ 4/\sqrt{\max(n_1,n_2)n_3}$   \\
F-TNN   & $ 3/\sqrt{\max(n_1,n_3)n_2}$             & $ 3/\sqrt{\max(n_1,n_2)n_3}$          \\
\bottomrule
\end{tabular}
\label{table-lambda}
\end{table}

\subsubsection{Color Image Recovery}
We select 4 images\footnote{The images named ``airplane'', ``fruits'', and ``baboon'' are of the size $512\times512\times3$ and available at {http://sipi.usc.edu/database/database.php}, while the image ``watch'' of the size $1024\times768\times3$ is available at {https://www.sitepoint.com/mastering-image-optimization-in-wordpress/}}, respectively named ``airplane'', ``watch'', ``fruits'', and ``baboon'', as ground truth clean images. Then, the salt-and-pepper noise is added to these images, affecting $\rho$ pixels. The parameter $\rho$ varies from 5\% to 10\%. Table \ref{table-rpca-image} presents the averaged PSNR, SSIM, and FSIM values of the results by different methods for the color image recovery. We can find that the performance of our method is the best with different $\rho_s$. We exhibit the visual results on the images ``airplane'' and ``watch'' in Fig. \ref{figure-rpca-image}. It can be obtained that all the tensor-based methods remove the salt-and-pepper noise while the performance of MRPCA is unsatisfactory. The residual images, which are absolute values of the difference between results and clean images, are magnified with a factor 2 for better visualization. From the residual images, we can see that our method preserves the structure and details of the color images well.

\begin{table}[!t]\renewcommand\arraystretch{1}\setlength{\tabcolsep}{5pt}\scriptsize
\renewcommand\arraystretch{1}\centering
\caption{Quantitative comparisons of the image recovery results of MRPCA \cite{candes2011robust}, SNN \cite{goldfarb2014robust}, TNN \cite{lu2019tensor}, DCTNN \cite{lu2019exact2}, and the proposed method. The \textbf{best} values and the \underline{second best} values are respectively highlighted by bolder fonts and underlines.}
\begin{tabular}{ccccccccc}                                                                     \toprule
$\rho$ &Index  &Observed& MRPCA &  SNN & TNN&  DCTNN &F-TNN\\           \midrule
\multirow{3}{*}{ 5\% }
& PSNR &  18.005  &  21.671  &  30.188  &  29.791  & \underline{31.735} & \bf 33.846 \\
& SSIM &  0.587  &  0.771  &  0.962  &  0.964  & \underline{0.979} & \bf 0.987 \\
& FSIM &   0.833  &   0.896  &   0.973  &   0.970  & \underline{0.982} & \bf 0.990 \\       \midrule
\multirow{3}{*}{10\% }
& PSNR &  14.987  &  19.245  &  27.932  &  29.140  & \underline{30.807} & \bf 31.937 \\
& SSIM &  0.450  &  0.664  &  0.917  &  0.957  & \underline{0.971} & \bf 0.975 \\
& FSIM &   0.744  &   0.844  &   0.954  &   0.965  & \underline{0.977} & \bf 0.984 \\
\bottomrule
\end{tabular}
\label{table-rpca-image}
\end{table}

\begin{figure}[!t]
\centering\tiny\setlength{\tabcolsep}{0.95pt}
\renewcommand\arraystretch{0.9}
\begin{tabular}{cccccccc}
&Observed& {\tiny MRPCA \cite{candes2011robust}} &{\tiny SNN \cite{goldfarb2014robust}}&{\tiny TNN \cite{lu2019tensor}} &{ \tiny DCTNN \cite{lu2019exact2}}& FTNN&Groudtruth \\
\multirow{3}{*}{\rotatebox[origin=c]{90}{$\rho=5\%$}}&
\includegraphics[width=0.128\linewidth]{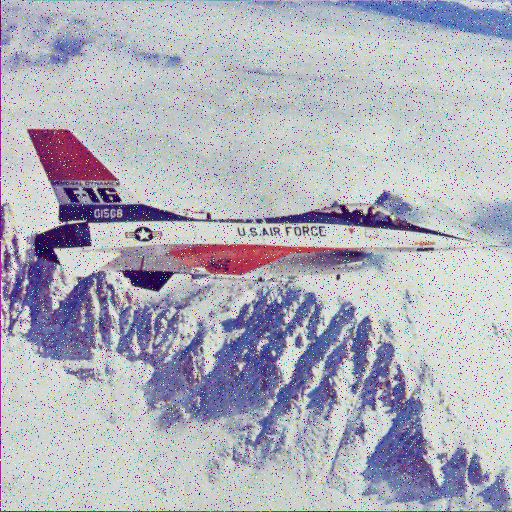} &
\includegraphics[width=0.128\linewidth]{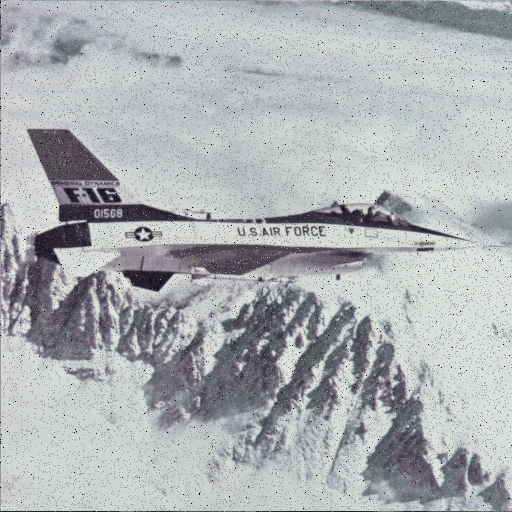} &
\includegraphics[width=0.128\linewidth]{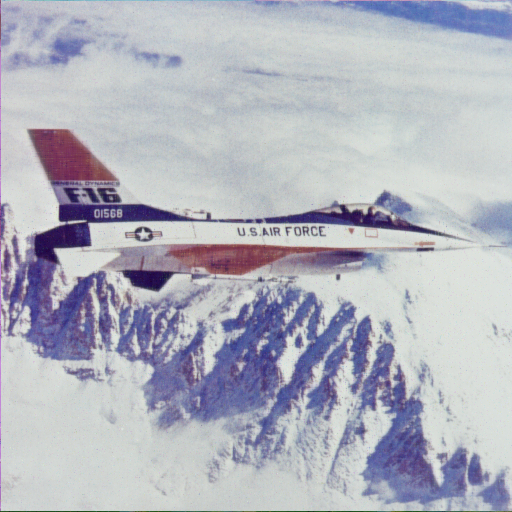} &
\includegraphics[width=0.128\linewidth]{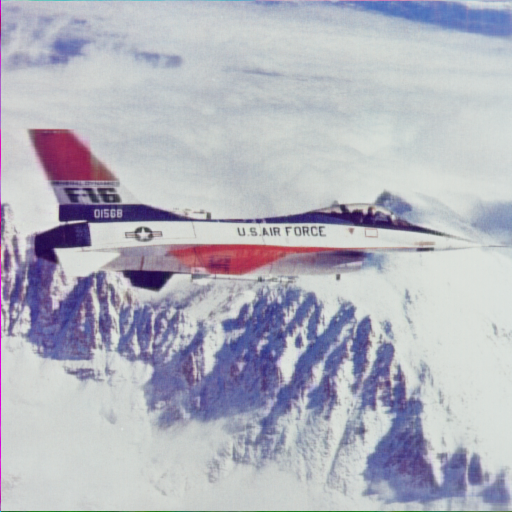} &
\includegraphics[width=0.128\linewidth]{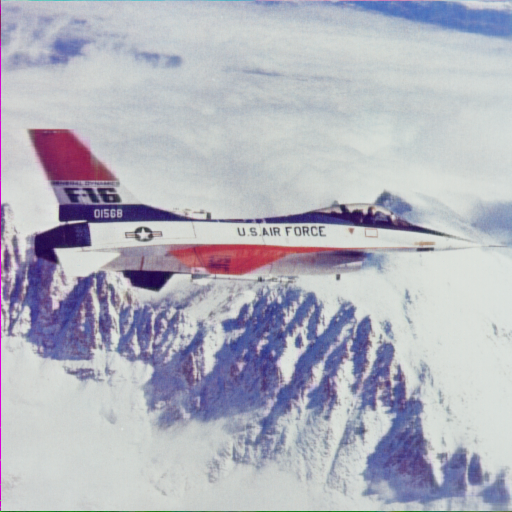} &
\includegraphics[width=0.128\linewidth]{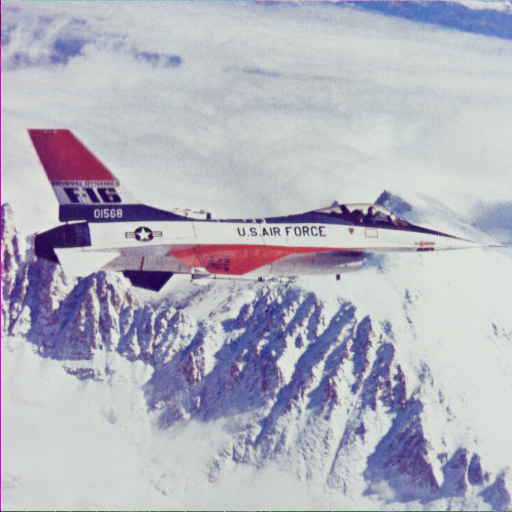} &
\includegraphics[width=0.128\linewidth]{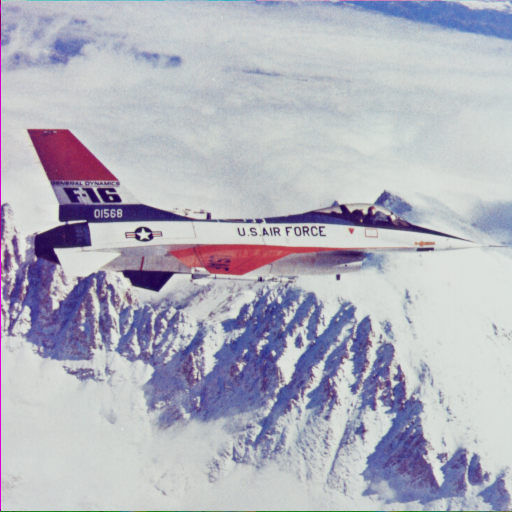} \\
&
\includegraphics[width=0.128\linewidth]{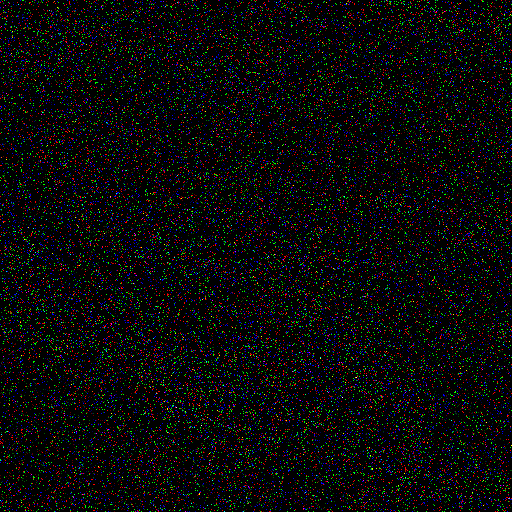} &
\includegraphics[width=0.128\linewidth]{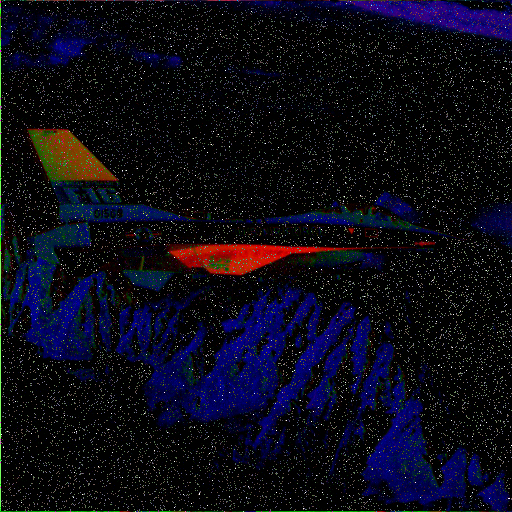} &
\includegraphics[width=0.128\linewidth]{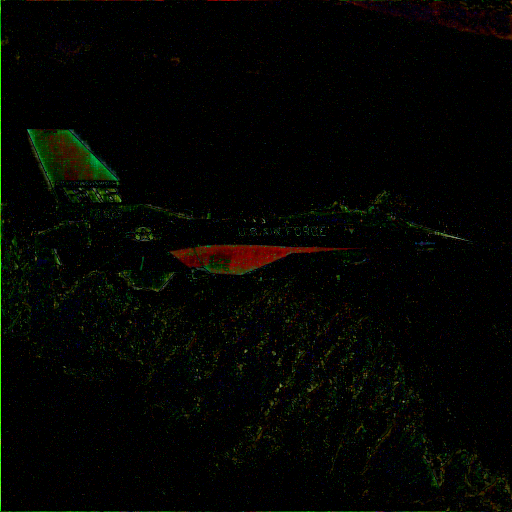} &
\includegraphics[width=0.128\linewidth]{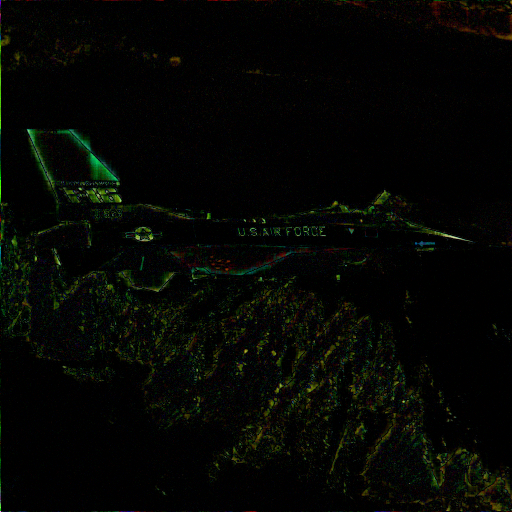} &
\includegraphics[width=0.128\linewidth]{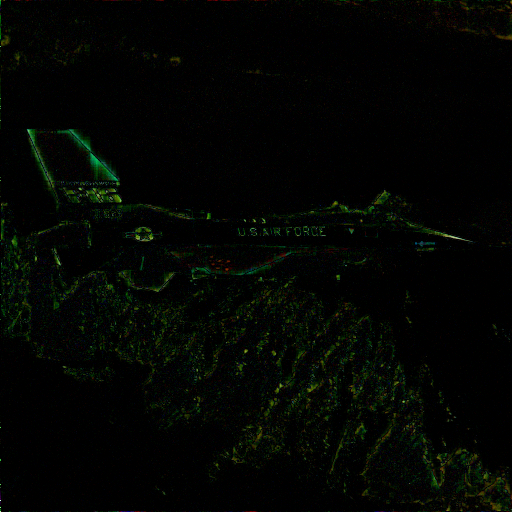} &
\includegraphics[width=0.128\linewidth]{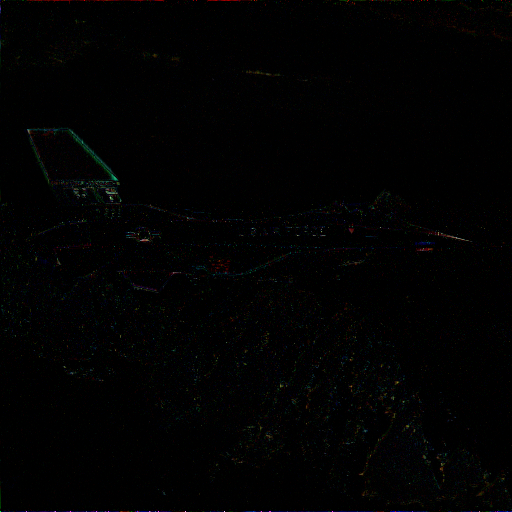} &
\includegraphics[width=0.128\linewidth]{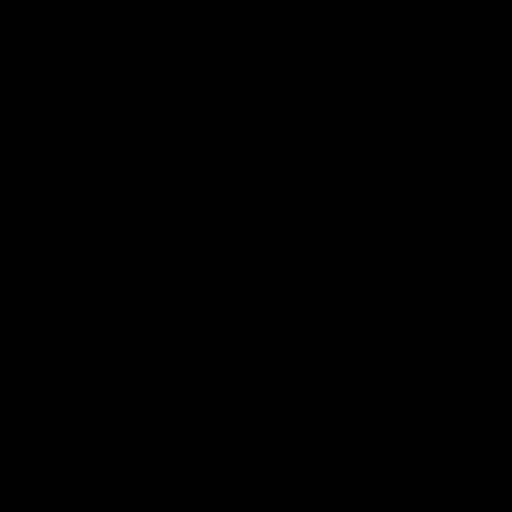} \\

\multirow{3}{*}{\rotatebox[origin=c]{90}{$\rho=10\%$}}&
\includegraphics[width=0.128\linewidth]{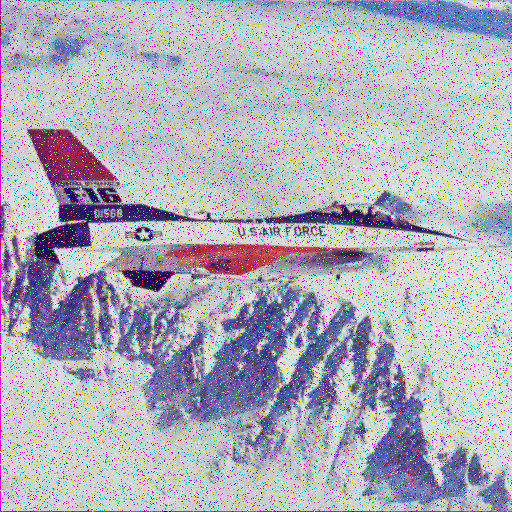} &
\includegraphics[width=0.128\linewidth]{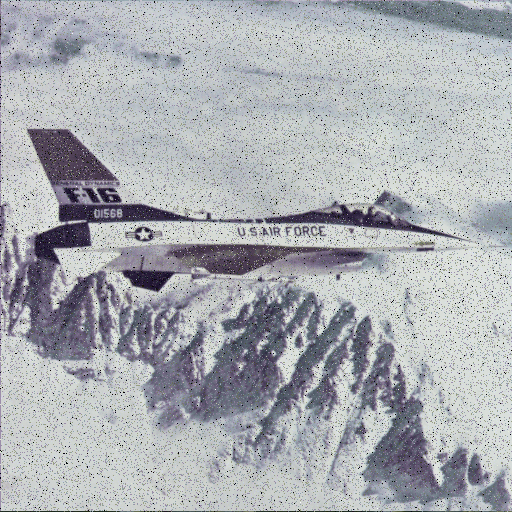} &
\includegraphics[width=0.128\linewidth]{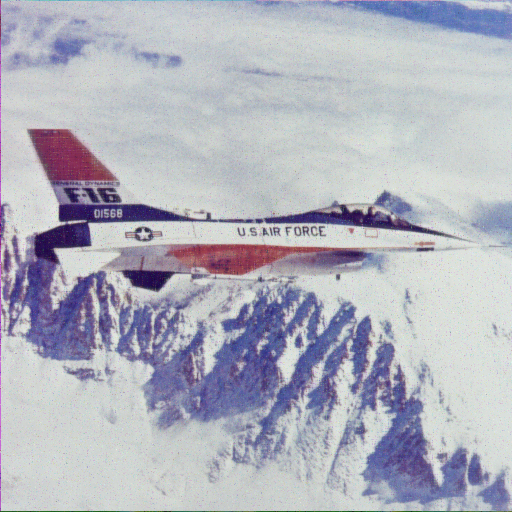} &
\includegraphics[width=0.128\linewidth]{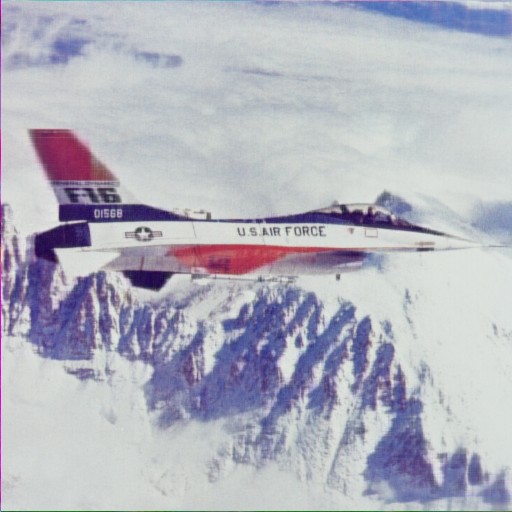} &
\includegraphics[width=0.128\linewidth]{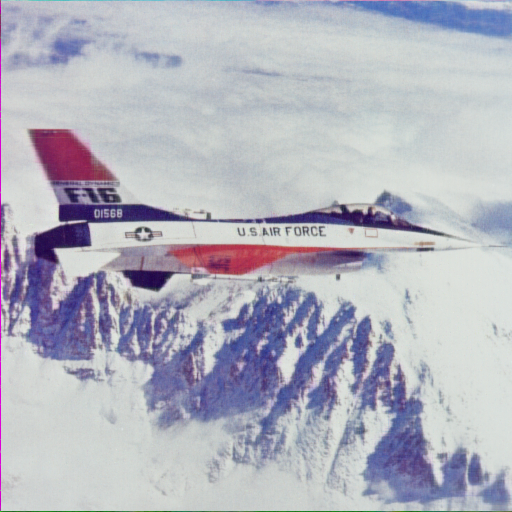} &
\includegraphics[width=0.128\linewidth]{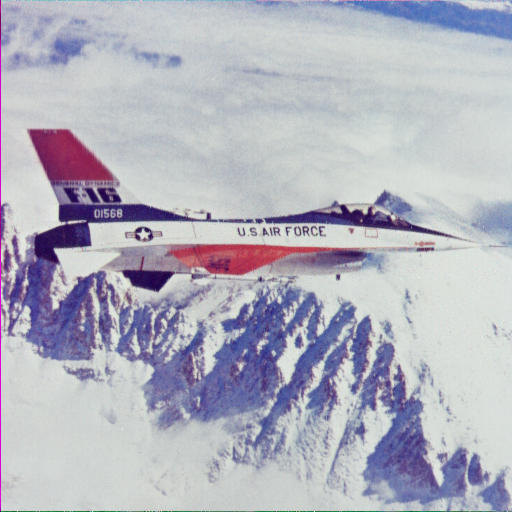} &
\includegraphics[width=0.128\linewidth]{figs/image222/airplane_GT.png} \\
&
\includegraphics[width=0.128\linewidth]{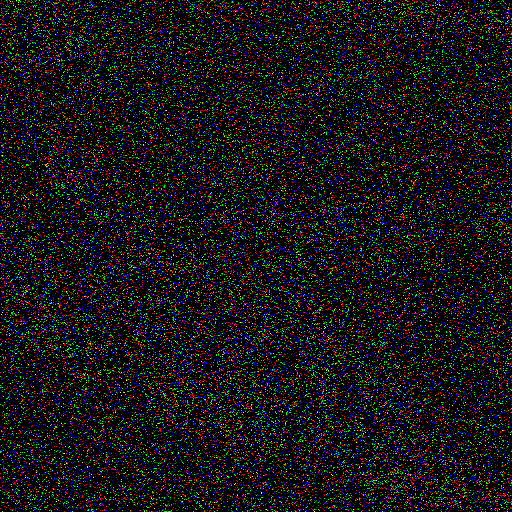} &
\includegraphics[width=0.128\linewidth]{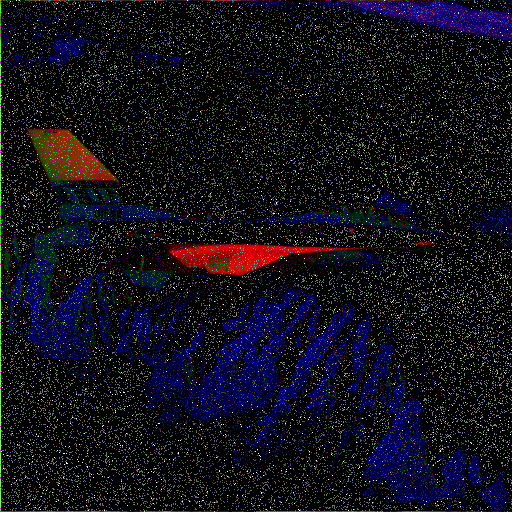} &
\includegraphics[width=0.128\linewidth]{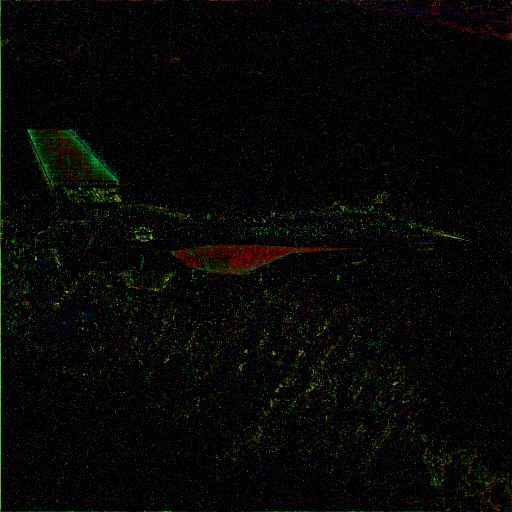} &
\includegraphics[width=0.128\linewidth]{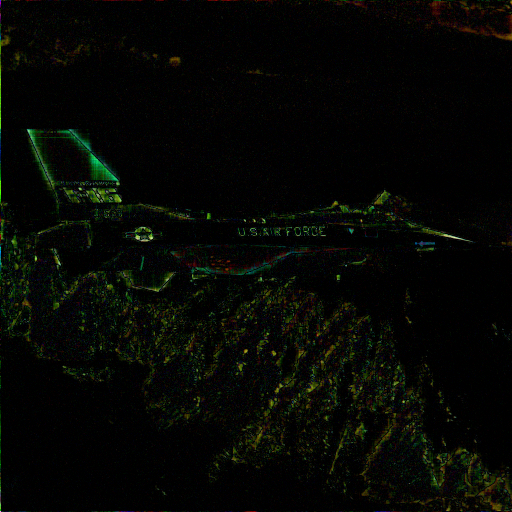} &
\includegraphics[width=0.128\linewidth]{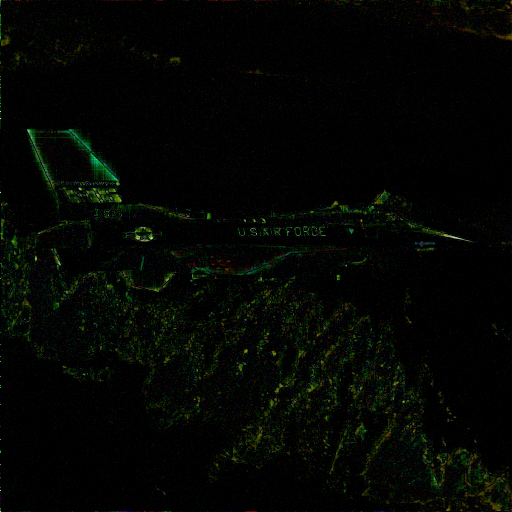} &
\includegraphics[width=0.128\linewidth]{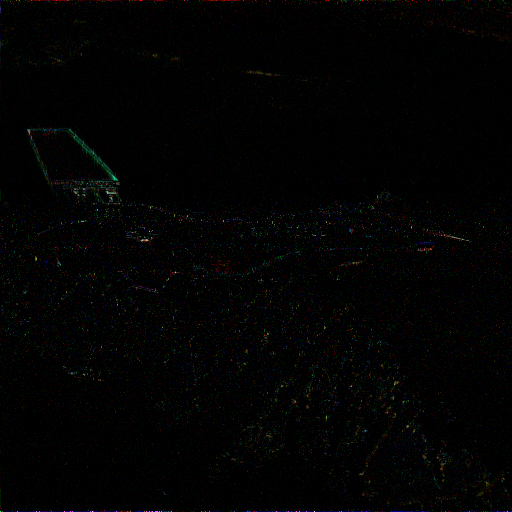} &
\includegraphics[width=0.128\linewidth]{figs/image222/airplane_GT_rse.png} \\

&Observed& {\tiny MRPCA \cite{candes2011robust}} &{\tiny SNN \cite{goldfarb2014robust}}&{\tiny TNN \cite{lu2019tensor}} &{ \tiny DCTNN \cite{lu2019exact2}}& FTNN&Groudtruth \\
\multirow{3}{*}{\rotatebox[origin=c]{90}{$\rho=5\%$}}&
\includegraphics[width=0.128\linewidth]{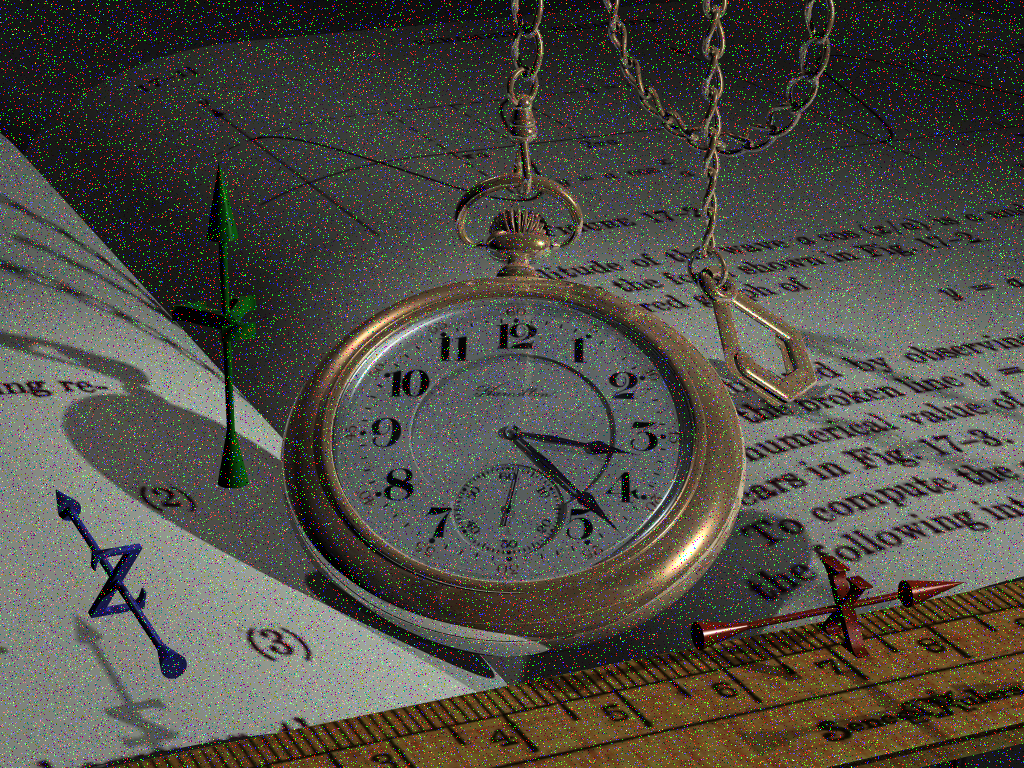} &
\includegraphics[width=0.128\linewidth]{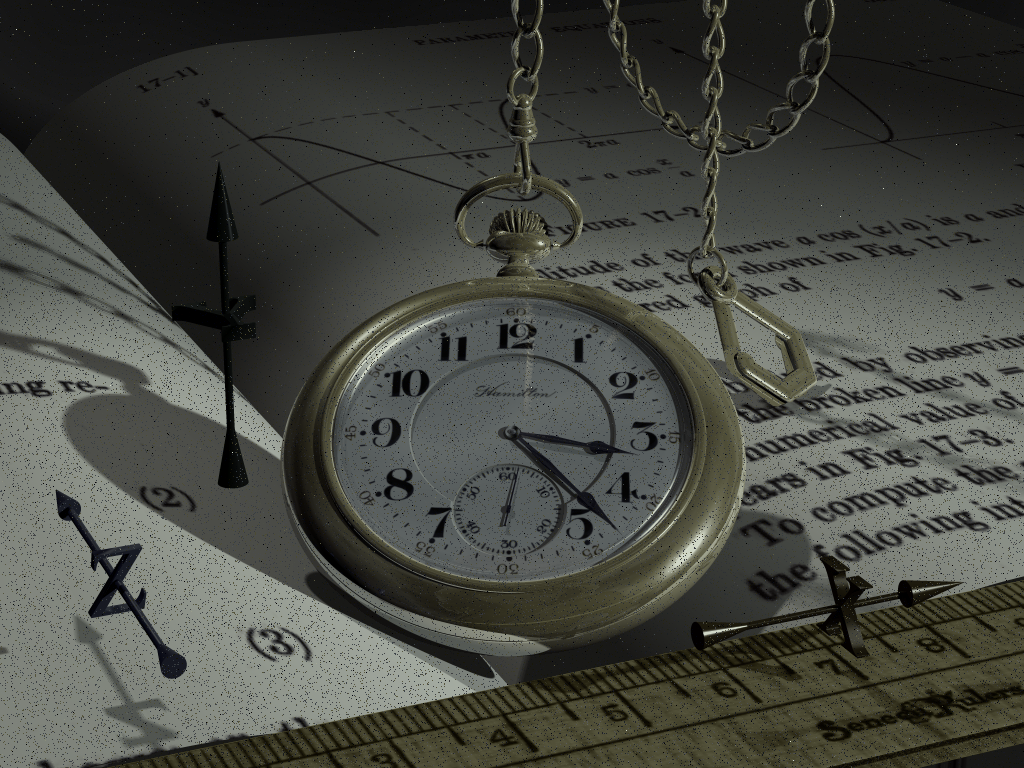} &
\includegraphics[width=0.128\linewidth]{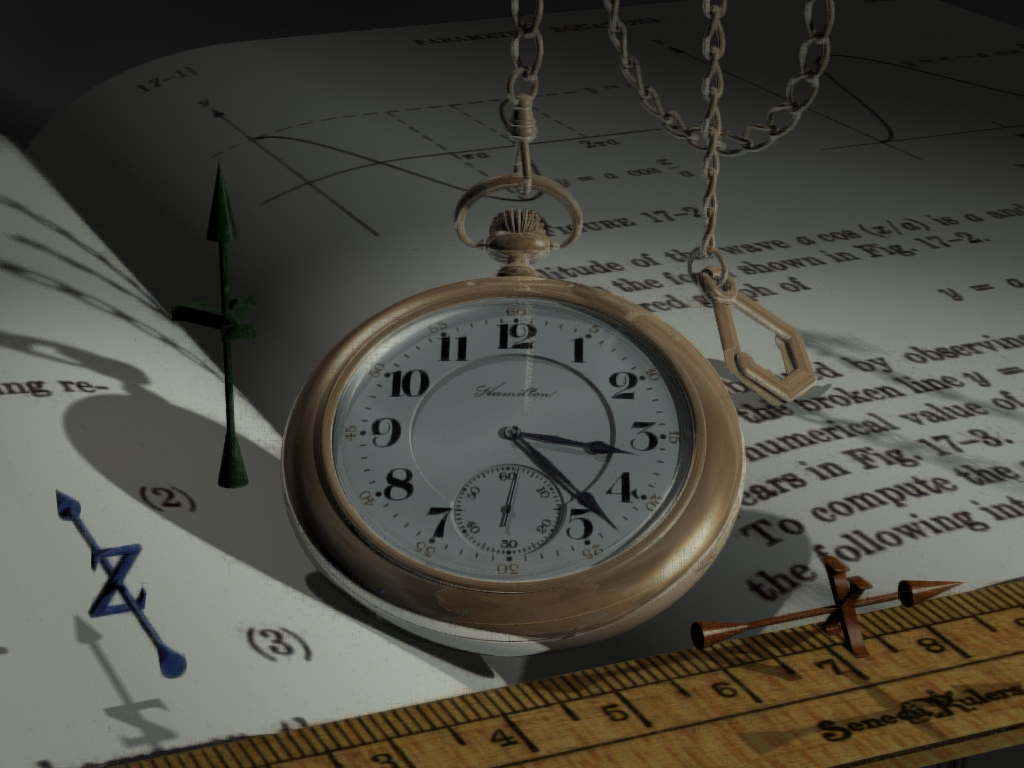} &
\includegraphics[width=0.128\linewidth]{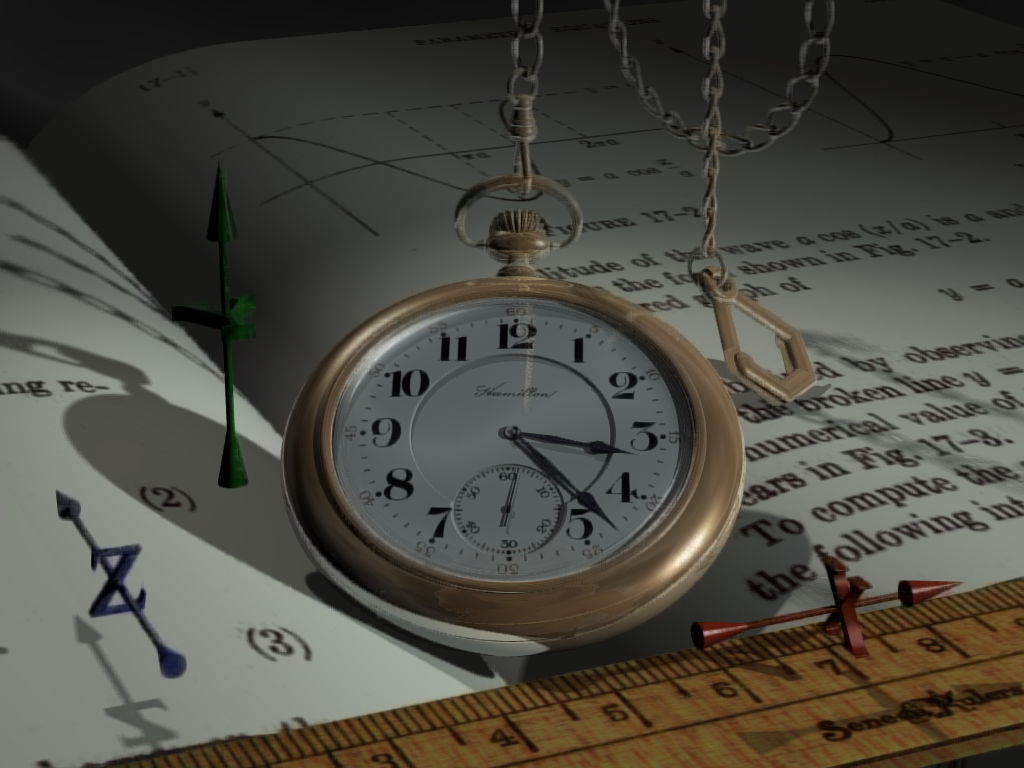} &
\includegraphics[width=0.128\linewidth]{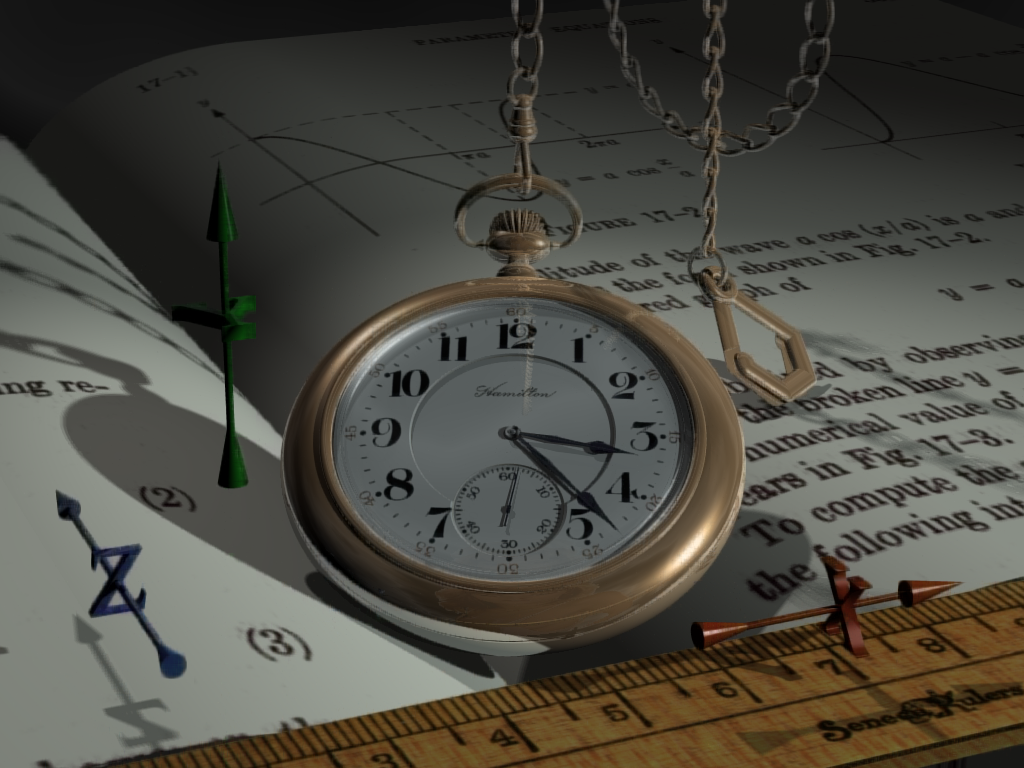} &
\includegraphics[width=0.128\linewidth]{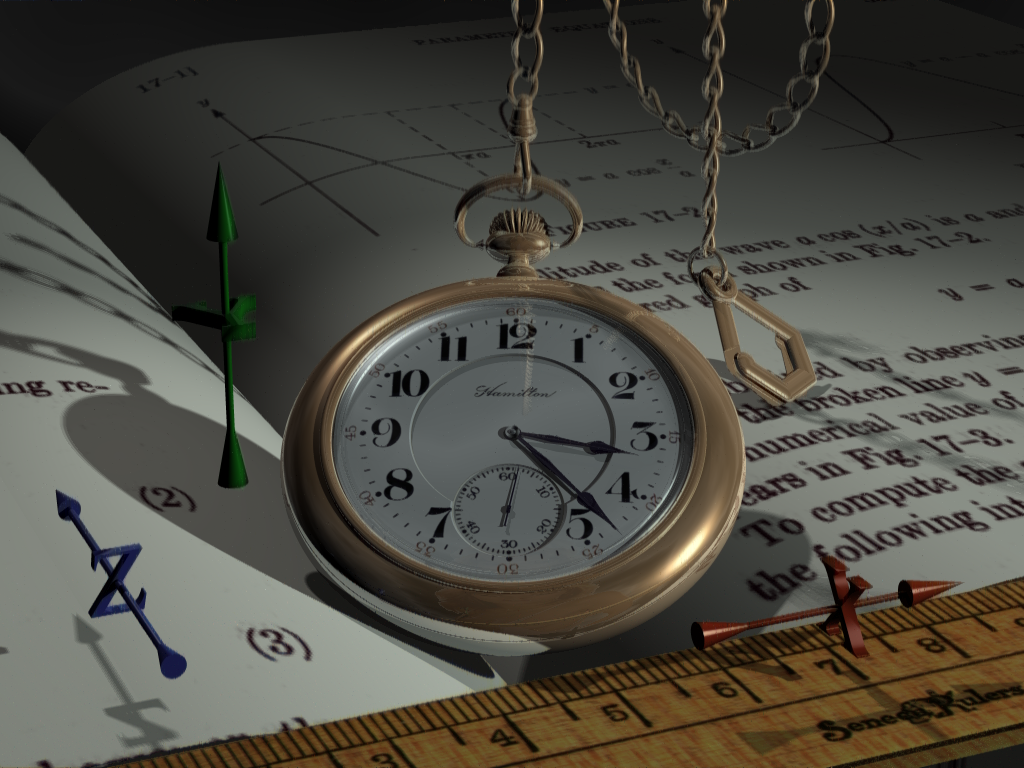} &
\includegraphics[width=0.128\linewidth]{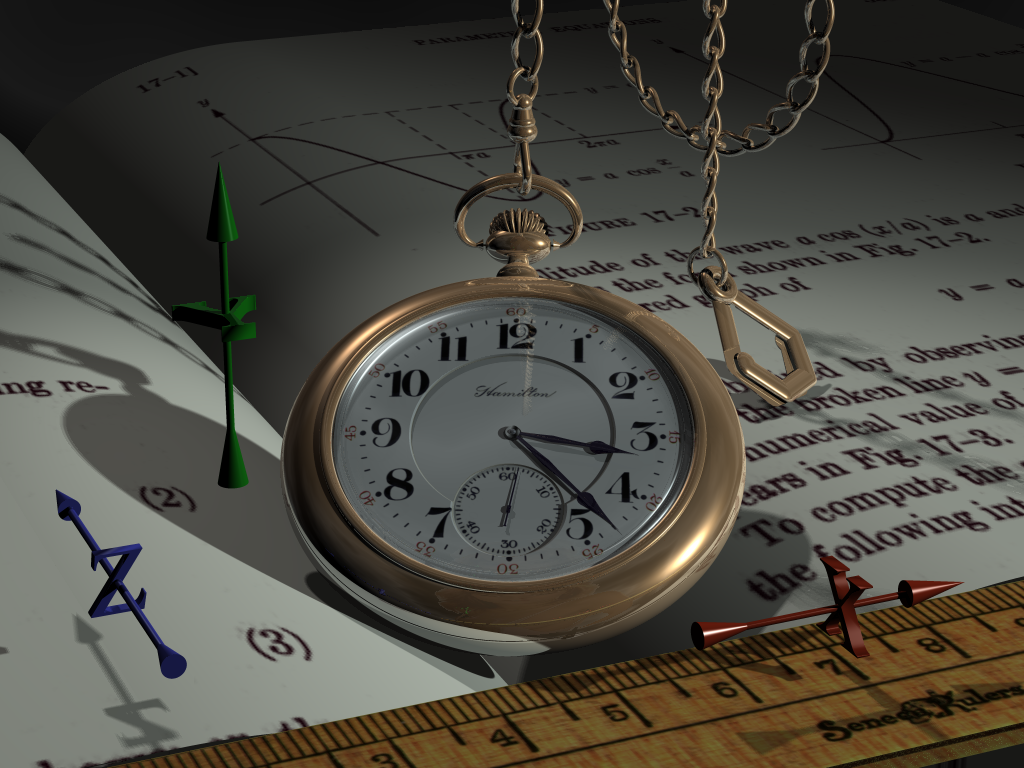} \\
&
\includegraphics[width=0.128\linewidth]{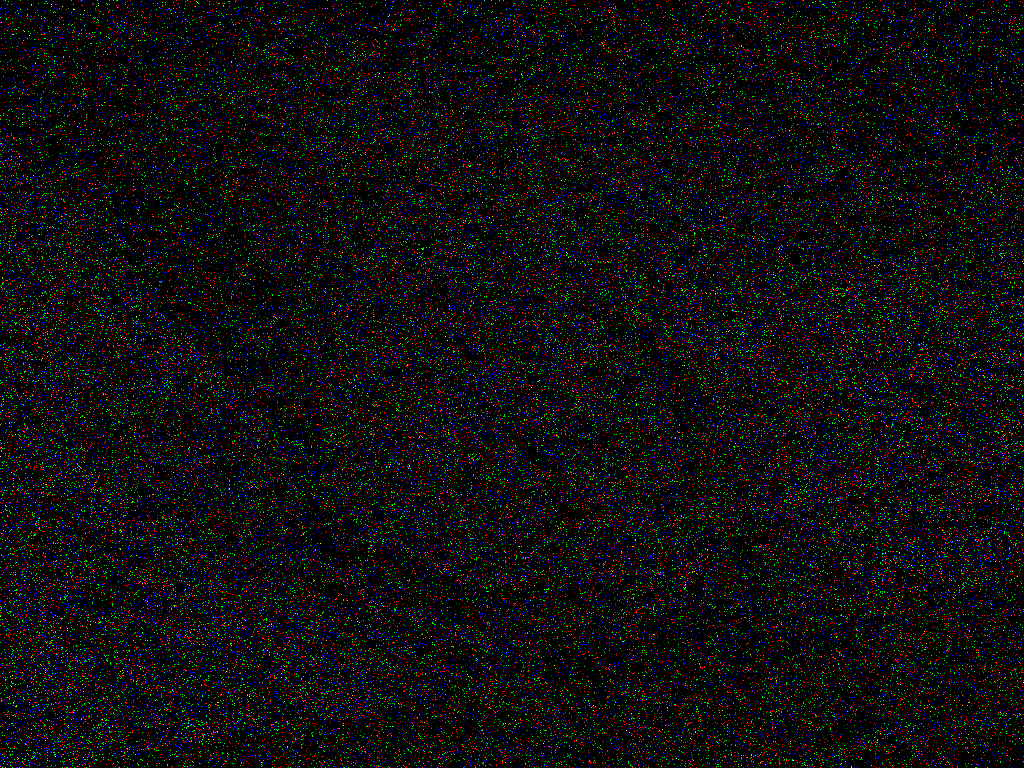} &
\includegraphics[width=0.128\linewidth]{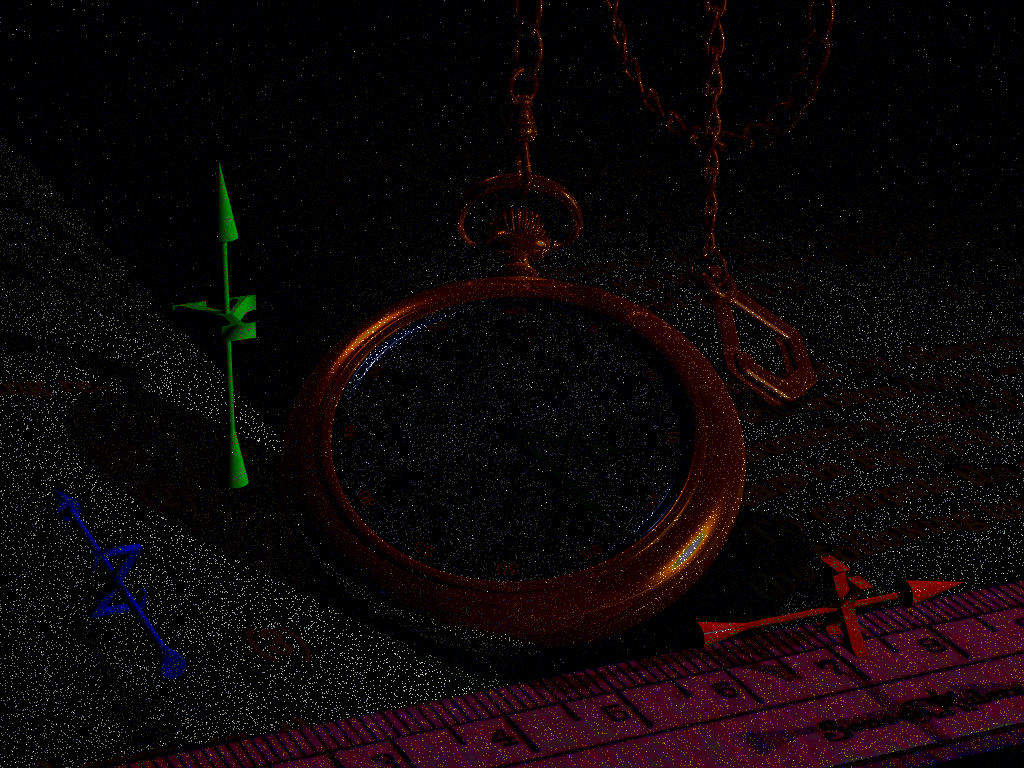} &
\includegraphics[width=0.128\linewidth]{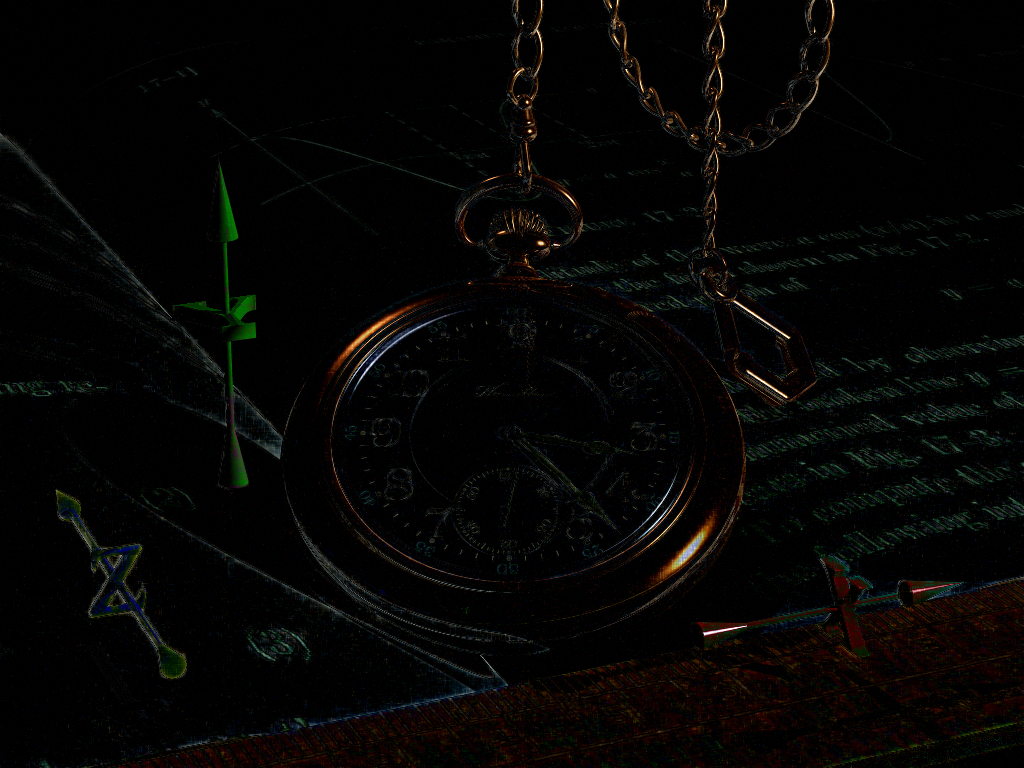} &
\includegraphics[width=0.128\linewidth]{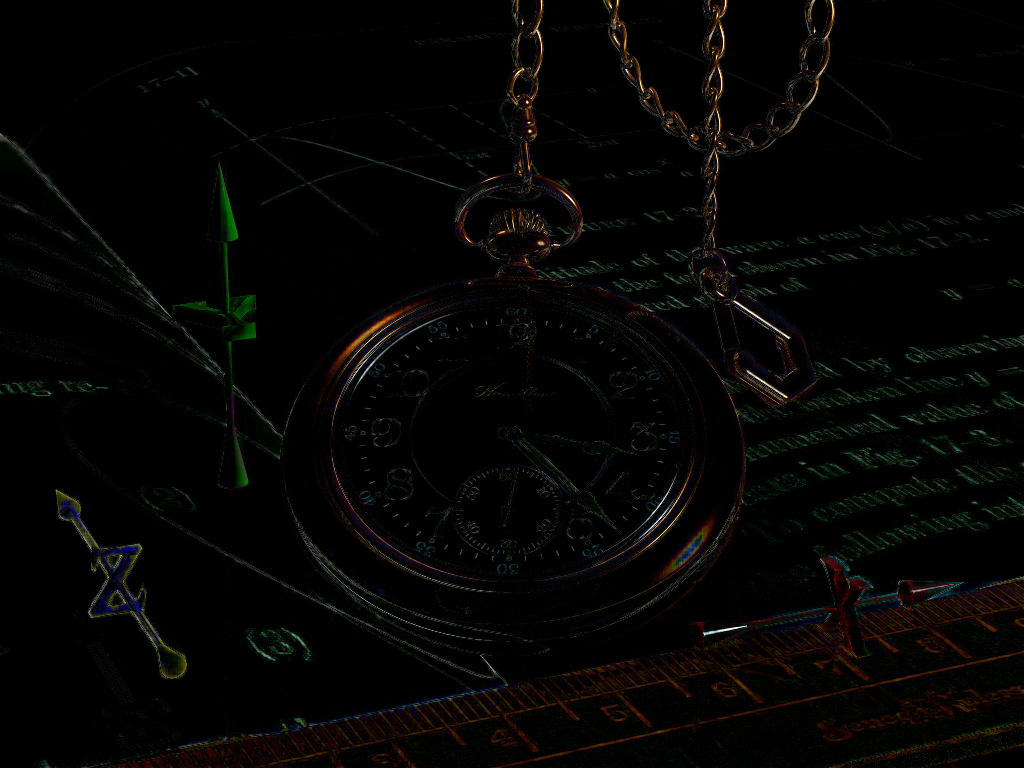} &
\includegraphics[width=0.128\linewidth]{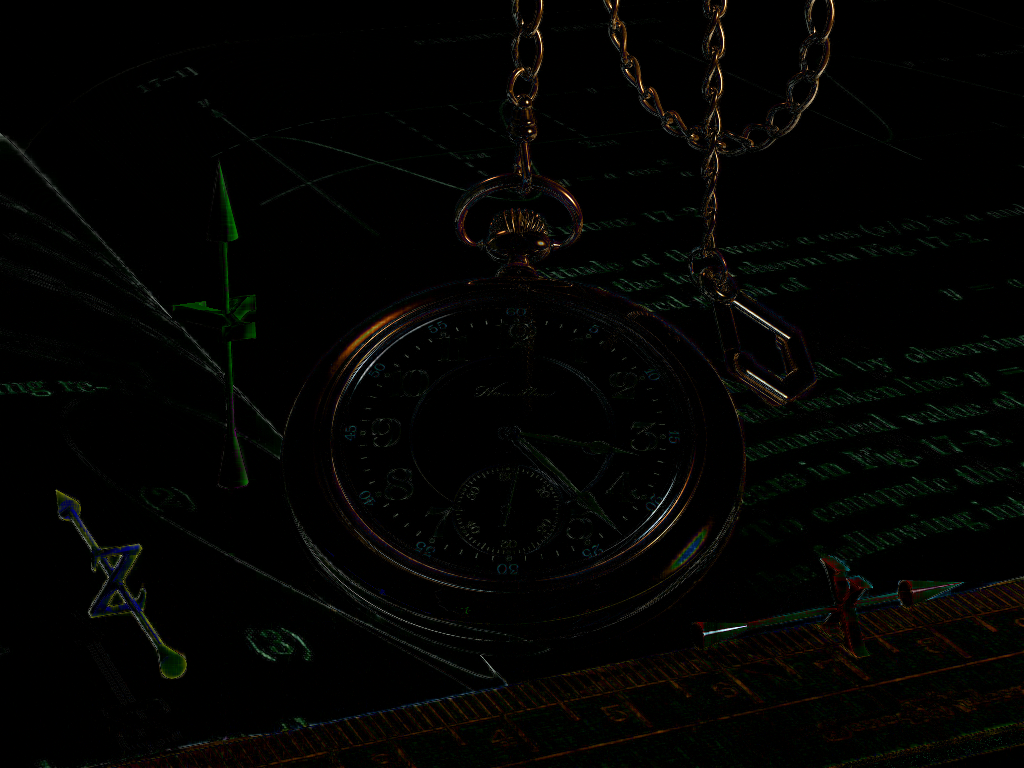} &
\includegraphics[width=0.128\linewidth]{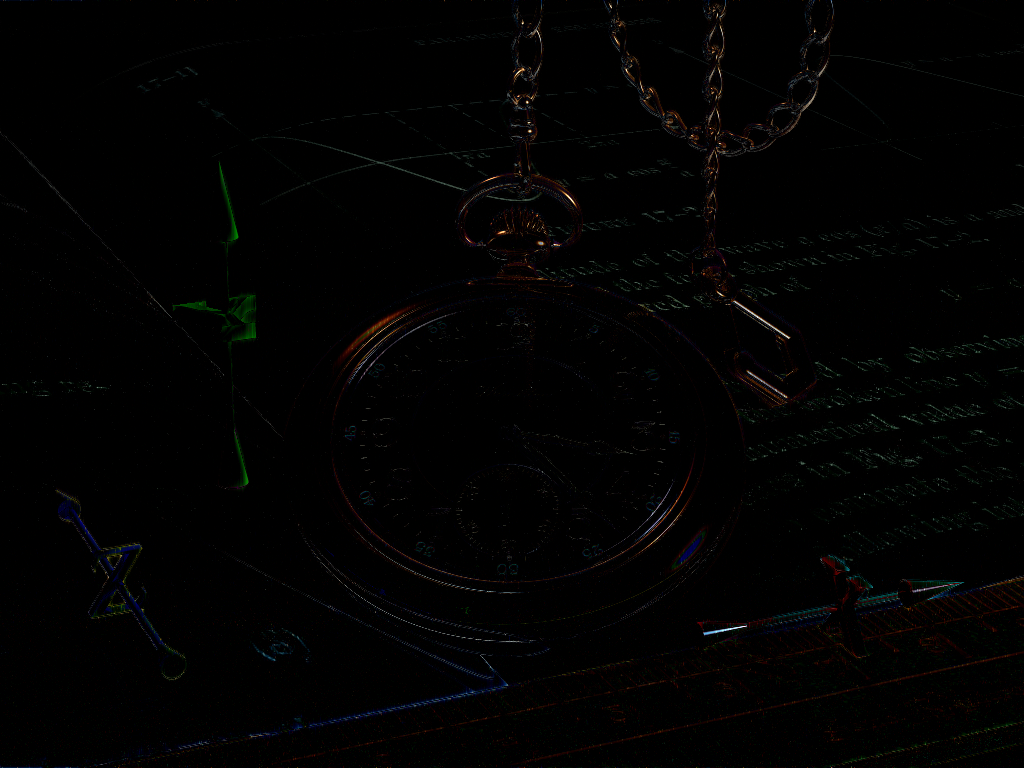} &
\includegraphics[width=0.128\linewidth]{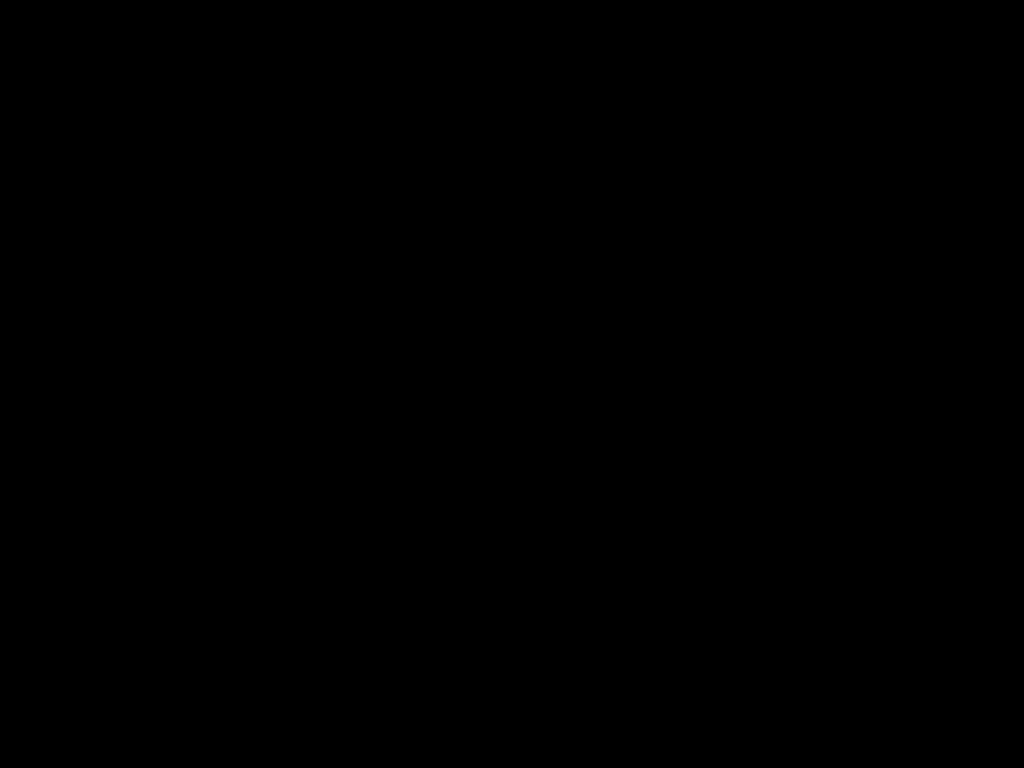} \\
\multirow{3}{*}{\rotatebox[origin=c]{90}{$\rho=10\%$}}&
\includegraphics[width=0.128\linewidth]{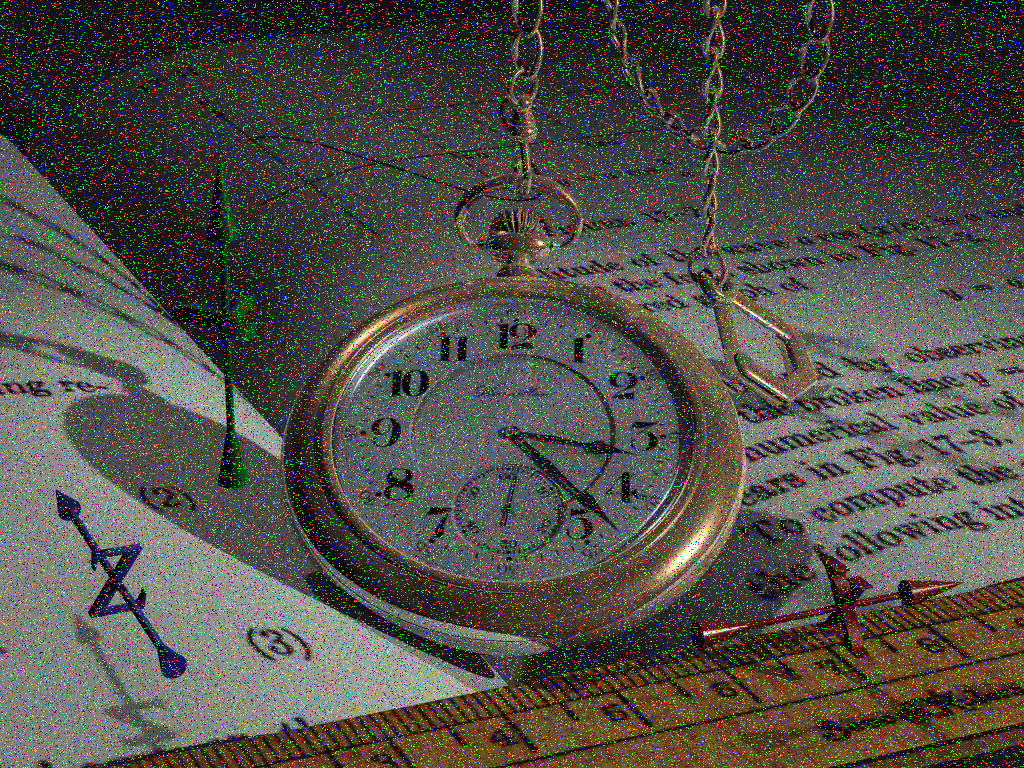} &
\includegraphics[width=0.128\linewidth]{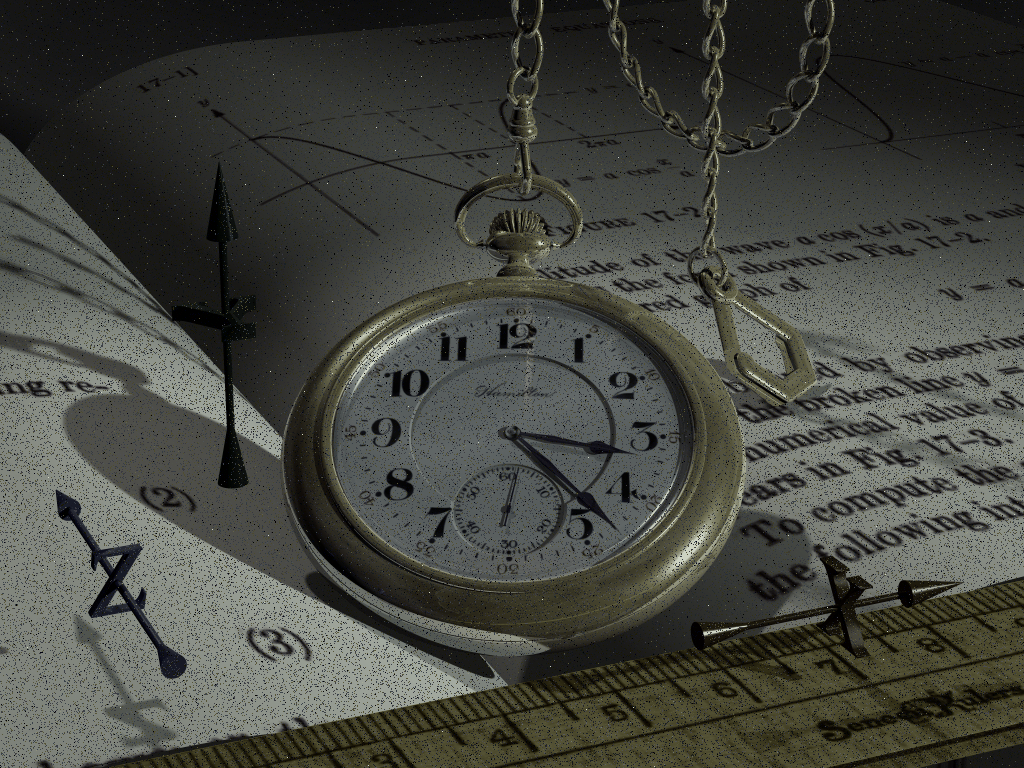} &
\includegraphics[width=0.128\linewidth]{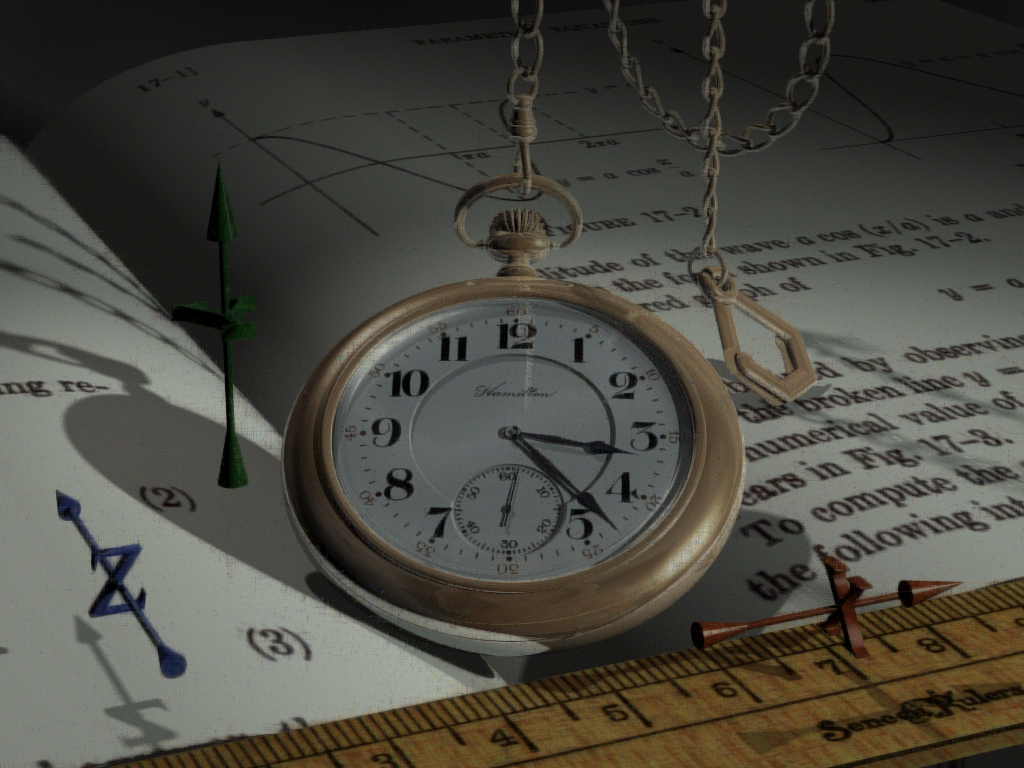} &
\includegraphics[width=0.128\linewidth]{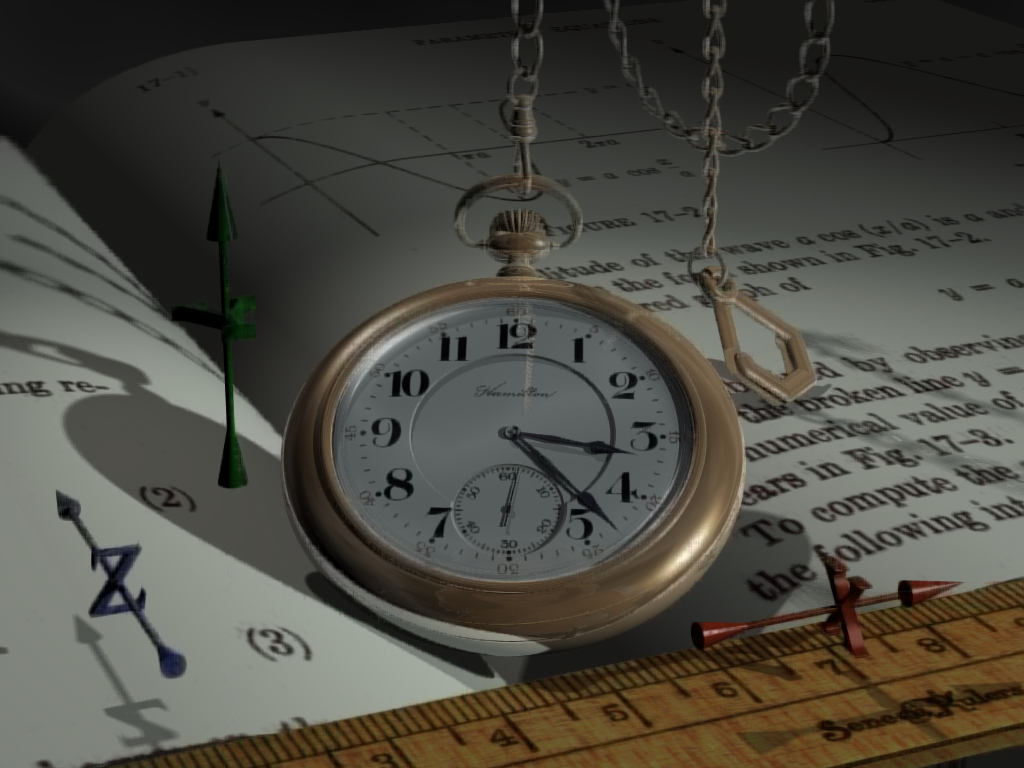} &
\includegraphics[width=0.128\linewidth]{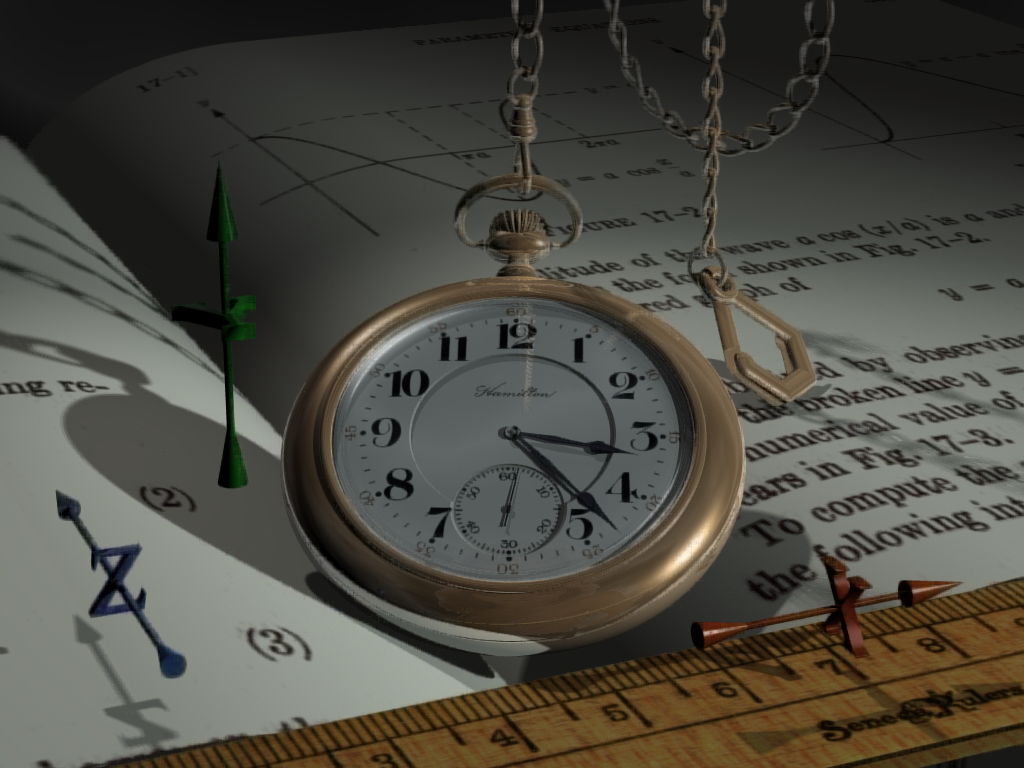} &
\includegraphics[width=0.128\linewidth]{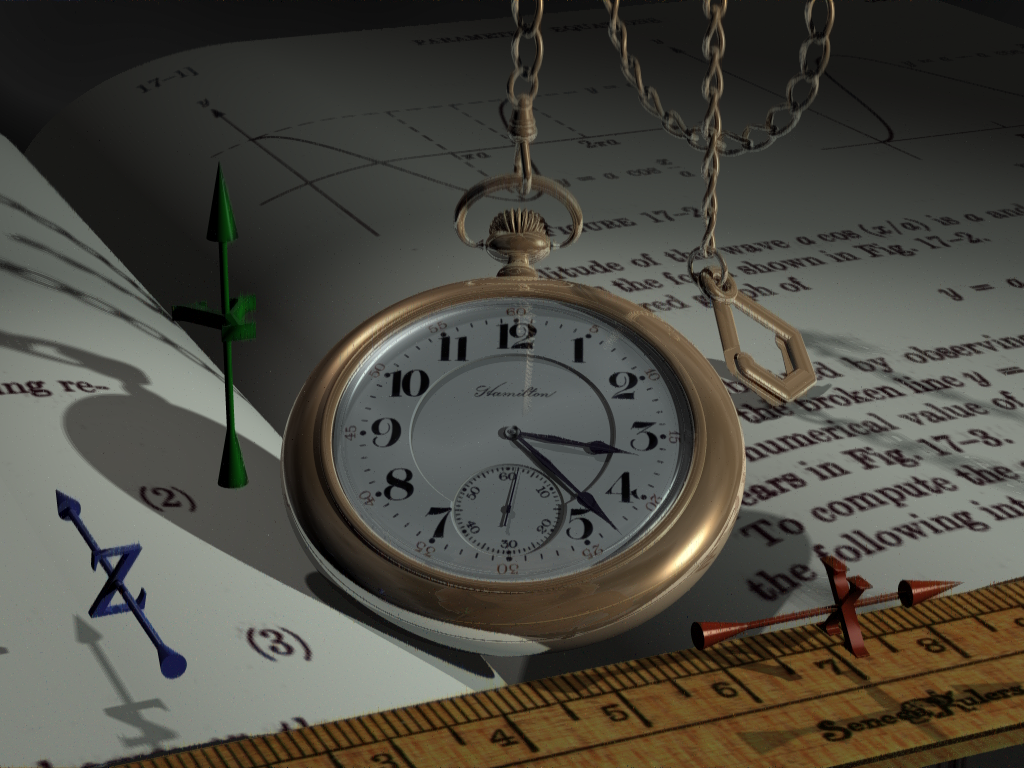} &
\includegraphics[width=0.128\linewidth]{figs/image222/watch_GT.png} \\
&
\includegraphics[width=0.128\linewidth]{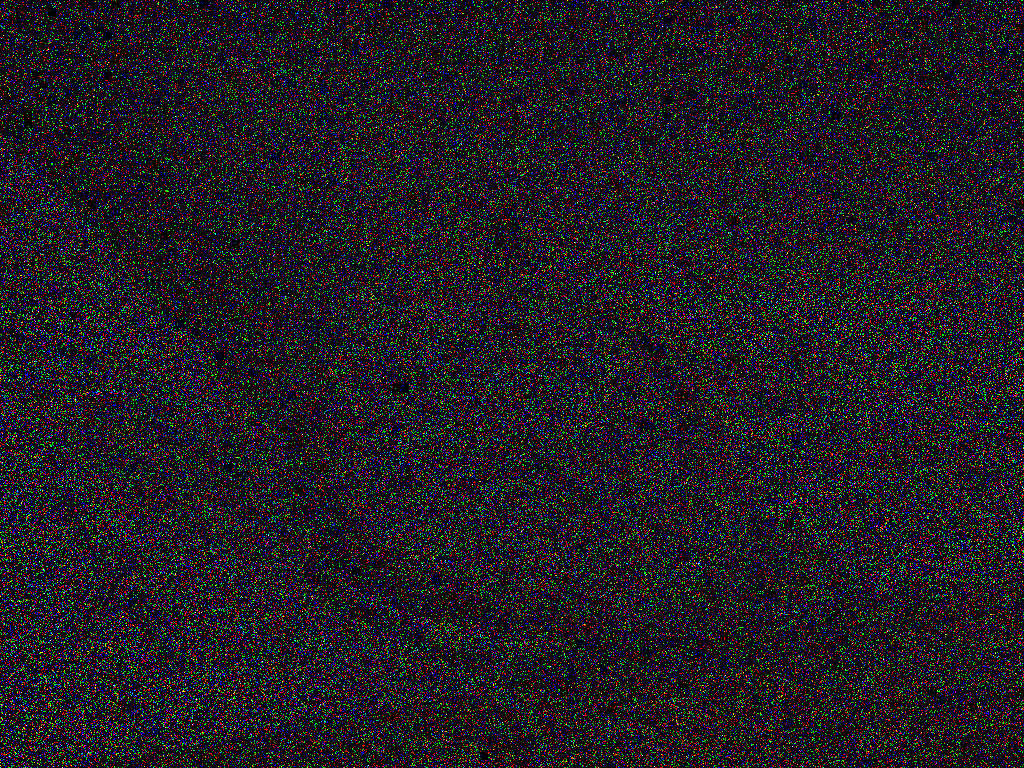} &
\includegraphics[width=0.128\linewidth]{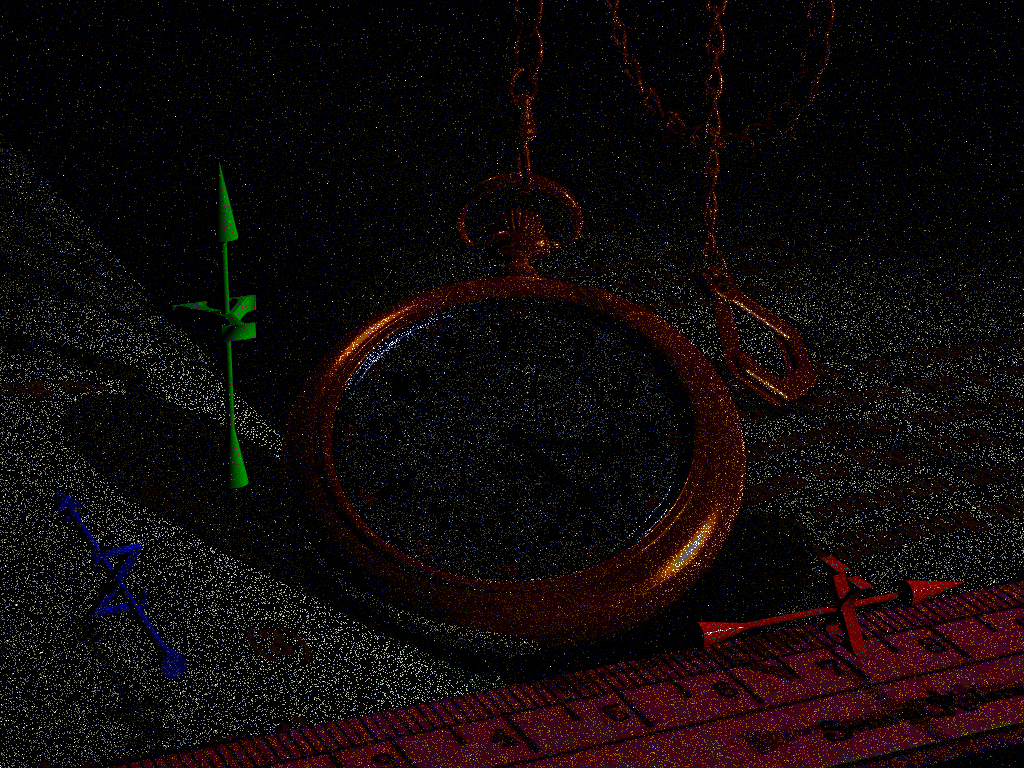} &
\includegraphics[width=0.128\linewidth]{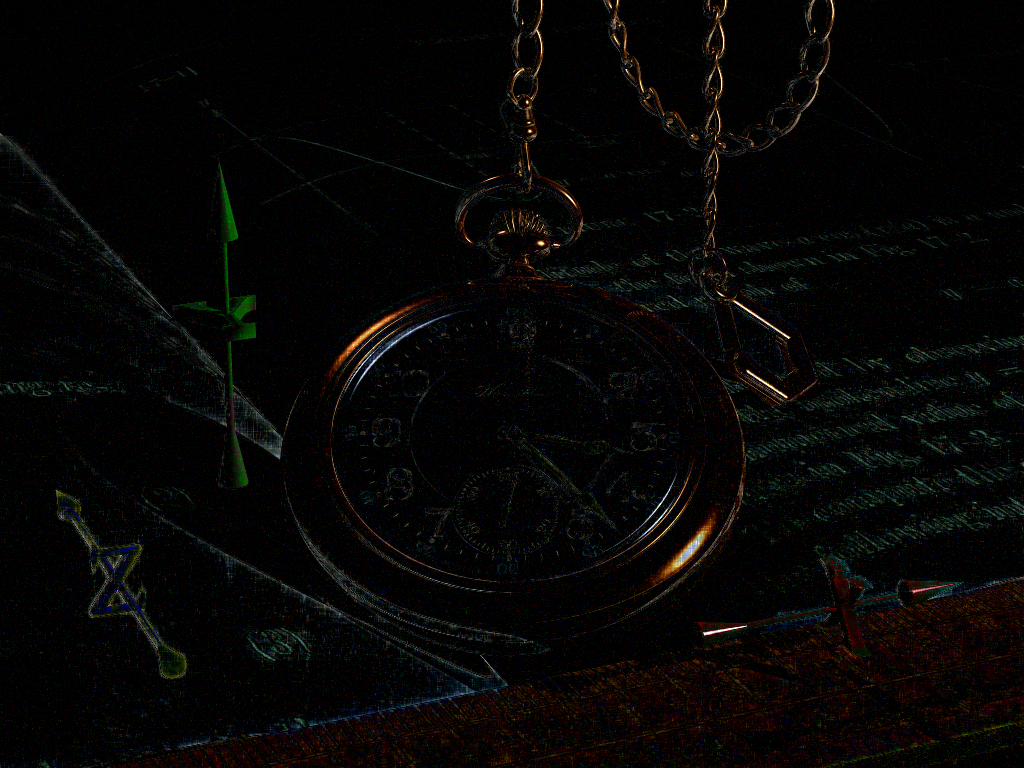} &
\includegraphics[width=0.128\linewidth]{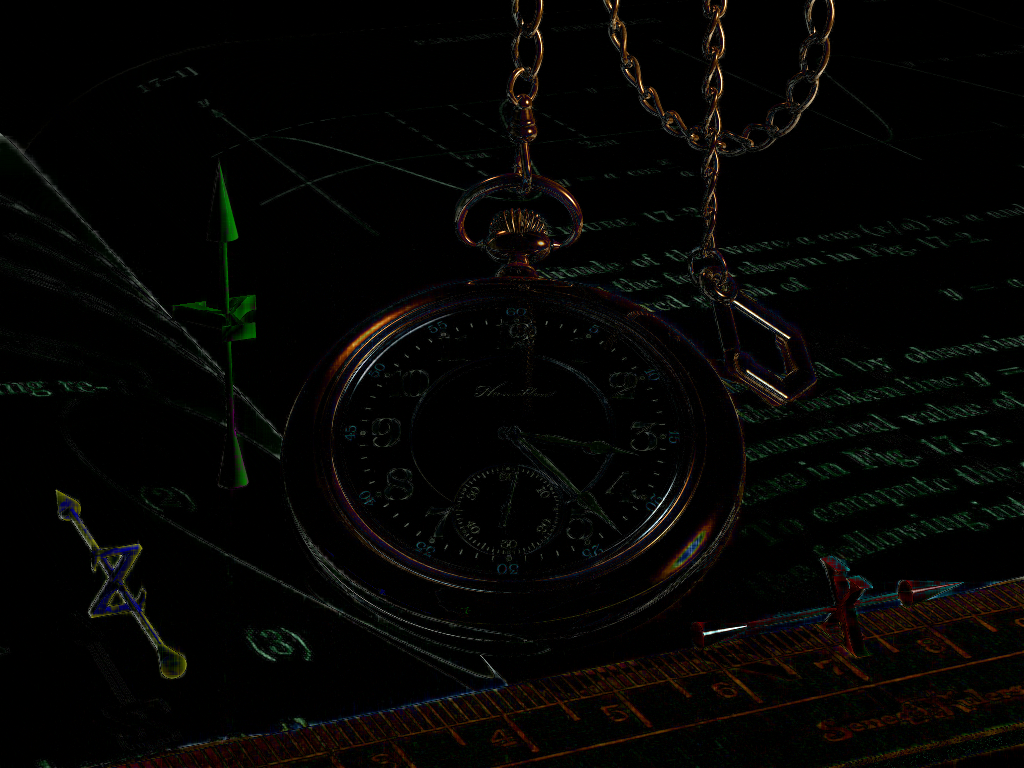} &
\includegraphics[width=0.128\linewidth]{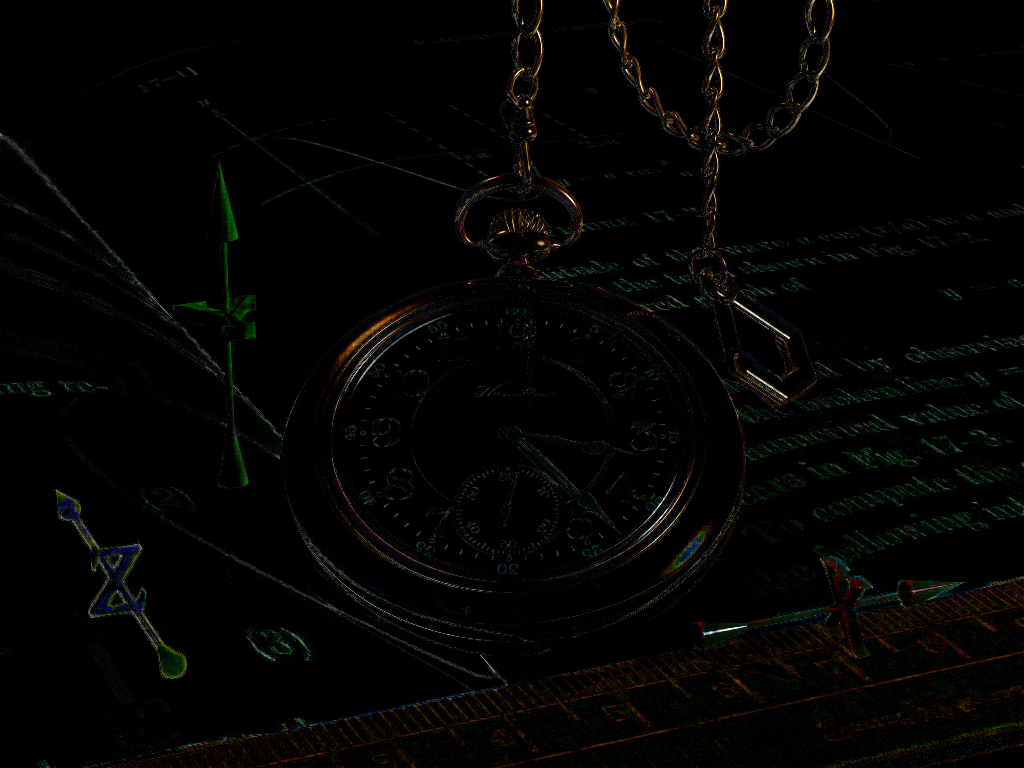} &
\includegraphics[width=0.128\linewidth]{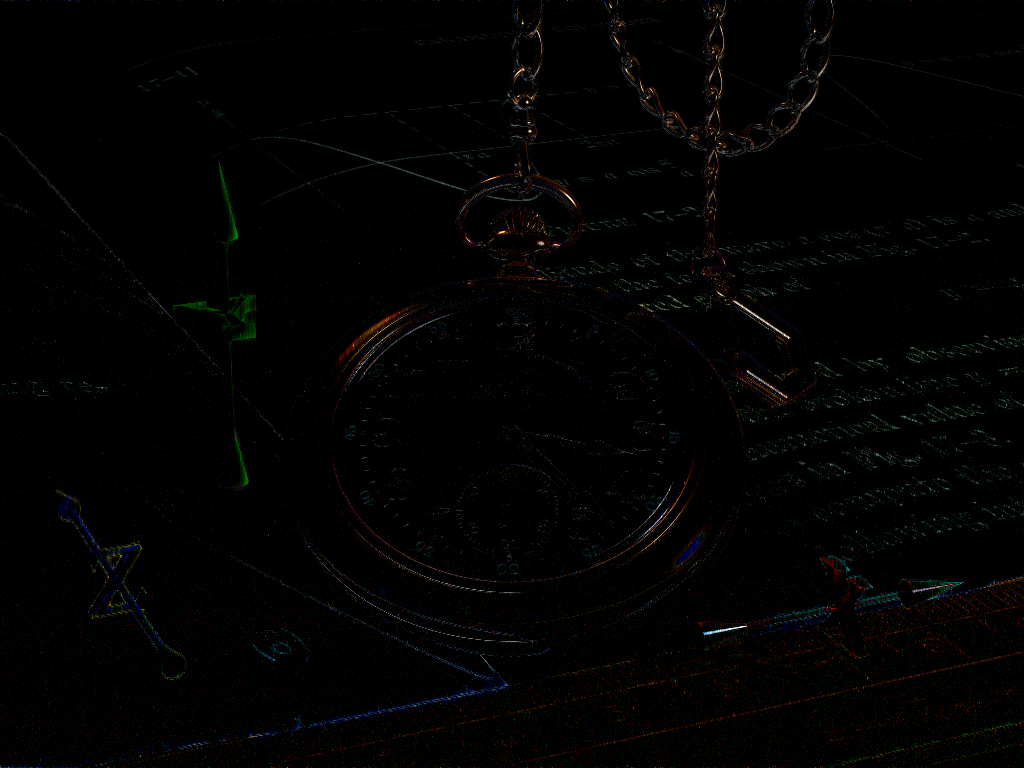} &
\includegraphics[width=0.128\linewidth]{figs/image222/watch_GT_rse.png} \\
\end{tabular}
\caption{The top four rows are the image recovery results and residual images on the image ``airplane'', and the bottom 4 rows are corresponding to the image ``watch''. }
\label{figure-rpca-image}
\end{figure}

\subsubsection{Background substraction}
\begin{figure}[!t]
\centering\scriptsize\setlength{\tabcolsep}{1pt}
\renewcommand\arraystretch{0.9}
\begin{tabular}{cccccc}
Observed& MRPCA \cite{candes2011robust} &SNN \cite{goldfarb2014robust}&TNN \cite{lu2019tensor} &DCTNN \cite{lu2019exact2}& FTNN\\
\includegraphics[width=0.15\linewidth]{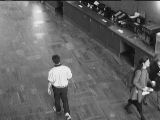} &
\includegraphics[width=0.15\linewidth]{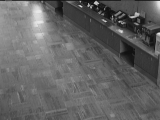} &
\includegraphics[width=0.15\linewidth]{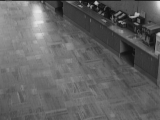} &
\includegraphics[width=0.15\linewidth]{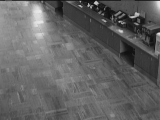} &
\includegraphics[width=0.15\linewidth]{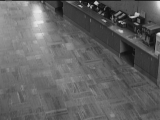} &
\includegraphics[width=0.15\linewidth]{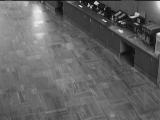} \\
&\includegraphics[width=0.15\linewidth]{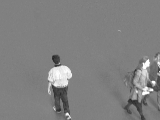} &
\includegraphics[width=0.15\linewidth]{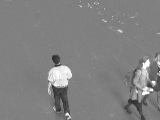} &
\includegraphics[width=0.15\linewidth]{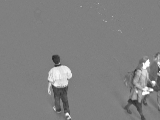} &
\includegraphics[width=0.15\linewidth]{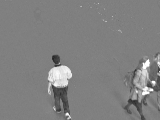} &
\includegraphics[width=0.15\linewidth]{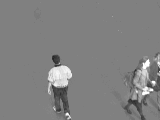} \\
Observed& MRPCA \cite{candes2011robust} &SNN \cite{goldfarb2014robust}&TNN \cite{lu2019tensor} &DCTNN \cite{lu2019exact2}& FTNN\\
\includegraphics[width=0.15\linewidth]{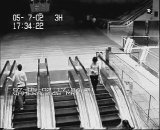} &
\includegraphics[width=0.15\linewidth]{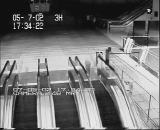} &
\includegraphics[width=0.15\linewidth]{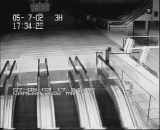} &
\includegraphics[width=0.15\linewidth]{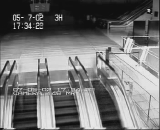} &
\includegraphics[width=0.15\linewidth]{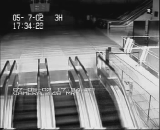} &
\includegraphics[width=0.15\linewidth]{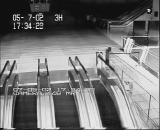} \\
&\includegraphics[width=0.15\linewidth]{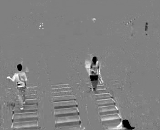} &
\includegraphics[width=0.15\linewidth]{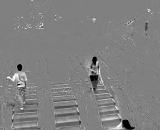} &
\includegraphics[width=0.15\linewidth]{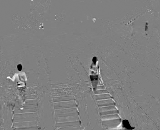} &
\includegraphics[width=0.15\linewidth]{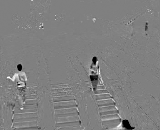} &
\includegraphics[width=0.15\linewidth]{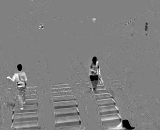} \\
Observed& MRPCA \cite{candes2011robust} &SNN \cite{goldfarb2014robust}&TNN \cite{lu2019tensor} &DCTNN \cite{lu2019exact2}& FTNN\\
\includegraphics[width=0.15\linewidth]{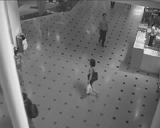} &
\includegraphics[width=0.15\linewidth]{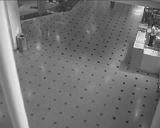} &
\includegraphics[width=0.15\linewidth]{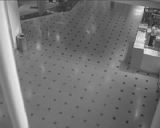} &
\includegraphics[width=0.15\linewidth]{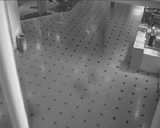} &
\includegraphics[width=0.15\linewidth]{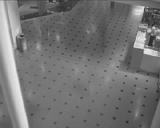} &
\includegraphics[width=0.15\linewidth]{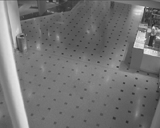} \\
&\includegraphics[width=0.15\linewidth]{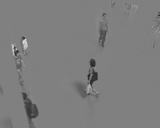} &
\includegraphics[width=0.15\linewidth]{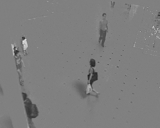} &
\includegraphics[width=0.15\linewidth]{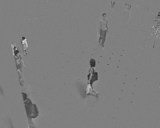} &
\includegraphics[width=0.15\linewidth]{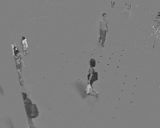} &
\includegraphics[width=0.15\linewidth]{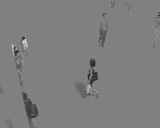} \\
Observed& MRPCA \cite{candes2011robust} &SNN \cite{goldfarb2014robust}&TNN \cite{lu2019tensor} &DCTNN \cite{lu2019exact2}& FTNN\\
\includegraphics[width=0.15\linewidth]{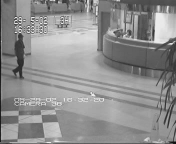} &
\includegraphics[width=0.15\linewidth]{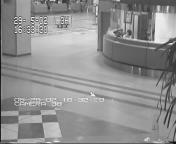} &
\includegraphics[width=0.15\linewidth]{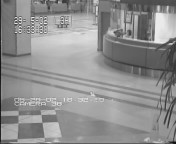} &
\includegraphics[width=0.15\linewidth]{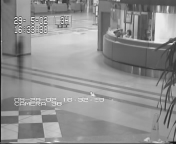} &
\includegraphics[width=0.15\linewidth]{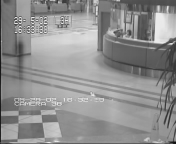} &
\includegraphics[width=0.15\linewidth]{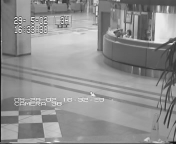} \\
&\includegraphics[width=0.15\linewidth]{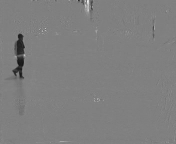} &
\includegraphics[width=0.15\linewidth]{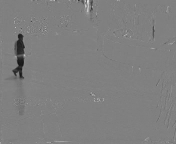} &
\includegraphics[width=0.15\linewidth]{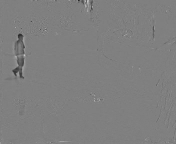} &
\includegraphics[width=0.15\linewidth]{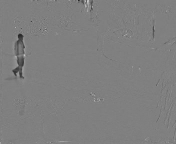} &
\includegraphics[width=0.15\linewidth]{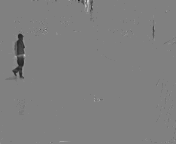} \\

\end{tabular}
\caption{Background substraction results by different methods. The left column lists one frame of the observed video. From top to bottom are respectively separation results, i.e., the background and the foreground, of the video ``Bootstrap1285'', ``Escalator2805'', ``ShoppingMall1535'', and  ``hall1368''. For better visualization, we add 0.5 to the foreground.}
\label{fig-rpca-video}
\end{figure}

Four video sequences, respectively named ``Bootstrap1285'', ``Escalator2805'', ``ShoppingMall1535'', and  ``hall1368'', are selected from Li's dataset\footnote{Data available at {http://vis-www.cs.umass.edu/~narayana/castanza/I2Rdataset/}}. After transforming the color frames to gray level ones, each video is of the size $130\times160\times40$. Results by all of the methods are displayed in Fig. \ref{fig-rpca-video}. We can see that our method and MRPCA perform well for the videos ``Bootstrap1285'' and ``ShoppingMall1535'', while some incorrectly extractions can be found in the foreground results by other three methods, the front desk in ``Bootstrap1285'' and the dot pattern of the ground in ``ShoppingMall1535'' for examples. For videos ``Escalator2805'' and ``hall1368'', all the methods incorrectly extract contents of the background to the foreground, more or less. Overall, the foregrounds extracted by our method are the purest.

\subsection{Discussions}
\subsubsection{Framelet setting}
In this part, taking the completion of MRI data (SR = $10\%$) as an example, we evaluate the performance of the proposed method with different Framelet transformation settings.
Firstly, including the piece-wise cubic B-spline (denoted as ``cubic''), we also adopted the Haar wavelet (denoted as ``Haar'') and the piece-wise linear B-spline (denoted as ``linear'') to generate the framelet transformation.
Meanwhile, we also set the decomposition levels from 1 to 5.
The quantitative metrics of the results obtained by the proposed method with different framelet settings are reported in Table \ref{MRIpara}.
From Table \ref{MRIpara}, we can find that the piece-wise cubic B-spline is the best choice.
As the decomposition level arise, the performance of the proposed method becomes better until level 5. Setting the level as 3 or 4 is a good choice.

\begin{table}[!t]
\renewcommand\arraystretch{0.9}\setlength{\tabcolsep}{4pt}\scriptsize\centering
\caption{The PSNR, SSIM and FSIM of the recovery results on the MRI data by the proposed method with different framelet settings. The \textbf{best} values  are highlighted by bolder fonts.}
\begin{tabular}{ccccccc}
\toprule

 Filters &Index & Level = 1 & Level = 2 & Level = 3 & Level = 4 & level = 5  \\ \midrule
 \multirow{3}{*}{Haar}
& PSNR & 21.176  & 23.327  & 24.183  & 24.366  & 24.372 \\
& SSIM & 0.537  & 0.647  & 0.680  & 0.685  & 0.685 \\
& FSIM & 0.755  & 0.801  & 0.817  & 0.821  & 0.821 \\       \midrule
\multirow{3}{*}{Linear}
& PSNR & 22.466  & 24.904  & 25.538  & 25.563  & 25.509 \\
& SSIM & 0.611  & 0.717  & 0.738  & 0.738  & 0.735 \\
& FSIM & 0.785  & 0.834  & 0.846  & 0.848  & 0.847 \\       \midrule
\multirow{3}{*}{Cubic}
& PSNR & 23.726  & 26.077  & \bf 26.287  & 26.104  & 25.970 \\
& SSIM & 0.673  & 0.761  & \bf 0.765  & 0.759  & 0.746 \\
& FSIM & 0.812  & 0.858  & \bf 0.863  & 0.862  & 0.858 \\       \bottomrule
\end{tabular}
\label{MRIpara}
\end{table}
\subsubsection{Convergency Behaviours}

\begin{figure}[!t]
\centering
\includegraphics[width=0.98\linewidth]{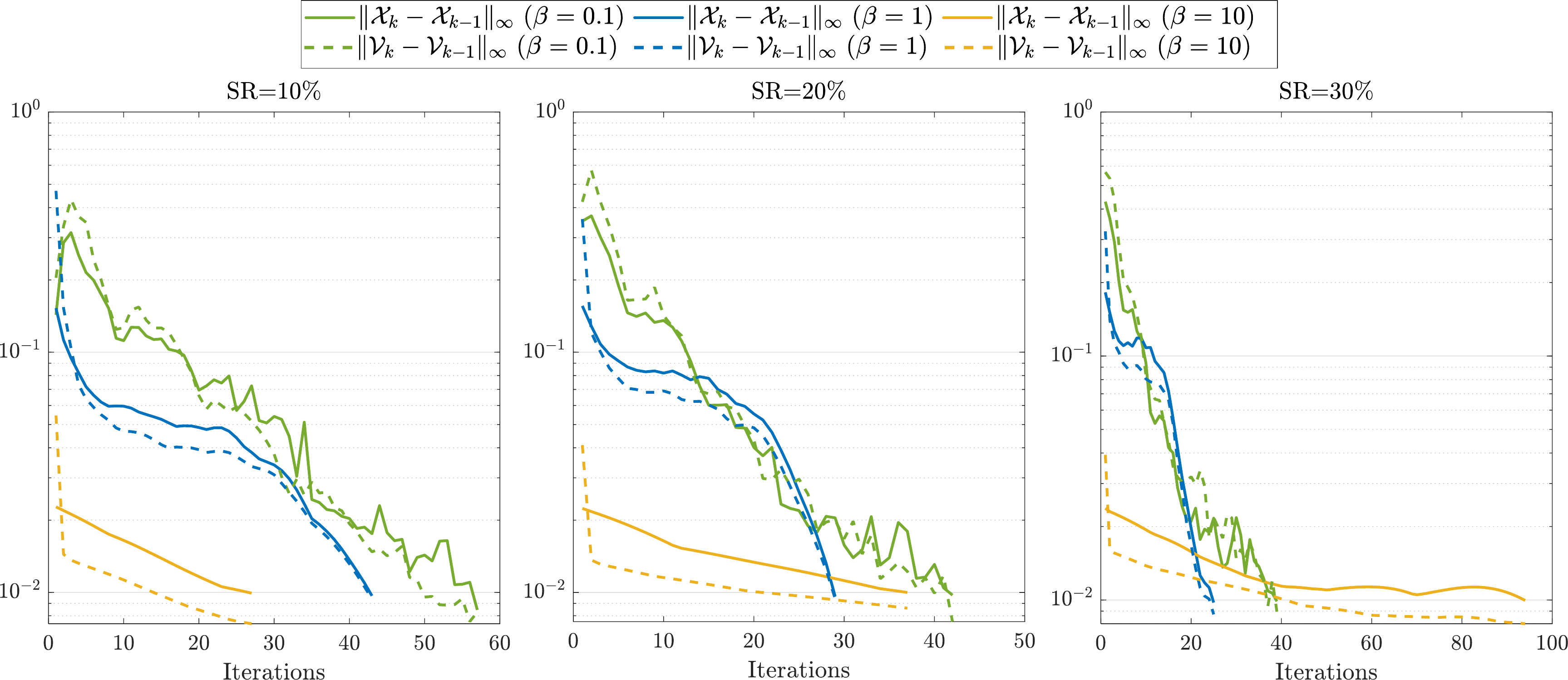}
\caption{The convergence behaviours of Algorithm \ref{alg}, with respect to different sampling rates and different $\beta$.}
\label{fig-convergency}
\end{figure}

Also, we take the completion of MRI data as an example to illustrate the convergency behaviours of our algorithm with respect to different sampling rates and different parameters. In the framework of ADMM, the parameter $\beta$, which is brought in by the augmented Lagrangian function, mainly affects the convergency behaviour of our method. Thus, we test our algorithm with $\beta = 10^{-1},1,10$. We plot $\|\mathbf{\mathcal{V}}^{k+1}-\mathbf{\mathcal{V}}^{k}\|_\infty$ and $\| \mathbf{\mathcal{X}}^{k+1} -\mathbf{\mathcal{X}}^{k} \|_\infty$ of each iteration in Fig. \ref{fig-convergency}. It can be seen that when $\beta = 10^{-1}$ and 1 our algorithm steadily converges. Although the behaviour of $\| \mathbf{\mathcal{X}}^{k+1} -\mathbf{\mathcal{X}}^{k} \|_\infty$ is not that stable when $\beta = 10$, our algorithm also converges rapidly.

\section{Conclusions}\label{Sec:Con}
In this paper, we propose to replace the Fourier transform by the framelet in the t-SVD framework.
Then, we formulate the framelet representation of the tensor multi-rank and tensor nuclear norm.
A low-rank tensor completion model and a tensor robust principal component analysis model are proposed by minimizing the framelet based tensor nuclear norm. We develop ADMM based algorithms to solve these convex models with guaranteed convergence.
We compare the performance of the proposed method with state-of-the-art methods via numerical experiments on the magnetic resonance imaging data, videos, color images, and multispectral images.
Our method outperforms many state-of-the-art methods quantitatively and visually.

{\footnotesize
\bibliographystyle{ieeetran}
\bibliography{ref}
}
\end{document}